\documentclass[preprint2]{pp7}
\usepackage{soul}
\usepackage{xcolor}
\usepackage[switch]{lineno}
\usepackage[utf8]{inputenc}

\input{pp7.h}

\begin{document}

\title{\textbf{\LARGE Organic chemistry in the first phases of Solar-type protostars }}

\author {\textbf{\large Cecilia Ceccarelli }}
\affil{\small\it Univ. Grenoble Alpes, CNRS, IPAG, 38000 Grenoble, France}
\author {\textbf{\large Claudio Codella }}
\affil{\small\it INAF, Osservatorio Astrofisico di Arcetri, 50125 Firenze, Italy}
\affil{\small\it Univ. Grenoble Alpes, CNRS, IPAG, 38000 Grenoble, France}
\author {\textbf{\large Nadia Balucani }}
\affil{\small\it Dip. di Chimica, Biologia e Biotecnologie,  Universit\`{a} di Perugia, 06123 Perugia, Italy}
\author {\textbf{\large Dominique Bockelée-Morvan }}
\affil{\small\it LESIA, Obs. de Paris, PSL Research University, CNRS, Sorbonne Univ., Univ. Paris Diderot, F-92195 Meudon, France}
\author {\textbf{\large Eric Herbst }}
\affil{\small\it Dept. of Chemistry, University of Virginia, PO Box 400319, Charlottesville, VA 22904, USA}
\author {\textbf{\large Charlotte Vastel }}
\affil{\small\it IRAP, Universit\'e de Toulouse, CNRS, CNES, UPS, (Toulouse), France}
\author {\textbf{\large Paola Caselli}}
\affil{\small\it Max-Planck-Institut f\"ur extraterrestrische Physik (MPE), 85748 Garching, Germany}
\author {\textbf{\large C\'ecile Favre }}
\affil{\small\it  Univ. Grenoble Alpes, CNRS, IPAG, 38000 Grenoble, Francee}
\author {\textbf{\large Bertrand Lefloch }}
\affil{\small\it  Univ. Grenoble Alpes, CNRS, IPAG, 38000 Grenoble, France}
\author {\textbf{\large Karin \"Oberg}}
\affil{\small\it Center for Astrophysics — Harvard \& Smithsonian 60 Garden St., Cambridge, MA 02138, USA}
\author {\textbf{\large Satoshi Yamamoto }}
\affil{\small\it Department of Physics, The University of Tokyo, Tokyo 113-0033, Japan}

\begin{abstract}
\baselineskip = 11pt
\leftskip = 1.5cm 
\rightskip = 1.5cm
\parindent=1pc
{\small 
  Planetary systems such as our own are formed after a long process where matter condenses from diffuse clouds to stars, planets, asteroids, comets and residual dust, undergoing dramatic changes in  physical and chemical state in less than a few million years.
  Several studies have shown that the chemical composition during the early formation of a Solar-type planetary system is a powerful diagnostic to track the history of the system itself.
  Among the approximately 270 molecules so far detected in the ISM, the so-called interstellar complex organic molecules (iCOMs) are of particular interest both because of their evolutionary diagnostic power and because they might be potential precursors of biomolecules, which are at the basis of terrestrial life.  
  This Chapter focuses on the evolution of organic molecules during the early stages of a Solar-type planetary system, represented by the prestellar, Class 0/I and  protoplanetary  disk\index{Protoplanetary disks} phases, and compares them with what is observed presently in Solar System comets.
  Our twofold goal is to review the processes at the base of organic chemistry during Solar-type star formation and, in addition, to possibly provide constraints on the early history of our own planetary system.
 \\~\\~\\~}
\end{abstract}  


\cleardoublepage 
\section{\textbf{INTRODUCTION: A HISTORICAL PERSPECTIVE}}\label{sec1:intro}

Since the discovery of the first diatomic molecule, CH, in the interstellar medium (ISM) at the end of the 1930's, new molecular discoveries have been made steadily over time.
Most of the ISM molecules have been discovered by the use of telescopes in the radio, millimeter (mm) and far-infrared (FIR)  wavelengths, where molecules have rotational and ro-vibrational lines.
As of the present, the number of identified ISM molecules is about 280, of which about 40\%  contain at least six atoms\footnote{\url{https://cdms.astro.uni-koeln.de/classic/molecules}.}.
Importantly, all these ''large'' molecules possess at least one carbon atom, a fact that already suggests a potential link between interstellar chemistry and terrestrial life.
Indeed, in the zoo of detected species, they are considered special and are called "interstellar Complex Organic Molecules" \citep[iCOMs or COMs:][]{Ceccarelli2017,Herbst2009}.   
The lower case "i" is meant to emphasize that the adjective "complex" only applies in the context of the ISM.
Table \ref{tab:sec1-iCOMs} lists the names and the chemical formulae of the iCOMs most commonly found in Solar-type star-forming regions.

Since the first discovery of formamide (NH$_2$CHO) by \cite{Rubin1971}, iCOMs have aroused the curiosity of astronomers.
The discovery that iCOMs are numerous and very abundant in massive star forming regions \citep{blake1987} started to represent a serious challenge for astrochemists \citep[e.g.][]{Charnley1992}.
Eventually, the discovery of numerous and abundant iCOMs in Solar-type protostars\index{Protostars} \citep{Cazaux2003} suggested a potential link between  interstellar chemistry and the emergence of life on Earth.
Understanding how iCOMs are formed and whether they are passed on to nascent planets, comets\index{Comets} and asteroids then became a major topic of astronomy.

At present, two major iCOM formation paths have been evoked in the literature, reviewed in Sec. \ref{sec2:chemistry}.
The first one, of which a first version was in vogue until about 2005, postulates that during the cold and quiescent phases of molecular cloud and prestellar core, the interstellar dust grains are enveloped by ice mantles which sublimate when a protostar appears at the centre and warms up the surrounding dust \citep{Millar1991,Charnley1992}. 
The mantle components are ejected into the gas-phase and undergo reactions that form new and more complex molecules.
However, in 2005, a breakthrough occurred with an experiment that showed that the electron recombination of protonated methanol breaks the species into smaller species \citep{Geppert2006} rather than producing methanol, which was the assumption of the epoch.
In addition, in 2004, theoretical chemical computations showed that an abundant and common iCOM, methyl formate (HCOOCH$_3$), could not be formed by any known gas-phase reaction \citep{horn2004}.
These negative results proved to be the death of the theory for most iCOMs.

\begin{table*}
    \begin{center}
    \caption{List of the iCOMs most commonly detected in Solar-type star-forming regions together with their binding energies (BE) and pre-exponential factors $\nu_{des}$, when available (see \S ~\ref{subsec:sec2-mantle-desorption}). 
    References provide the source of the BE and $\nu_{des}$ values.}
    \label{tab:sec1-iCOMs}
    \begin{tabular}{l|c|c|c|l}
        \hline
        Name & Formula & BE & $\nu_{des}$ & Ref.$^a$\\
             &         & (K) & ($\times 10^{12}$ s$^{-1}$) & \\
        \hline
        Methanol & CH$_3$OH & 3770--8618 & $\sim 3\times 10^5$ & 1, 2\\
        Formic acid$^b$ & HCOOH & 5382--10559 & $\sim 10^5-10^6$ & 1\\
        Acetaldehyde & CH$_3$CHO & 2809--6038 & $\sim 1\times 10^5$ & 3\\
        Methyl formate & HCOOCH$_3$ & -- & -- & -- \\
        Glycolaldehyde & HCO(CH$_2$)OH & -- & -- & -- \\
        Acetic acid & CH$_3$COOH & -- & -- & -- \\
        Dimethyl ether & CH$_3$OCH$_3$ & -- & -- & -- \\
        Acetone & CH$_3$COCH$_3$ & -- & -- & -- \\
        Ethanol & CH$_3$CH$_2$OH & $\sim 7000$ & $\sim 4\times 10^6$ & 4\\
        Propanal & CH$_3$CH$_2$CHO & -- & -- & -- \\
        Ethylene glycol & (CH$_2$OH)$_2$ & $\sim 7100$ & $\sim 10^5-10^6$ & 5 \\
        Methyl cyanide & CH$_3$CN & 4745--7652 & $\sim 2\times 10^5$  & 1 \\
        Ethyl cyanide & CH$_3$CH$_2$CN & -- & -- & -- \\
        Cyanoacetylene & HC$_5$N & -- & -- & -- \\
        Formamide & NH$_2$CHO & 5793--10960 & $\sim 4\times10^6$ & 1, 6 \\
        \hline
    \end{tabular}
    \end{center}
    {\small NOTES: 
    $^a$ References: 1, \citet{Ferrero2020}; 2, \citet{Minissale2022}; 3, \citet{Ferrero2022}; 4, Private communication from J. Enrique-Romero; 5, \citet{Bianchi2022-svs13}; 6, \citet{Chaabouni2018}.
    $^b$ Formic acid is included in the list, although it is not an iCOM by strict definition, because it contains three heavy atoms and is an important organic molecule.}
\end{table*}

The view of several astrochemists became that, if iCOMs cannot form in the gas-phase, they must be synthesized on the grain surfaces.
After all, the most abundant molecule, H$_2$, is known to be a grain-surface product \citep[e.g.][]{vandeHulst1946,Hollenbach1970,vidali2013}.
Following this idea, \cite{Garrod2006} developed a new theory, in which grain-surface chemistry can produce iCOMs during the warm-up phase (see Sec. \ref{sec2:chemistry} for more detail).
Several laboratory experiments starting in the 90s had already shown that, if an ice with a composition similar to that observed in the ISM by IR absorption observations \citep[see the review by][]{Boogert2015} is illuminated by UV photons \citep{schutte1992,bernstein1995,bernstein1999} or energetic particles \citep{Strazzulla1997,Palumbo1999}, several very complex molecules are formed on the ice.
The most recent experiments show that even amino acids and nucleobases can be found after irradiation by UV photons or energetic particle irradiation \citep[e.g.][]{munoz-caro2002,DeMarcellus2015,Oba2019}.
Following this idea, several studies have been carried out to better constrain the pathways to iCOMs induced by photolysis and radiolysis \citep[see e.g. the review by][]{Oberg2016}.
Finally, recent experimental studies have shown that  iCOMs and even glycine could be formed via non-energetic processes on CO-rich ices irradiated by thermal H atoms \citep[e.g.][]{Qasim2019,Ioppolo2021,Fedoseev2022-acetaldehyde}.
\begin{figure*}[th]
 \epsscale{1.6}
 \plotone{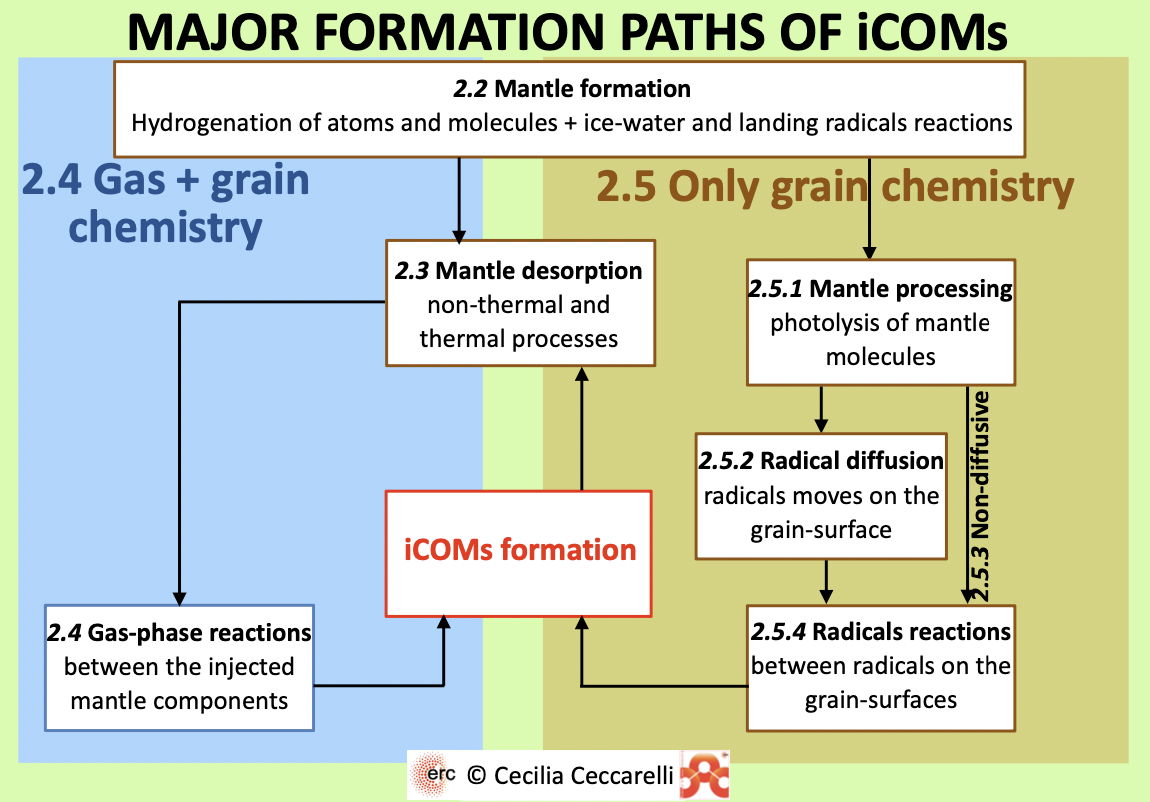}
 \caption{\small Sketch of the two major paths invoked in the literature for the formation of iCOMs: \textsc{Gas + grain chemistry} (\S ~\ref{subsec:sec2-gas-phase}) and \textsc{Only grain chemistry} (\S ~\ref{subsec:sec2-onlygrainchem}).
   Both predict the formation of icy mantles enveloping the interstellar dust grains (\S ~\ref{subsec:sec2-mantle-form}), prevalently during the cold molecular cloud and prestellar core stages (Sec. \ref{sec3:psc}). 
   The mantle constituents are the result of the hydrogenation of atoms and simple molecules, such as CO, and the oxidation (\S .~\ref{subsubsec:sec2-mantle-hydrogenation}).
   The iCOM methanol, for example, is formed in this first step.
   In addition, after the first layers of water-ice are formed, radicals landing on the grain surfaces can react with the ice-water molecules (\S ~\ref{subsubsec:sec2-mantle-iceradical}).
   The iCOM ethanol, for example, can be  formed in this way.
   After this first step, the two paths differ as follows.\\
   \textit{Left blue panel:} 
   In the \textsc{Gas + grain chemistry} path, the  mantle components are either partially or completely injected into the gas-phase by thermal and non-thermal desorption processes (\S ~\ref{subsec:sec2-mantle-desorption}), where they undergo gas-phase reactions that form iCOMs (\S ~\ref{subsec:sec2-gas-phase}).
   This path can occur either via mantle thermal desorption (\S ~\ref{subsubsec:sec2-thermdes}) in hot corinos\index{Hot corinos} (Sec. \ref{sec4:protostars}) and inner protoplanetary disks\index{Protoplanetary disks} (Sec. \ref{sec5:disks}), or via mantle non-thermal desorption (\S ~\ref{subsubsec:sec2-nonthermdes}) in prestellar cores (Sec. \ref{sec3:psc}), outer protoplanetary disks\index{Protoplanetary disks} (Sec. \ref{sec5:disks}) and molecular outflows\index{Molecular outflows} (Sec. \ref{sec6:outflows}).\\
   \textit{Right brown panel:} 
   In the \textsc{Only grain chemistry} path, the icy mantles are processed by UV (photolysis) and CR (radiolysis) irradiation while being formed (\S ~\ref{subsubsec:sec2-mantle-process}). 
   When the dust temperature increases, radicals on the mantles become mobile and diffuse (\S ~\ref{subsubsec:sec2-radical-diffusion}) and, when they meet, they combine forming iCOMs (\S ~\ref{subsubsec:sec2-radical-reactions}).
   Alternatively, non-diffusive processes can also combine radicals into iCOMs (\S ~\ref{subsubsec:sec2-nondiffusive}).
   Likewise, some radicals are also predicted to meet and react in non-diffusive processes during the cold phase.}
 \label{fig:sec2-chem-scheme}
\end{figure*}

After more than 15 years, the "grain-surface does all" theory is starting to have its own problems.
First, since 2012, observations have shown that iCOMs are numerous and relatively abundant (see Sec. \ref{sec3:psc}) in  cold ($\sim10$ K) prestellar cores, where a warm-up phase is absent.
Therefore, these  observations pose a significant challenge to the original version of the \cite{Garrod2006} theory and to astrochemistry in general.
At the same time, new laboratory experiments have proved that some neutral-neutral gas-phase reactions, previously thought to be highly inefficient at the low temperatures of the ISM, actually become very fast, thanks to the quantum tunneling of pre-reactive complexes \citep{Shannon2013}.
Finally, deep searches of the chemical literature have demonstrated that important gas-phase reactions are missing in the astrochemical networks \citep{Balucani2015} and new quantum chemical calculations have been enriching the databases with new gas-phase routes \citep{Barone2015,Skouteris2018}.
The game is once again afoot. 
Very likely, both  grain-surface and gas-phase chemistry collaborate to yield the zoo of iCOMs now known to be present in Solar-type star forming regions.

In addition, whatever the formation route, in cold and dense\index{Dense cores!chemistry} regions, such as in prestellar 
cores\index{Prestellar cores}  and
protoplanetary disks\index{Protoplanetary disks}, some mechanism has to be able to remove iCOMs from the grain mantles and inject them into the gas-phase, where they are detected.
This is a challenge for our understanding of the iCOM chemistry and the current models.
Several non-thermal processes have been proposed in the literature, (see Sec. \ref{sec2:chemistry}), which depend on several poorly known parameters.
A crucial one, common to almost all non-thermal processes, is the binding energy of the iCOMs, namely the energy necessary to release the frozen molecule into the gas.
New experiments and theoretical calculations are also finally providing information on these processes and parameters, which are absolutely necessary for the interpretation of the astronomical observations.

In this Chapter, we review the state of the art, by first introducing the basics about iCOM chemistry (Sec. \ref{sec2:chemistry}) and then describing and discussing each step of the early phases of Solar-type star formation: prestellar cores (Sec. \ref{sec3:psc}), protostars\index{Protostars} (Sec. \ref{sec4:protostars}), protoplanetary disks\index{Protoplanetary disks} (Sec. \ref{sec5:disks}), molecular outflows\index{Molecular outflows}, which offer a unique chemistry laboratory (Sec. \ref{sec6:outflows}), and comets\index{Comets}, which permit us to forge the link between the present Solar System and the inheritance of interstellar organic chemistry (Sec. \ref{sec7:comets}).
A final section (Sec. \ref{sec8:conclusions}) concludes the Chapter.

\section{\textbf{iCOM FORMATION: GAS-PHASE AND GRAIN-SURFACE CHEMISTRY}} \label{sec2:chemistry}
\subsection{Overview}

Two major paths have been invoked in the literature for the formation of iCOMs in the ISM, as summarized in Fig. \ref{fig:sec2-chem-scheme} and here named \textsc{Gas + grain chemistry} and \textsc{Only grain chemistry}, respectively.

The first step (\S ~\ref{subsec:sec2-mantle-form}), common to both theories, mainly occurs during the cold molecular cloud and prestellar core phases (Sec. \ref{sec3:psc}), with the formation of icy mantles covering the interstellar grains.
Atoms and small molecules remain glued to the grains when they hit them and, there, they are hydrogenated and oxidized (\S ~\ref{subsubsec:sec2-mantle-hydrogenation}) \citep[e.g.][]{tielens1982}.
In this way, water, CO$_2$, methanol, ammonia and methane become the major components of the icy mantles, as observed in the Near Infrared (NIR) \citep[e.g.][]{Boogert2015}.
As soon as the first layers of water ice\index{Water ice} are formed, (small) radicals landing on the grain surfaces can also react with the iced water molecules to form some iCOMs, such as ethanol \citep{Perrero2022-ethanol}.
After this first step, the two paths diverge.

In the \textsc{Gas + grain chemistry} path (\S ~\ref{subsec:sec2-gas-phase}), depending on the temperature, thermal and non-thermal processes release the grain mantle components into the gas-phase, partially or completely (\S ~\ref{subsec:sec2-mantle-desorption}).
Once in the gas-phase, the mantle components react with the other gaseous species and form, among other species, iCOMs other than those formed in the first step \citep[e.g.][]{Vasyunin2013,Balucani2015,Skouteris2018}.

In the \textsc{Only grain chemistry} path (\S ~\ref{subsec:sec2-onlygrainchem}), the grain mantles are processed by UV (photolysis) and cosmic-rays (CR) (radiolysis) irradiation, forming numerous and abundant radicals in their bulk (\S ~\ref{subsubsec:sec2-mantle-process}).
These radicals become mobile when the dust temperature increases to about 20--30 K (\S ~\ref{subsubsec:sec2-radical-diffusion}).
When the radicals meet on the surfaces they can react and form iCOMs (\S ~\ref{subsubsec:sec2-radical-reactions}) \citep[e.g.][]{Garrod2006,Garrod2008,Ruaud2015,Aikawa2020}.
Alternatively, non-diffusive processes can also cause the radicals to meet on the grain surface and react to form iCOMs (\S ~\ref{subsubsec:sec2-nondiffusive}) \citep{Jin2020,Garrod2022}.
Finally, the iCOMs formed on the grain surfaces are released into the gas-phase, where they are observed, by thermal and non-thermal processes (\S ~\ref{subsec:sec2-mantle-desorption}).

Probably, both gas-phase and grain-surface chemistry are at work in the ISM, dominating the formation of different iCOMs in different environments and at different times.

In the following, we will review the major processes which enter into the formation of iCOMs as well as their presence in the gas-phase, where they are observed: the grain mantle formation (\S ~\ref{subsec:sec2-mantle-form}) and desorption processes (\S ~\ref{subsec:sec2-mantle-desorption}), the gas-phase (\S ~\ref{subsec:sec2-gas-phase}) and  grain-surface (\S ~\ref{subsec:sec2-onlygrainchem} chemistry, as well as theories alternative to the two reported in Fig. \ref{fig:sec2-chem-scheme} (\S ~\ref{subsec:sec2-other-formation-theories}).
The overview of all the above mentioned processes and their interaction is schematised in Fig. \ref{fig:sec2-chem-scheme}, where the paragraphs in which they are discussed are also reported.
Our goal is to provide the general ideas behind the astrochemical models that are used to interpret the astronomical observations as well as the uncertainties caused by our limited knowledge of the involved basic processes.

\subsection{Mantle formation} \label{subsec:sec2-mantle-form}
The grain mantles are mostly made up of water, CO and CO$_2$ with less abundant hydrogenated species such as ammonia, formaldehyde and methanol, as shown by IR solid-state astronomical observations \citep{Boogert2015}.

\subsubsection{Hydrogenation and oxidation of atoms/molecules} \label{subsubsec:sec2-mantle-hydrogenation}
In general, there is ample consensus on the fact that atoms and simple molecules (e.g. CO) are hydrogenated and oxygenated when they freeze-out onto the grain surfaces.
The simple iCOM methanol is believed to be prevalently formed in this way, via the successive addition of H atoms to the frozen CO \citep[e.g.][]{Watanabe2002,Rimola2014}.
However, even though several laboratory experiments have been carried out and showed the feasibility of the process, the details which enter into the astrochemical models are not completely understood.

Two major questions are still open: 
(1) how the H atoms diffuse on the icy grain surfaces and
(2) what is the barrier of the CO + H and H$_2$CO + H reactions.
For example, models often assume that H atoms diffuse on the icy grain surfaces via hopping \citep[e.g.][]{Garrod2013,Aikawa2020}, while recent sophisticated quantum mechanical (QM) studies by \citet{Senevirathne2017} show that below $\sim10$ K tunneling dominates in amorphous ice.
Actually, the relative contribution of H hopping versus tunneling is very difficult to quantify, since it depends on the ice surface structure \citep{hama2012}.
In addition, in the (many) astrochemical models adopting the \citet{Garrod2013} recipes and parameters, the diffusion energy of H is underestimated with respect to the theoretical results by \citet{Senevirathne2017}, with a consequent overestimate of the H diffusion.
Likewise, in the same models, the energy barriers of the CO + H and H$_2$CO + H reactions are assumed to be rectangular with a width derived from gas-phase computations \citep{Garrod2013}.
However, when compared with the theoretical QM calculations that takes into account the icy surfaces, the rates of CH$_3$OH formation are underestimated by orders of magnitude \citep{Song2017}.

\subsubsection{Radical-water ice reactions} \label{subsubsec:sec2-mantle-iceradical}
Recent theoretical studies have shown that some iCOMs can be formed on  water-ice-rich surfaces by reactions of radicals, landing from the gas-phase into the grain surface, with the water molecules of the ice itself.
\citet{Rimola2018} studied the case of the formation of formamide by the reaction of CN with H$_2$O to form HNCOH, followed by an H atom addition.
More recently, \citet{Perrero2022-ethanol} showed that ethanol can be formed by the reaction of CCH with one water molecule of ice, plus successive hydrogenation by landing H atoms.
Finally, \cite{Molpeceres2021} showed that C atoms landing on the icy grain surfaces react with water molecules and form formaldehyde.
The study of the radical-ice water route of formation is at its infancy and may give an interesting alternative path to some iCOM formation not involving radical-radical combination on the grain surfaces.

\subsection{Mantle desorption}\label{subsec:sec2-mantle-desorption}

In interstellar clouds, molecules adsorbed onto mantles of icy material can desorb either thermally (\S ~\ref{subsubsec:sec2-thermdes}) or non-thermally (\S ~\ref{subsubsec:sec2-nonthermdes}). 

\subsubsection{Thermal desorption}\label{subsubsec:sec2-thermdes}

The thermal desorption of a species bound to a substrate can be approximately described by the Polanyi-Wigner equation \citep{Kolasinski2002}.  
The first-order solution pertains to adsorbate molecules that are not strongly associated with each other.  
This occurs in most cases, when they are adsorbed on a surface made of other species (e.g. water or CO). 
Under these conditions, the solution for the rate of desorption $k_{\rm des}$ is given by:
\begin{equation}\label{eq:desorption}
    k_{\rm des} = \nu_{des} \exp({-E_{\rm des}/T}) ~\sim~ \nu_{des} \exp({-BE/T}) 
\end{equation}
where $E_{\rm des}$ is the desorption energy, which is approximately the binding energy (BE), given in units of temperature, and $\nu_{des}$ is the pre-exponential factor, which depends on the adsorbate and surface \citep{Minissale2022}.
We will discuss in some detail both quantities because they have an important impact on the astrochemical models.

Thermal desorption is rarely important at 10 K except for atomic hydrogen and a few other light physisorbed species where  van der Waals and dispersion forces keep the species bound to the surface  because the desorption energy is typically large enough to keep the adsorbate bound.  
As the temperature increases, the more volatile adsorbates begin to desorb, and by a temperature of typically 100--120 K, thermal desorption has become an important process for a variety of molecular species.

An intermittent form of thermal desorption is powered by CR which, when they strike the grains, heat them to a temperature that could cause the sublimation of species frozen in the grain mantles \citep{Leger1985,Hasegawa1993,Roberts2007,Sipila2021}.

\vspace{-0.3cm}
\paragraph{Overview of the BE and $\nu_{des}$ derivation}
BE and $\nu_{des}$ can be either measured via laboratory experiments or estimated via QM computations.
\citet{Minissale2022} provide a recent and detailed review of both techniques.
Here we briefly discuss the limitations on the published data used in astrochemical models.

In laboratory experiments, BE and $\nu_{des}$ are \textit{simultaneously} derived from the so-called TPD (Thermal Programmed Desorption) experiments.
The species is deposited or co-deposited over a cold (typically 10 K) surface, which is then heated at a constant rate and the sublimation rate into the gas-phase measured as a function of time.
Therefore, TPD experiments strictly measure rates of desorption.
BE and $\nu_{des}$ are then derived from the desorption rate curve by using Eq. \ref{eq:desorption} \citep[e.g.][]{HeVidali2016}.

Theoretical calculations compute the energy necessary for the adsorbate species to leave the surface using the TST (Transition State Theory) methodology.
The theoretical works usually only provide the BE value, without reporting $\nu_{des}$, even though it is a product of the calculations \citep[e.g.][]{Ferrero2022}.

In general, the pre-exponential factor can be analytically calculated, following the recipe given by \citet{tait2005} \citep[see the discussion in ][]{Minissale2022,Ferrero2022}.
For relatively large molecules, such as acetaldehyde, $\nu_{des}$ is $\sim 10^{18}$ s$^{-1}$ \citep{Ferrero2022}.

Both recent experiments and theoretical computations have shown that, actually, there is not a unique BE for each species, but rather a distribution of BEs that reflects that the bond of the adsorbed species depends on the site and the species orientation \citep[e.g.][]{Ferrero2020,bovolenta2020,Tinacci2022-ammonia}.
Astrochemical models often, but not always \citep[e.g.][]{grassi2020}, use one single value for the BE, and assume $\nu_{des}\sim 10^{12}$ s$^{-1}$, following the recipe by \citet{Hasegawa1993}, regardless of the source of the BE evaluation.
However, as it is the case for acetaldehyde, $\nu_{des}$ could be as large as $10^{18}$ s$^{-1}$, depending on the species and the method used to derive the BE \cite[e.g.][]{Ferrero2022}.
A review of the $\nu_{des}$ is reported in \citet[][see also their Fig. 14]{Minissale2022}.

\vspace{-0.3cm}
\paragraph{BE and $\nu_{des}$ of iCOMs}
In general, only the BE and $\nu_{des}$ of a few species have been measured in the laboratory \citep[e.g.][]{collings2004,HeVidali2015,HeVidali2016,Corazzi2021,Ferrero2022} or computed \citep[e.g.][]{Wakelam2017BE,das2018,Ferrero2020,Ferrero2022, Tinacci2022-ammonia}.
The computations refer to the bond energy of one water molecule \citep{Wakelam2017BE}, a more complex situation with up to six water molecules \citep{das2018}, to models of amorphous water ice\index{Water ice} clusters \citep{Ferrero2020,Ferrero2022}, and to models of $\sim200$ water-ice grains \citep{Germain2022,Tinacci2022-ammonia}, the most reliable. 
The iCOMs where BE and $\nu_{des}$ have been either measured or evaluated on an ice cluster of more than 20 water molecules are listed in Tab. \ref{tab:sec1-iCOMs}. 
Clearly, there is a massive lack of data in this respect and this is a first serious warning for the astrochemical models. 

\begin{figure}[htp]
    \centering
    \includegraphics[width=8cm]{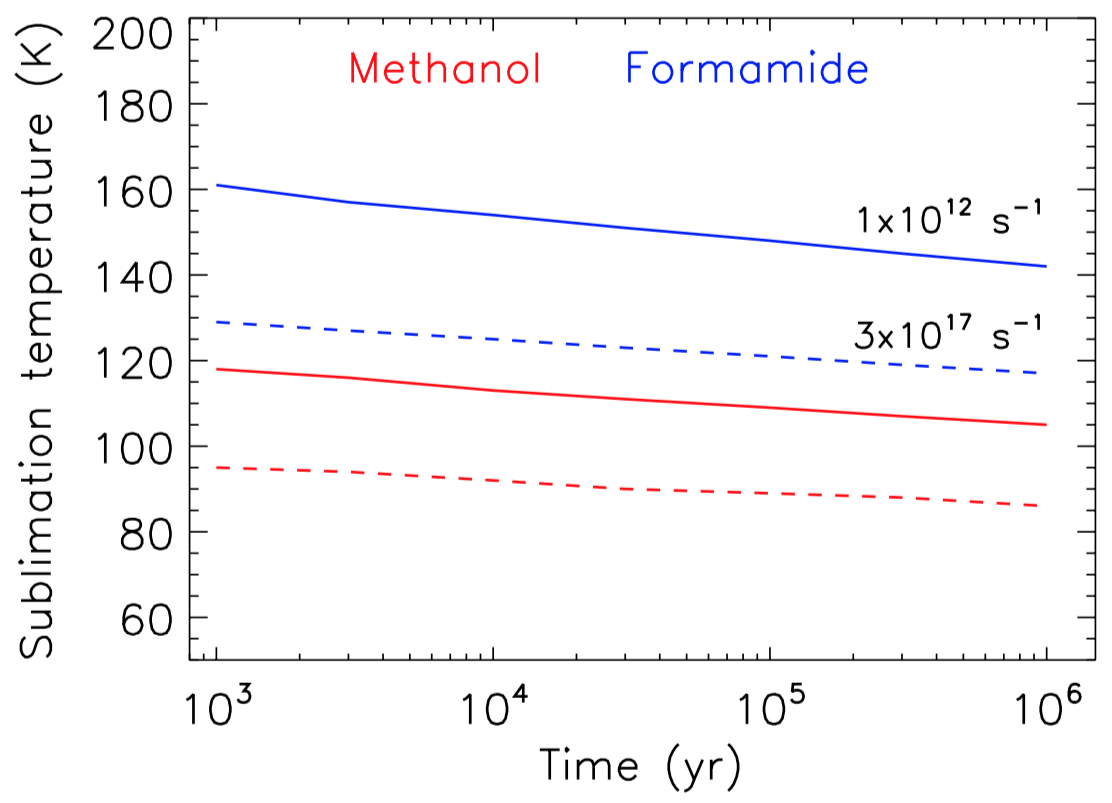}
    \caption{Sublimation temperature of frozen methanol (red) and formamide (blue), as a function of the heating time, for a pre-exponential factor (Eq. \ref{eq:desorption}) equal to $1\times10^{12}$ s$^{-1}$ (obtained using the \citet{Hasegawa1993} formula: solid line) and $3\times10^{17}$ s$^{-1}$ (obtained using the \citet{tait2005} formula: dashed line), respectively (see text).}
    \label{fig:sec2-Tdes}
\end{figure}

To illustrate the impact of the assumed pre-exponential factor value $\nu_{des}$ (Eq. \ref{eq:desorption}), Fig. \ref{fig:sec2-Tdes} shows the sublimation temperature of frozen methanol (whose BE is taken equal to 6200 K) and formamide (BE = 8400 K) as a function of the heating time, which often, but not always, corresponds to the age of the studied astronomical object. 
Two values are assumed for $\nu_{des}$ (see discussion above).
To obtain the sublimation temperature we used the half-life time (=ln(2)/$k_{\rm des}$) of the frozen species.
Taking the example of methanol, if the heating time is $10^3$ yr then the sublimation temperature would be $\sim 120$ K, while if it is $10^6$ yr it is $\sim 105$ K.
Likewise, the difference due to the pre-exponential factor assumed $1\times 10^{12}$ and $3\times 10^{17}$ s$^{-1}$ is $\sim 20-30$ K.

In the context of iCOM formation on the grain surfaces by the combination of radicals (HCO, CH$_3$, NH$_2$ et cetera), the BE of the latter is also a crucial parameter because it defines the range within which the radicals remain on the surface and can, therefore, react with each other.
Actually, the diffusion energy, which regulates whether radicals can move on the grain surfaces, is also a function of BE.
At the time of this Chapter writing, there are no laboratory measurements of the binding energies of radicals, because they are extremely difficult to obtain.
Theoretical calculations on models of amorphous water ice\index{Water ice} exist for a few molecular radicals: O$_2$, OH, HCO, CH$_3$ and NH$_2$ \citep[e.g.][]{Ferrero2020}.
%

\subsubsection{Non-Thermal desorption}\label{subsubsec:sec2-nonthermdes}
There are a number of different mechanisms for non-thermal desorption at low temperatures, only some of which are reasonably well studied.  
Common to all of these mechanisms is a source of energy to augment the low temperature in the grain mantle. 
Several mechanisms are discussed here. 

\vspace{-0.3cm}
\paragraph{Chemical or reactive desorption}
Chemical or reactive desorption occurs when a surface reaction possesses sufficient exothermicity  to overcome the binding energy of the products.  
This process is analogous to what chemists refer to as unimolecular dissociation, which is governed by the well-known RRKM (Rice–Ramsperger–Kassel–Marcus) theory, or its simpler version, the RRK model \cite[e.g.][]{DiGiacomo2015}. 
An approximation to the fraction of products released into the gas-phase through chemical desorption based on the RRK approximation predicts approximately 1\% \citep{Garrod2007}. 

More recent and more sophisticated studies appearing in the literature have challenged this simplified and usual assumption.
For example, \cite{pantaleone2020} have simulated the chemical desorption of the reaction H + CO $\rightarrow$ HCO via ab initio molecular dynamics simulations. 
These authors found that the reaction energy ($\sim 1.4$ eV) is largely absorbed by the icy grains in less than 1 ps, leaving the newly formed HCO radical with too little energy to break the bonds with the water surface.
As a consequence, very little or no chemical desorption is expected to occur in this case, as was also found in the experimental studies by \citet{Minissale2016a}.
On the other hand, \cite{pantaleone2021} studied the H + H $\rightarrow$ H$_2$ reaction and found that the reaction energy ($\sim 5$ eV) is also largely absorbed by the water ice\index{Water ice} but the remaining fraction ($\sim 2$ eV) allows the newly formed H$_2$ to desorb, thanks to its low BE.
Therefore, each system behaves in a different way and the result depends on the reaction energy, the efficiency of the water ice\index{Water ice} to absorb it  and the BE of the formed species.
Finally two systematic experimental studies have provided a semi-empirical theory, in which the chemical desorption depends on the species and on the substrate \citep{Minissale2016a,Yamamoto2019}.

With respect to iCOMs, methanol is the only species for which estimates of the chemical desorption efficiency exist \citep{Minissale2016a,Chuang2018,pantaleone2020}.
Experiments showed that the chemical desorption efficiency caused by the addition of H atoms to  frozen CO and the partially hydrogenated species HCO, H$_2$CO, CH$_3$O and CH$_3$OH, is very low ($\le$2 \%) on water-rich surfaces, \citep{Minissale2016a,Chuang2018}, in agreement with theoretical predictions \citep{pantaleone2020}.
Chemical desorption for other iCOMs forming on the grain surfaces are at present unknown.

\vspace{-0.3cm}
\paragraph{Photodesorption}
Another non-thermal mechanism often invoked in the literature is photodesorption via UV radiation.
This process has been studied in the laboratory although this direct process can be confused with photodissociation, in which the smaller products are the species that desorb.  
The direct process appears to occur with a 
yield of $\sim 10^{-2}$ for CO \citep{Fayolle2011}.
In these experiments, the photodesorption spectrum tends to mimic the gas-phase spectrum \citep{Fayolle2011}, namely the gas-phase absorption spectrum of CO as a function of frequency and the desorption spectrum of CO from a surface look the same in frequency with the desorption spectrum having greater linewidths.
A lower average efficiency of $\sim 10^{-5}$ for methanol is thought to occur because of the simultaneous photodissociation of the species \citep{Bertin2016,Martin-Domenech2016}.

\vspace{-0.3cm}
\paragraph{CR-induced desorption}
Recent experiments have shown that sputtering by heavy CR can potentially produce gaseous methanol, originally frozen on the grain mantles \citep{dartois2015,dartois2019}.
Please note that this process is not to be confused with the CR-induced thermal desorption described in \S ~\ref{subsubsec:sec2-thermdes}, in which the CR serve only to increase the temperature.

\vspace{-0.3cm}
\paragraph{Sputtering}
In molecular shocks\index{Shocks}, frozen species can be injected into the gas-phase because of sputtering caused by the ion-neutral drift velocity \citep{Draine1995,flower1995}.
Several models exist in the literature to predict the amount of sputtered species, especially when the latter is SiO, a common shock\index{Shocks!chemistry} tracer \citep[e.g.][]{flower1995,caselli1997}.
In addition to the treatment of the physical and dynamical structure of the shocked\index{Shocks} gas, the crucial parameter in these models is the so-called threshold energy of each species, which is the energy of the projectile needed  to extract the species from the grain mantle.
There is more than one treatment to evaluate the threshold energy and, not surprisingly, all depend on the BE of the species on the surface \citep[e.g.][]{May2000}.

\begin{figure*}[tb]
    \centering
    \includegraphics[width=15cm]{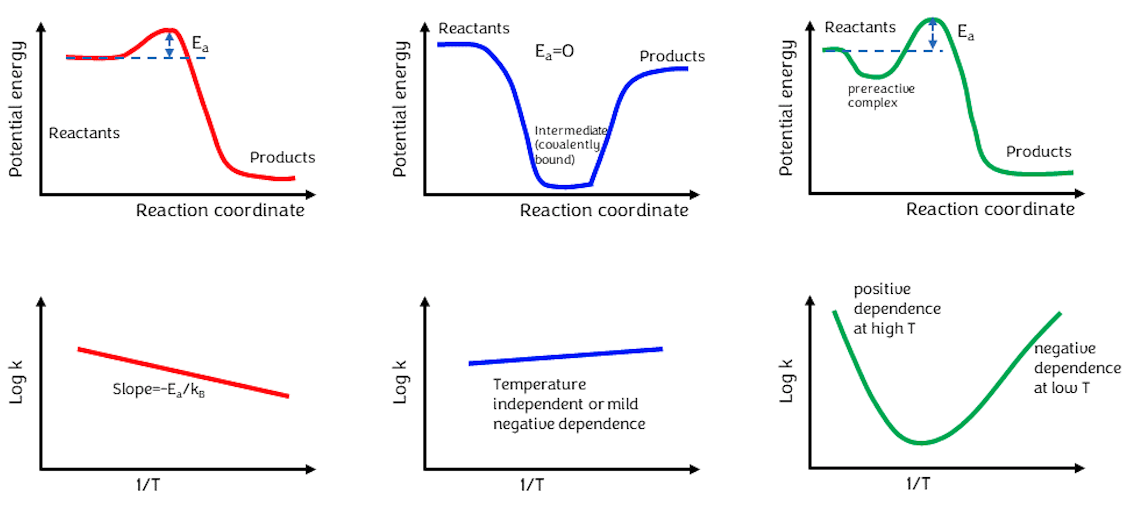}
    \caption{Schematic view of the potential energy versus the reaction coordinate (upper panels) and rate constant versus the inverse of the temperature (1/T) (lower panels) of the three major classes of bimolecular gas-phase reactions.
    The rate coefficients of bimolecular reactions mostly obey the Arrhenius equation, $k$(T) = $A ~exp(-{\rm E_a}/{\rm T})$, where the temperature dependence of the rate coefficient is represented by a constant factor A (only slightly dependent on temperature) multiplied by an exponential function bearing the dependence on the temperature T. 
    If we plot log($k$) as a function of inverse temperature for a reaction characterized by a positive activation energy (${\rm E_a}$), the slope is given by --${\rm E_a}$ (lower left panel). 
    In the case of reactions with a null activation energy, the rate coefficient is temperature independent or shows a mild negative dependence on T (middle panels). 
    Only recently, there has been experimental evidence that the presence of a weak non-covalent pre-reactive bound complex before the barrier leads to a bimodal trend with T: at high temperature, the system experiences the presence of the barrier and an increase of T corresponds to an increase in the value of $k$. However, as T goes below a certain value, the increased lifetime of the pre-reactive complex will favor the reaction via quantum tunnelling through the activation energy barrier and the rate coefficient has a positive trend with decreasing T.}
    \label{fig:sec2-gasphase=scheme}
\end{figure*}

Among the iCOMs, only methanol has been the focus of specific studies \citep{May2000,Jimenez2008} even though it is predicted by the Paris-Durham shock\index{Shocks!chemistry} model, which is publicly available\footnote{\url{https://ism.obspm.fr/shock.html}} \citep{lesaffre2013}.
While there are studies on the threshold of species such as SiO and CO on silicate surfaces \citep[e.g.][]{May2000}, no specific experimental or theoretical studies exist about methanol on water- or CO- rich ices, to the best of our knowledge.
The methanol threshold energy is assumed to be equal to that of water in the Paris-Durham shock\index{Shocks!chemistry} model.

\subsection{Gas-phase formation of iCOMs}\label{subsec:sec2-gas-phase}

\subsubsection{Overview}
Once the precursor molecules formed in the icy mantle of interstellar grains have desorbed, molecular complexity can increase further through gas-phase reactions. 
In addition, the main destruction routes included in astrochemical models occur in the gas-phase.

In the low densities of  Solar-type star forming regions, the most important gas-phase reactions that lead to an increase in chemical complexity are bimolecular reactions.
This is because three-body collisions are too rare except in the innermost and densest regions of protoplanetary disks\index{Protoplanetary disks} (Sec. \ref{sec5:disks}). 
The same is true also for their destruction routes. 
In addition, given the low temperatures, only reactions which are exothermic are possible. 
Exothermicity, however, is insufficient because bimolecular reactions will not take place unless the reactants collide with a certain amount of energy called the "activation energy" (${\rm E_a}$) of the reaction. 
The activation energy can be seen as the energy involved in breaking or weakening the original bonds 
before the new ones are formed and only those collisions with energies equal to or greater than the activation energy will result in a reaction. 
At low temperatures, sometimes, tunneling under the activation energy could be important.
For this reason, bimolecular reactions of interest in astrochemistry always involve either a radical or an ion; that is, transient species with unpaired electrons, the reactions of which are usually characterized by a very small ($\leq 1$ kJ/mol) or null activation energy. 

Two major classes of reactions can form iCOMs in the gas-phase: ion-neutral and neutral-neutral reactions.
In Fig. \ref{fig:sec2-gasphase=scheme}, the potential energy profiles of typical bimolecular reactions (upper panels) and the corresponding Arrhenius plots (that is, the variation of the rate coefficient with the temperature, lower panels) are shown. 
If a reaction is characterized by an activation energy, ${\rm E_a}$, the rate coefficient increases with increasing temperature. 
If the activation energy is null (barrierless reactions), the rate coefficient is either temperature independent or exhibits a slightly negative dependence. 
For this reason, until recently only reactions with null ${\rm E_a}$ have been considered in astrochemical models. 
However, recent experiments based on the CRESU\footnote{Cin\'etique de R\'eaction en Ecoulement Supersonique Uniforme, which stands for Reaction Kinetics in a Uniform Supersonic Expansion.} technique have shown a very interesting effect: the presence of a pre-reactive complex before the system surpasses the activation energy can make the reaction efficient also at temperatures that are too low to overcome the barrier, via the tunneling effect (see below). 
Note that Fig. \ref{fig:sec2-gasphase=scheme} refers mainly to neutral-neutral systems; the rate constants of ion-neutral  reactions have more restrictive possibilities.  
Reactions with non-polar neutrals are temperature-independent, while reactions with polar ions are inversely dependent on the square root of temperature.

We will briefly describe ion-neutral and neutral-neutral reactions in the following subsections.
Finally, \S ~\ref{subsec:sec2-react-networks} will discuss the state-of-art of astrochemical gas-phase reactions networks.

\subsubsection{Ion-neutral reactions} \label{subsubsec:sec2-gas-ionneut}
As briefly mentioned in the Introduction, interstellar gas-phase chemistry was initially based mostly on ion-neutral reactions, because of their large rate coefficients at the low temperatures of the ISM \citep{Herbst1973,Watson1974,Dalgarno1976}. 
The final step in the complex ion-neutral reaction chains was supposed to be the conversion of the ionic species into their neutral counterparts, via electron-ion recombination reactions with the release of one H atom and the preservation of the ion molecular skeleton. 
Later on, however, it was shown that electron-ion recombination reactions are strongly dissociative because of the large amount of energy released in the process and that the major product channels most often consists of  small neutral moieties \citep{Geppert2006}. 
Product channels dominated by the release of one hydrogen atom, on the other hand, rarely account for more than a few percent of products.
%

\subsubsection{Neutral-neutral reactions} \label{subsubsec:sec2-gas-neutneut}
The role of neutral-neutral reactions was only recognized in the 90's after the introduction of an experimental method that allowed investigation of the kinetics of bimolecular reactions at very low temperatures, based on the CRESU technique\citep{Smith2000}. 

At present, the characterization of neutral-neutral reactions relies on three complementary approaches:
(1) the CRESU technique, to derive low temperature global rate coefficients; 
(2) collision-free experiments, to derive the nature of the primary reaction products and their branching ratios; and
(3) QM theoretical computations, to derive potentials and rates for chemical reactions.
This combined effort is necessary because the CRESU technique is able to characterize gas-phase reactions under the appropriate temperature conditions, but cannot reproduce the low number density environment typical of the ISM, while collision-free beam experiments are able to reproduce the low number density environment, but not the low temperature conditions, \citep[e.g.][]{Casavecchia2009,Casavecchia2015}, with the  exception of the Bordeaux apparatus developed by \citet{Costes2010}.

Finally, QM theoretical calculations are essential not only to support the interpretation of the experimental data, but also to assist the extrapolation of the experimental results. 
In addition, when no experimental data are available, a theoretical characterization of a gas-phase reaction can provide a realistic estimate of the reaction rate coefficients and product branching ratios \citep[e.g.][]{Sleiman2018,Ocana2018,Ocana2019,Nguyen2019}.

Reactions characterized by a pre-reactive complex and an activation energy have recently attracted much attention (see Fig. \ref{fig:sec2-gasphase=scheme}). 
In particular, the reactions of the abundant OH radical with several oxygenated iCOMs fall in this category and their investigation has had a profound influence in the understanding of interstellar chemistry. 
The effect is double: on the one hand, the reaction of oxygenated iCOMS with OH must be considered as an efficient destruction route of those species because of the abundance of the OH radical. 
On the other hand, the radicals formed by the H-abstraction process can react further and be involved in the formation of other iCOMs \citep[e.g.][]{Shannon2013,Balucani2015}.

The most important case of this family is the reaction between OH and CH$_3$OH. 
This reaction is known to have an experimental activation energy ($\sim10$ kJ mol$^{-1}$) \citep[e.g.][]{Shannon2013,Roncero2018}. 
In spite of that, below 200 K the rate coefficient changes its slope with the temperature and starts increasing with decreasing temperature (see Fig. \ref{fig:sec2-gasphase=scheme}). 
The $k$ values below 100 K can be larger by a factor of 10$^3$ with respect to the room temperature value. 
The explanation for such a strong effect has been rationalized by invoking the role of a pre-reactive complex \citep{Shannon2013,Roncero2018}, which is formed by the strong H-bond interaction between the OH radical and the -OH functional group of methanol. 
At very low temperature, the pre-reactive complex, characterized by a relatively deep potential well, can live for a time long enough to have a significant probability to tunnel through the barrier rather than re-dissociate back to reactants \citep{Heard2018}.


Concerning the reactions for which no experimental data exist but QM theoretical computations have been carried out, illustrative successful examples are the reaction NH$_2$ + H$_2$CO \citep{Barone2015,Skouteris2017}, the only formation gas-phase route of formamide known so far, and the reaction O + CH$_3$CH$_2$OH \citep{Skouteris2018} on which an entire reaction chain to form very common iCOMs is derived, including glycolaldehyde and acetaldehyde.


Finally, barrierless reactions mostly occur via the formation of one (or more) covalently bound intermediate(s) (see middle panel of Fig. \ref{fig:sec2-gasphase=scheme}). 
In the absence of three-body collisions, the intermediate cannot be stabilized by dispersing part of its internal energy and, if exothermic exit channels are open, its high internal energy content leads to its fragmentation into one or more sets of bimolecular products. 
In astrochemistry, however, radiative association might occur, i.e., the intermediate can lose part of its internal energy by emitting a photon \citep{Herbst1985}. 
Normally, these processes are not competitive because the typical reaction time is of the order of picosecond while spontaneous emission occurs within milliseconds. 
However, if no exothermic channels exist and the intermediate can only re-dissociate back to reactants, radiative association can occur with a significant rate coefficient. 
An important case is represented by the reaction CH$_3$ + CH$_3$O forming dimethyl ether \citep{Vasyunin2013,Balucani2015}, for which theoretical computations predict large reaction constants \citep{Tennis2021}.

%
\begin{figure*}[th]
 \epsscale{1.5}
 \plotone{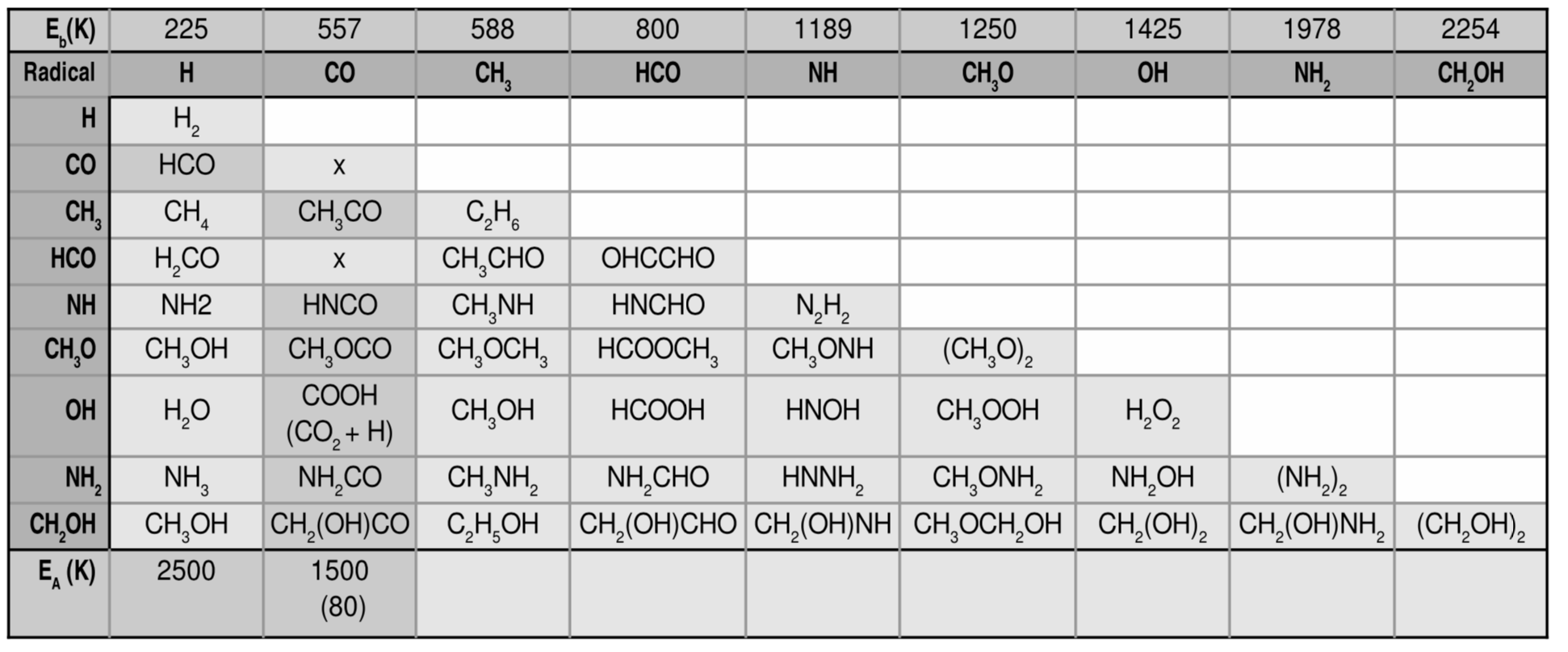}
 \caption{\small Scheme of the iCOMs proposed to form on  grain-surfaces via the combination of radicals present on  icy granular mantles. 
 The darker gray cells indicate the presence of activation barriers, whose value is assumed to be that reported in the last row. 
 Adapted from \cite{Garrod2008}.
   }
 \label{fig:sec2-radrad-reactions}
\end{figure*}
\subsection{Only grain chemistry formation of iCOMs}\label{subsec:sec2-onlygrainchem}

This theory, inspired by numerous laboratory experiments \citep[e.g.][]{oberg2009,Linnartz2015,kaiser2015}, was first proposed by \citet{Garrod2006} and is based on three fundamental steps: (i) the formation of radicals by UV photons induced by CR irradiation (\S ~\ref{subsubsec:sec2-mantle-process}), (ii) the diffusion of radicals when the dust temperature increases to 20--30 K, so that radicals become mobile (\S ~\ref{subsubsec:sec2-radical-diffusion}), and (iii) the combination of radicals into iCOMs (\S ~\ref{subsubsec:sec2-radical-reactions}).
Subsequently, always inspired by laboratory experiments \citep[e.g.][]{Theule2013,Qasim2019,Ioppolo2021}, a new theory, called non-diffusive grain surface chemistry, was developed where the step (ii) of the previous one is not necessary (\S ~\ref{subsubsec:sec2-nondiffusive}). 

\subsubsection{Mantle processing} \label{subsubsec:sec2-mantle-process}

During the formation of the grain mantles, which are enriched in water, ammonia, methanol, formaldehyde and methane, CR-induced UV photons penetrate the mantles and photo-dissociate the frozen species creating radicals.
For example, OH and CH$_3$ are formed from the photo-dissociation of methanol.
The rate at which the radicals are formed is usually assumed to be equal to that in the gas-phase, which is probably an overestimation.
For example, based on modeling of ice formation following the \cite{Garrod2008} scheme and recipes, \citet{kalvans2018} estimated that, in order to reproduce the observed solid-phase carbon oxides and ammonia, the rate of radical formation should be a factor 3 lower than that in the gas-phase.

\subsubsection{Radical diffusion} \label{subsubsec:sec2-radical-diffusion}
The motion of radicals involves hopping over potential barriers, the so-called diffusion energy (DE) barriers, normally treated as a fraction of the desorption barrier \citep{Hasegawa1992}.
Therefore, the diffusion of radicals depends on their BE and the fraction DE/BE.
We already noted that the BEs of only a few radicals have been computed and published in the literature (\S ~\ref{subsec:sec2-mantle-desorption}).
To the best of our knowledge, there are no measurements nor computations of the fraction DE/BE in radicals involved into the formation of iCOMs.
Various studies on non-radical species and atoms have reported DE/BE ratios from 0.3 to 0.6 \citep{Karssemeijer2014,Minissale2016b,He2018,Zaverkin2022}.
Finally, the coupled theoretical and experimental study by \citet{Ghesquiere2015} suggests that small molecules trapped in water ice\index{Water ice} can diffuse at temperatures larger than $\sim 60$ K driven by the diffusion of the water molecules in the ice.
However, at lower temperatures, neither molecules or radicals seem to diffuse in the mantle bulk \citep{Theule2020} (against the assumption of the majority of astrochemical models).

\subsubsection{Non-diffusive grain-surface chemistry} \label{subsubsec:sec2-nondiffusive}
Laboratory experiments have shown that active chemistry also occurs on granular surfaces via thermal, non-diffusive processes \citep[e.g.][]{Theule2013}, in particular in CO-rich ice surfaces \cite[e.g.][]{Fedoseev2015-glyco,Qasim2019,Ioppolo2021,Fedoseev2022-acetaldehyde}.
Recently, based on earlier treatments using stochastic simulations for the production of CO$_2$ \citep{Chang2014}, \citet{Jin2020} developed a detailed theory of non-diffusive motion, which can occur via a variety of mechanisms, the simplest of which is a ``three-body'' process in which two species on the surface initially undergo a diffusive or Eley-Rideal type motion  even at low temperatures and form a single molecule.   
If this single molecule is formed adjacent to a reactive species, a rapid non-diffusive association can occur since no diffusive barrier is involved. 

As an example, consider the formation of CO$_2$ on a surface. 
The standard diffusive process to form CO$_2$ occurs between surface CO and O, a reaction that possesses a non-negligible activation energy and is rather slow.  
Now consider O and H atoms that move on the surface to form a single OH radical, with only a small barrier.  
The OH can then react quickly with a nearby CO molecule, a non-diffusive process to form CO$_{2}$ that has a small activation energy and turns out to be much faster than the simple diffusion mechanism between CO and O.
To transfer the mathematics from stochastic simulations to standard kinetic treatments \citep{Herbst2021}, one removes any barriers to diffusive motion on the surface and replaces them by the probability that a reactive third body lies adjacent to the newly-formed single molecule and can react without any chemical activation energy.
The reaction is treated as instantaneous.  
Other non-diffusive mechanisms include photodissociation leading to fragments, which are formed adjacent to a reactive species. 
These and other non-diffusive surface mechanisms have been included in a simulation of low-temperature chemistry and often found to form iCOMs more rapidly than via standard diffusive chemistry \citep{Jin2020}.
Moreover the approach can also be used successfully to simulate low temperature laboratory experiments leading to iCOMs, for example glycine \citep{Ioppolo2021}.

\subsubsection{Radical reactions} \label{subsubsec:sec2-radical-reactions}
A crucial basic assumption of the \textsc{Only grain chemistry} path is that, when two radicals meet on the grain surface, they react in a barrierless manner and combine to form an iCOM \citep{Garrod2008}.
In this way, several iCOMs are predicted to be formed, according to the scheme shown in Fig. \ref{fig:sec2-radrad-reactions}, as an example.
However, recent QM computations have shown that radical-radical reactions on  water ice\index{Water ice} surfaces can have barriers and can also have competitive channels such as H-abstraction processes that form simpler molecules. \citep[e.g.][]{Enrique-romero2016,Rimola2018,Enrique-Romero2021,Enrique-Romero2022,Molpeceres2022}.
As a consequence, the combination of radicals may not be the dominant channel or even not occur at all on the interstellar grain surfaces.
An emblematic and well studied case is that of acetaldehyde, which is unlikely to form on the grain surfaces via the association of HCO and CH$_3$ according to QM simulations \citep{Enrique-romero2016,Enrique-Romero2021} as opposed to that predicted by the \cite{Garrod2008} scheme (Fig. \ref{fig:sec2-radrad-reactions}).
Experiments where radical diffusion is very accurately controlled show that indeed this is the case: acetaldehyde is not formed on the cold grain surfaces \citep{Gutierrez-Quintanilla2021} (\cite{MartinDomenech2020} did see it form, albeit at a low efficiency). 
However, other formation reactions leading to species such as ethane, methylamine, and ethylene glycol do not show barriers \citep{Enrique-Romero2022}.
Also, a recent theoretical work considered the possibility for acetaldehyde to be formed by the reaction of a C atom landing on a water-methanol icy surface and encountering a methanol molecule \citep{Chouikha2022}, a mechanism that could be efficient in very peculiar conditions where carbon is not all locked into CO molecules but methanol is already formed on the grain mantles.

In conclusion, each system needs to be studied case by case to understand what probably happens on the interstellar grain mantles and in what conditions.

\subsection{Other routes of iCOM formation}\label{subsec:sec2-other-formation-theories}

\subsubsection{Rapid Radical Association}
Another theory for the iCOM formation appeared in the literature is called "Rapid Radical Association" (hereafter RRA) \citep{Rawlings2013}.
This theory proposes that iCOMs are formed by three-body gas-phase reactions between radicals in warm high density gas. 
This environment exists, for a very short period of time, following the sudden and total sublimation of grain ice mantles driven by the catastrophic recombination of trapped hydrogen atoms, and other radicals, in the ice.
These "explosions" essentially cause short-lived (100 ns) episodes of high density, high temperature gas which leads to efficient three-body reactions.

\subsubsection{Radiolysis}
Another granular mechanism to form iCOMs invoked in the literature, especially at low temperatures, is radiolysis, by which is meant the bombardment of icy mantles by CR in interstellar clouds, or by high energy protons or electrons in laboratory simulations \citep{Paulive2021}.  
If the icy mantles contain carbonaceous species such as CO or CO$_2$, bombardment leads to iCOM production via a complex sequence of reactions involving the formation of ions and secondary electrons, which in turn form energetic radicals, which lead to iCOMs. 
Although numerous experiments have been performed \citep[e.g.][]{Arumainayagam2019}, it is only recently that theoretical treatments for the low-temperature interstellar medium have been developed  \citep{Shingledecker2018} \citep[see also][]{Reboussin2014}.
That said, QM calculations providing an atomistic and precise view of the process are not available in the literature.

\subsection{State-of-the-art astrochemical reaction networks}\label{subsec:sec2-react-networks}
Two major gas-phase networks are publicly available and used by astrochemical models: KIDA\footnote{\url{http://kida.obs.u-bordeaux1.fr/}} \citep{Wakelam2015} and UMIST\footnote{\url{http://udfa.ajmarkwick.net/}} \citep{McElroy2013}.
These networks contain about 8000 reactions that involve about 500 species.

Unfortunately, as indicated in the two databases, the large majority of the included reactions has never been studied in laboratory experiments or by theoretical computations.
Even in the cases of reactions that have been investigated,  the experimental conditions rarely reproduce those typical of Solar-type star forming regions, either regarding the temperature or the density. In the absence of laboratory data, rate coefficients and their temperature dependence are mainly estimated with some chemical intuition or by drawing analogies with similarly known processes.
When the data are available outside the temperature or density range of relevance, they are used as such or are extrapolated. 

However, both approaches can be seriously wrong because: 
1) small details in the molecular structure can induce a huge change in the chemical behaviour and reasoning by analogy can cause severe mistakes;
2) the extrapolation of the temperature dependence of rate coefficients outside the investigated range can be very risky because a change in their reaction mechanism can alter the temperature dependence in 
non-Arrhenius reactions, while a too high density can lead to the
stabilization of reaction intermediates or affect the reaction outcome via secondary or three-body collisions.

Writing down schemes of unknown reactions can also be dramatically wrong when not even the enthalpy of formation of reactants and products is known. 
This is particularly true when unstable species such as ions or radicals, or other uncommon species are involved. 
For instance, \citet{Tinacci2022-clean} carried out a systematic work to "clean" the gas-phase reaction network \textsf{GRETOBAPE} \footnote{\url{https://aco-itn.oapd.inaf.it/aco-public-datasets/theoretical-chemistry-calculations}}, which is based on the KIDA network, from the most obvious source of error: the presence of endothermic reactions not recognized as such. 
Using the enthalpy of formation derived by electronic structure calculations \citep{Woon2009,Tinacci2021} about 5\% of the reactions in the original network resulted in endothermic reactions erroneously reported as barrieless or with too low activation energies.

Only one public database exists so far with a list of grain-surface reactions, posted on the KIDA website.
The list is based on the \cite{Garrod2008} and \cite{Garrod2013} reaction network and is presented in \cite{Ruaud2015}.

\begin{figure*}[th]
 \epsscale{1.5}
 \plotone{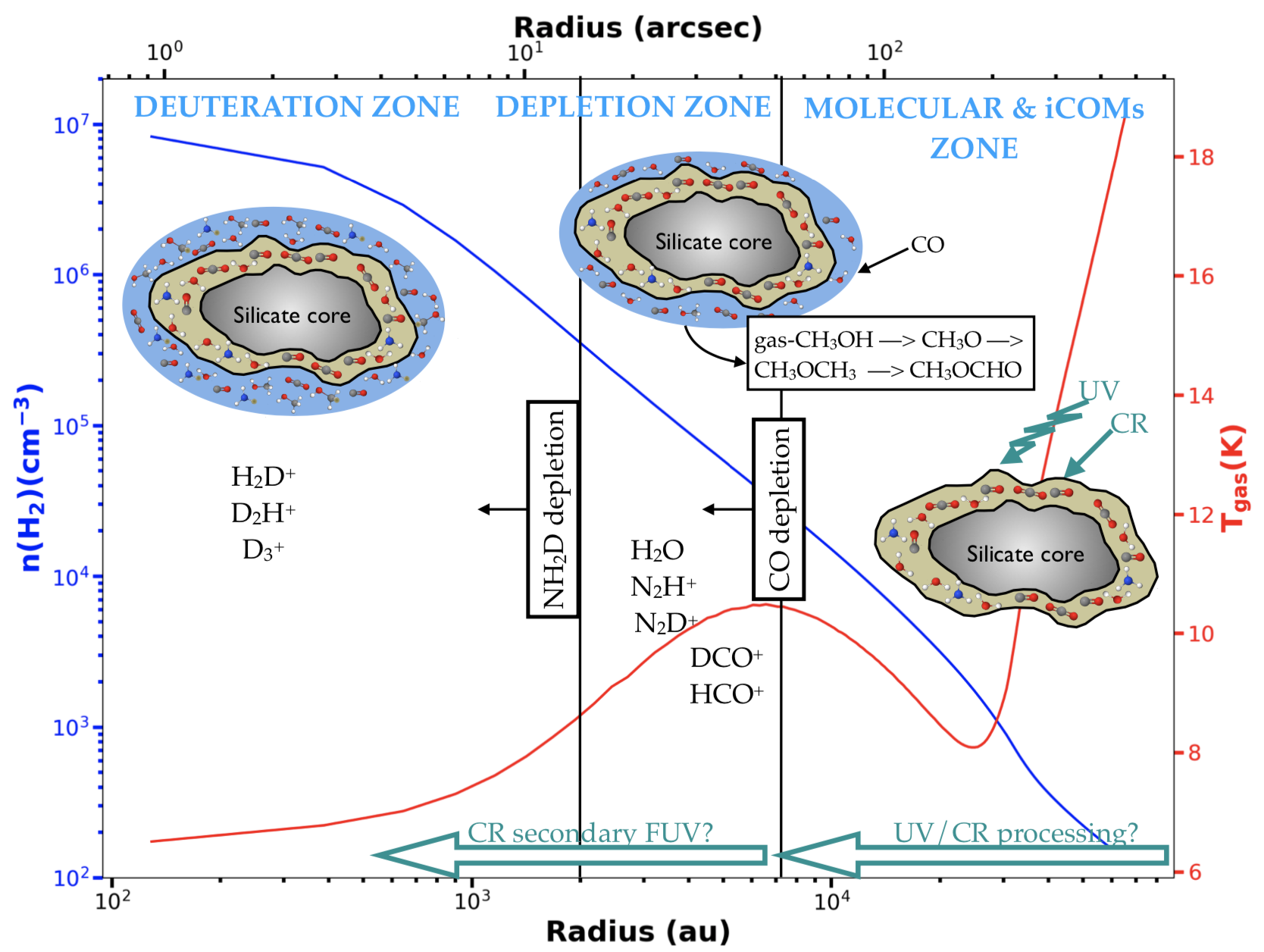}
 \caption{\small Sketch of the physical-chemical structure of the prototypical prestellar core L1544 (see text and \citet{Keto2010} for the temperature and density profiles). 
 The X-axis reports the distance from the center of the prestellar core in au (bottom: assuming a distance of 140 pc) and arcsec (top);
 The Y-axis reports the density (left: blue curve) and gas/dust temperature (right: red curve). 
 The two vertical black lines show the location where CO and NH$_2$D start to catastrophically freeze-out onto the grain mantles, disappearing from the gas-phase.
 They identify the radius of the CO and NH$_2$D depletion zones. 
 The dust grains are illustrated as silicate cores covered with the grain surface chemical composition (yellow) coated with the icy grain mantle (in blue; thicker in the cold dense\index{Dense cores} interiors of the core). 
 The temperature bump at $\sim 7000$ au corresponds to the so-called photo-desorption layer, a chemical zone between the FUV starlight pervading interstellar space and the FUV produced by CR \citep[see, for example,][]{Hollenbach2009,Caselli2012b}.
   }
 \label{fig:sec3-PSCstructure}
\end{figure*}

\section{\textbf{PRESTELLAR CORES: COLD DOES NOT MEAN POOR IN ORGANICS}}\label{sec3:psc}

\subsection{Overview of the physical and chemical structure}\label{subsec:sec3-overview}

The first step in the Solar-type star formation is represented by the starless cores\index{Starless cores}, dense\index{Dense cores}  condensations \citep{Benson1989} in the filaments of molecular clouds \citep[e.g.][]{Andre2019a,Konyves2015, Ladjelate2020,Fiorellino2021}. 
Prestellar cores\index{Prestellar cores} represent a particular sub-group of the starless cores\index{Starless cores}. 
They are gravitationally-bound objects \citep{Andre2014}, with evidence of contraction motions \citep{Keto2015,Lee2011} caused by the fact that matter slowly accumulates towards the center under gravitational contraction counteracted by magnetic pressure and low levels of turbulence \citep{McKee2007}. 
Prestellar cores\index{Prestellar cores} are relatively rare, as their life times are short \citep{Konyves2015}. 
The best studied ones are in the Solar neighbourhood, at distances below 200 pc. 
Their sizes are about 10$^4$ au, with centrally concentrated structures \citep{Ward-Thompson1999}, reaching central visual extinctions higher than 50 mag and densities higher than 10$^5$ cm$^{-3}$ \citep{Crapsi2005,Keto2008}, where the temperature drops to about 7\,K \citep[e.g.][]{Crapsi2007,Pagani2007}. 
They can show evidence of accretion of material from the parental cloud 
\citep[e.g.][]{Lee1999} as well as central contraction \citep{Caselli2012b}. 

In general, prestellar cores\index{Prestellar cores} are ideal regions to study the initial steps of chemistry, with no strong temperature gradient, no shocks\index{Shocks}, absence of internal heating and, therefore, no outflows\index{Molecular outflows} and protostellar feedback. 
The general physical-chemical structure of the prototypical prestellar core\index{Prestellar cores}, L1544\index[obj]{L1544}, is shown in the scheme of Fig. \ref{fig:sec3-PSCstructure}.
The density increases towards the center while the temperature decreases and reaches very low values ($\leq$10 K).
This physical structure defines three major zones from the chemical point of view: \\
\textit{1-} The molecular and iCOM zone, where several molecules and, specifically, iCOMs are observed.\\
\textit{2-} The depletion zone, where molecules such as CO freeze-out onto the grain mantles and disappear from the gas-phase.\\
\textit{3-} The deuteration\index{Deuteration} zone, where only H-bearing and D-bearing species remain gaseous.

We will briefly describe the three zones, starting from the first two.
We will, on the contrary, discuss in detail the organic chemical complexity in the prestellar\index{Prestellar cores} and starless cores\index{Starless cores}, respectively.

\subsection{Depletion and deuteration zones}\label{subsec:sec3-depl-duet}

In the central regions of prestellar cores\index{Prestellar cores}, the large densities and low temperatures lead to a zone where, with the exception of those containing only H and D atoms, gaseous species freeze-out onto the cold grain surfaces, forming thick icy mantles. 
In particular, CO depletion \citep[e.g.][]{Caselli1999} goes hand in hand with a high deuterium fractionation, meaning a high D/H ratio compared to the cosmic D/H value of $1.6 \times 10^{-5}$ \citep{Linsky2007}, in the many detected deuterated\index{Deuteration} molecules in the gas-phase \citep{Caselli2002b,Bacmann2003,Crapsi2005,Ceccarelli2007}. 
This process can be explained by fast reactions involving the abundant ion H$_{3}^{+}$, a product of the CR ionisation of H$_{2}$, with HD, which is the main reservoir of D-atoms in cold molecular gas. 
The D-atoms are then transferred to H$_{2}$D$^{+}$, D$_{2}$H$^{+}$ and even D$_{3}^{+}$ \citep[e.g.][]{Millar1989,Vastel2006}, before being transferred to gas-phase species (e.g. N$_{2}$D$^{+}$, DCO$^{+}$ and NH$_{2}$D) and, via dissociative recombination, to the D atoms landing on the dust grain surfaces.
The latter lead to an increase of the D/H atomic ratio, by several orders of magnitude above the D cosmic abundance, and an enhanced deuteration\index{Deuteration} of the molecules synthesised on the grain surfaces (e.g. H$_{2}$CO, CH$_{3}$OH, H$_2$S and H$_{2}$O). 
For example,  deuterated\index{Deuteration} methanol, CH$_2$DOH, has been detected and mapped toward the prototypical prestellar core\index{Prestellar cores} L1544\index[obj]{L1544} \citep{Bizzocchi2014,Chacon2019}.
The deuterated\index{Deuteration} ions, H$_{2}$D$^{+}$ and D$_{2}$H$^{+}$, have been detected in only few prestellar cores\index{Prestellar cores}, as these observations are particularly difficult \citep{Caselli2003,Caselli2008,Vastel2004,Parise2011,Harju2017,Brunken2014}.

While CO depletion is important in the central 7000 au region of L1544\index[obj]{L1544}, some N-bearing species such as N$_2$H$^+$ and NH$_3$ seem to trace the core center defined by the mm and sub-mm observations of the dust continuum emission \citep[e.g.][]{Tafalla2004,Spezzano2017}. 
This was unexpected since early measurements of the binding energy (BE) of N$_{2}$ and CO provided similar values \citep{Bisschop2006}.
However, more recent laboratory measurements \citep{Fayolle2016,Nguyen2018} and theoretical QM calculations \citep{Ferrero2020} show that the two species have a distribution of BEs which could possibly explain the observed segregation.
In less dense regions, the different timescales for the formation of CO versus N$_2$ and NH$_3$ (which also is formed from N$_2$ in the gas-phase), about a factor of ten, could also explain the observed difference in the CO and NH$_3$ emission.
In agreement with these new estimates of the N$_{2}$ and CO BEs, high-J transition  observations of N$_{2}$D$^{+}$ have provided evidence of depletion of this molecular ion toward a prestellar core\index{Prestellar cores} center \citep{Redaelli2019}, without resolving the depletion region. 
More recently, with the high sensitivity and spatial resolution obtained with ALMA\index{ALMA}, the presence of a complete depletion zone, where  gaseous ammonia also disappears, has been discovered towards L1544\index[obj]{L1544} within a 1700 au radius \citep{Caselli2022}, as well as towards H-MM1 \citep{pineda2022}.
This implies that most ($\sim 99.99$\%) species heavier than He are locked onto the dust-grain ice mantles \citep{Caselli2022}. 

\subsection{Organic chemistry and iCOMs in L1544} \label{subsec:sec3-l1544}

As mentioned above, one of the breakthrough novelties in organic chemistry since the publication of the Protostars \& Planets VI volume, is the  detection of iCOMs in prestellar cores\index{Prestellar cores}.
As we will discuss later, the presence of gaseous iCOMs in these cold objects poses a great challenge to the astrochemical models for various reasons and, consequently, provides strong constraints on how these molecules are formed and why they are present in the gas-phase.
Here we will review the observations, starting with the case of L1544\index[obj]{L1544}, which is the best studied prestellar core\index{Prestellar cores}, then discuss the other prestellar\index{Prestellar cores} and starless cores\index{Starless cores}.

As already hinted at above, among the known prestellar cores\index{Prestellar cores}, L1544\index[obj]{L1544} stands out.
It is one of the densest\index{Dense cores!chemistry} and most dynamically evolved ones, towards which H$_2$D$^+$ and H$_{2}$O, for example, were detected for the first time in the cold ISM \citep{Caselli2003,Caselli2012b}.
The physical and dynamical structure of L1544\index[obj]{L1544} has been computed by \citet{Keto2010} and \citet{Caselli2012b}, who predict a central temperature of $\sim 7$ K, as actually measured by Crapsi et al. (2007), and a peak density $\sim 10^{7}$ cm$^{-3}$ (see Fig. \ref{fig:sec3-PSCstructure}).  

The L1544\index[obj]{L1544} chemical composition and structure has been the target of several observational projects, in particular the unbiased spectral survey ASAI (Astrochemical Surveys At IRAM\index{IRAM}) \citep{Lefloch2018}\footnote{\url{https://www.oan.es/asai/}}.
The survey covered the 70--110 GHz range and led to the detection of a rich chemistry in L1544\index[obj]{L1544} and a better understanding of its molecular and iCOM zone. 
For instance, more than 20 S-bearing species have been detected in this survey and their modeling led to the conclusion of a high elemental sulphur depletion in the gas-phase \citep[e.g.][]{Vastel2018a,Cernicharo2018}. 
Similarly, many N-bearing species have been detected, which helped to better constrain the N chemistry \citep[e.g.][]{Vastel2015,Vastel2019}.

\begin{figure*}
    \centering
    \includegraphics[width=14cm]{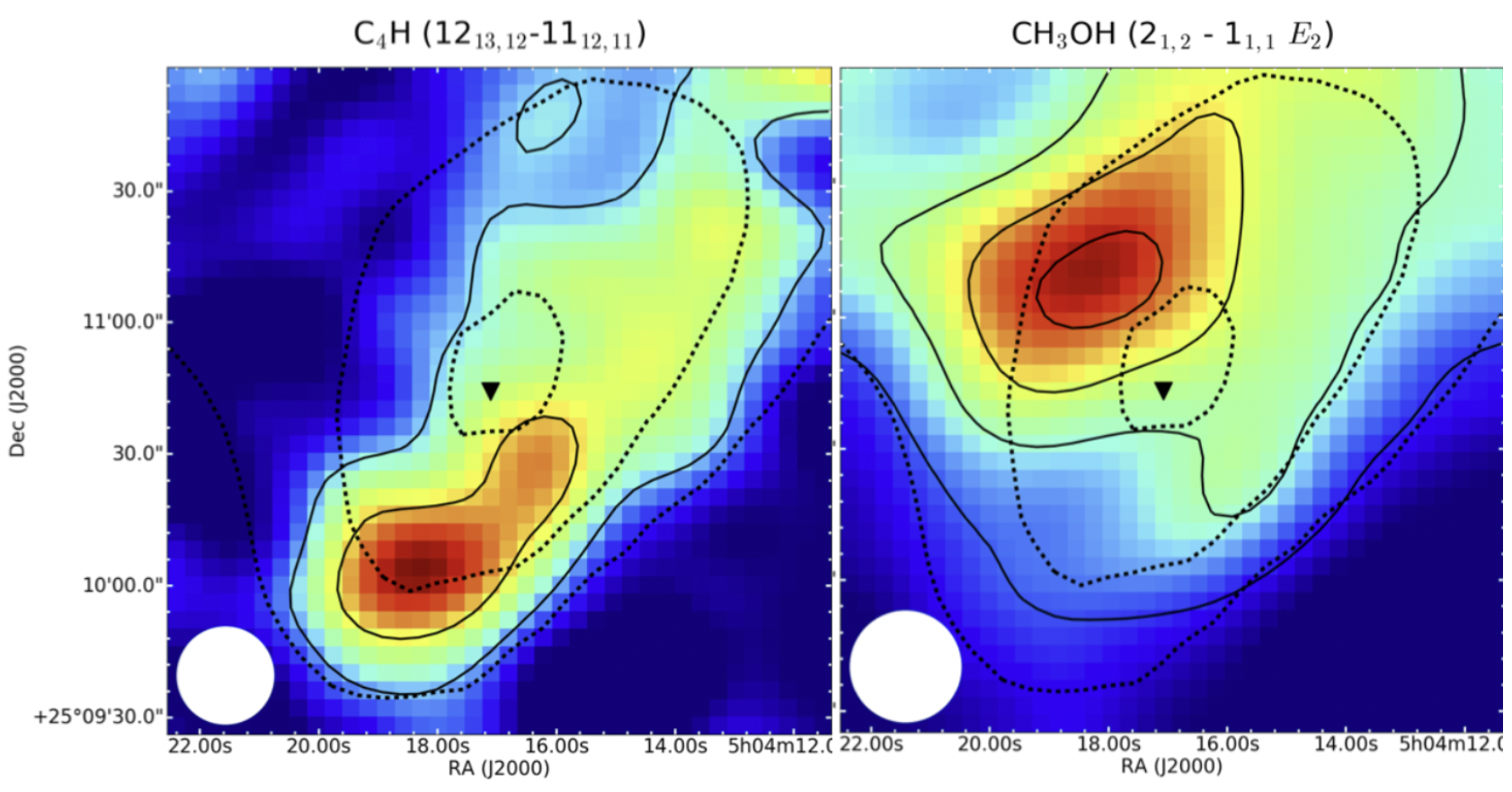}
    \caption{Map of the line emission of butadiynyl (C$_4$H: left panel) and methanol (CH$_3$OH: right panel) showing the different spatial distribution of iCOM (methanol) and C-chain (butadiynyl) species toward the L1544 prestellar core. 
    The solid lines represent contours of the molecular integrated emission. 
    The black triangle marks the dust peak and the dashed lines its extent. 
    The white circles represent the beam of the observations.
    Adapted from \cite{Spezzano2017}. }
    \label{fig:sec3-L1544map}
\end{figure*}

More relevant to this review, several iCOMs have been detected towards L1544\index[obj]{L1544}, such as methanol, acetaldehyde, dimethyl ether, methyl cyanide and vinyl cyanide \citep{Vastel2014,Bizzocchi2014,Jimenez2016,Vastel2019}. 
The study of methanol, which presents the brightest lines among the detected iCOMs, is very instructive to understand the methanol spatial origin.
First, from the non-LTE analysis of the methanol lines, \cite{Vastel2014} found that the bulk of the emission originates in the external ($\sim 8000$ au from the center), relatively dense ($\sim 2\times10^4$ cm$^{-3}$) and cold ($\sim 10$ K) layers of L1544\index[obj]{L1544}, marked as molecular and iCOM zone in Fig. \ref{fig:sec3-PSCstructure}.
The average abundance of methanol in the external layer is about $\sim 6\times 10^{-9}$ with respect to H$_2$.
The maps of the methanol emission by \cite{Bizzocchi2014} and \cite{Spezzano2016} and, at higher spatial resolution, by \cite{Punanova2018} confirmed this interpretation and add an important information on the methanol distribution.
The methanol emission is not uniformly distributed around the center of L1544\index[obj]{L1544}, but it is peaked towards the northeast, which is more shielded from interstellar UV photons than the southwest border of the L1544\index[obj]{L1544} condensation (Fig. \ref{fig:sec3-L1544map}).
In the same vein, \cite{Jimenez2016} showed that the abundance  of other  O-bearing iCOMs  is larger towards the methanol peak than towards its center by a factor 2--10 (within the uncertainties), again showing a definitive role of UV in the reducing gaseous iCOMs abundance.
The abundance of iCOMs is about 10--100 times smaller than that of methanol.
Table \ref{tab:sec3-psc-abu} lists the measured abundance of iCOMs in L1544\index[obj]{L1544}.

\begin{table}[]
    \centering
    \begin{tabular}{c|c|c}
        \hline
        Species  & \multicolumn{2}{c}{Column Density} \\
                 & \multicolumn{2}{c}{($\times 10^{12}$ cm$^{-2}$)} \\
                 & Center & Methanol peak \\
        \hline
        CH$_3$OH & 30 &  -- \\
        CH$_3$CHO & 0.5-1.2 & 3.2\\
        HCOOCH$_3$ & 4$\pm$4 & 2.3$\pm$1.4 \\
        CH$_3$OCH$_3$ & 1.5$\pm$0.2 & 0.77$\pm$0.16 \\
        t-HCOOH   & 0.5 & -- \\
        CH$_2$CHCN & 1.2$\pm$0.8 & 0.6$\pm$0.5 \\
        \hline 
    \end{tabular}
    \caption{Column densities of iCOMs detected towards L1544.
    The uncertainties, if reported, are those quoted in the original works: \cite{Vastel2004} and \cite{Jimenez2016}.}
    \label{tab:sec3-psc-abu}
\end{table}

While the northeast methanol peak is enriched with iCOMs, the southeast part of the L1544\index[obj]{L1544} prestellar core\index{Prestellar cores} is enriched in hydrocarbons and C-bearing chains.
Specifically, \cite{Spezzano2016} and \cite{Spezzano2017} found that the c-C$_2$H$_3$ and C$_4$H abundance peaks towards the southeast, where there is a sharp edge and, consequently, it is more exposed to the interstellar radiation field (Fig. \ref{fig:sec3-L1544map}).
Remarkably, in this region, several carbon chains, such as C$_4$H, C$_6$H and C$_6$H$^{-}$ are abundant \citep{Gupta2009}.
The C-chain chemistry is so efficient that even long cyanopolyynes with up to nine carbon atoms (HC$_9$N) have been detected \citep{Bianchi2022-cyanopolyynes}.

The ensemble of these observations seems to provide a very clear message.
iCOMs and hydrocarbons/C-chains follow quite alternative routes for the organic chemistry in prestellar cores\index{Prestellar cores}.
In the regions where iCOMs are abundant, hydrocarbons/C-chains are scarce and vice versa.
The presence of strong UV illumination seems, at the moment, the major player in determining the divergence.
We will discuss the consequences of this finding in Sec. \ref{sec4:protostars}.

\subsection{iCOMs in other prestellar cores}\label{subsec:sec3-icoms-psc}
The presence of iCOMs is not a prerogative of L1544\index[obj]{L1544}.
Although difficult to detect because their lines are weak, both because of the relatively low abundance and gas temperature, a few iCOMs have been detected in several prestellar cores\index{Prestellar cores}.
Notably, the survey by \cite{Scibelli2020} towards 31 starless cores\index{Starless cores}, of which many are prestellar cores\index{Prestellar cores} in the Taurus Molecular Cloud\index{Taurus Molecular Cloud}, showed that methanol is present in all the targeted prestellar cores\index{Prestellar cores} and acetaldehyde in 70\% of the sample.
Likewise, \citet{Punanova2022} mapped the methanol emission in the Taurus\index{Taurus Molecular cloud} L1495 filament and found emission associated with seven starless cores\index{Starless cores}.
Therefore, iCOMs are likely a common product of the cold chemistry.

As in L1544\index[obj]{L1544}, the methanol emission in the \cite{Scibelli2020} sources originates in a cold (T = 6--8 K) gas, and the derived column density ($\sim 10^{13}$ cm$^{-2}$) is very similar to that found in L1544\index[obj]{L1544}  \citep[$\sim 3\times10^{13}$ cm$^{-2}$:][]{Vastel2014}.
Assuming that acetaldehyde emits in the same gas as methanol, the CH$_3$CHO/CH$_3$OH is 0.02--0.2, with most sources in the middle of the distribution \citep{Scibelli2020}\footnote{However, we emphasize that the error bars are relatively large because of the few lines detected.}.
The comparison with the CH$_3$CHO/CH$_3$OH$\sim 0.02$ in 
L1544\index[obj]{L1544} \citep{Vastel2014} would suggest a certain similarity among the prestellar cores\index{Prestellar cores}, when the uncertainties are taken into account, even though this conclusion needs additional observations to rest on a solid basis.

In a few additional prestellar cores\index{Prestellar cores}, iCOMs other than methanol and acetaldehyde have been detected.
We will briefly describe their cases in the following.

\textit{L1689B}\index[obj]{L1689B} is a well-studied prestellar core\index{Prestellar cores} in the $\rho$ Ophiuchus star-forming region, which shows a moderate CO depletion \citep{Bacmann2002,Jessop2001}, probably sign of youth or a large CR/FUV irradiation.
L1689B\index[obj]{L1689B} is one of the first prestellar cores\index{Prestellar cores} where three iCOMs have been detected: acetaldehyde, dimethyl ether and methyl formate \citep{Bacmann2012}. 
The acetaldehyde column density is very similar to that found in L1544\index[obj]{L1544} and the \cite{Scibelli2020} cores, 1--3$\times 10^{13}$ cm$^{-2}$.
Likewise, dimethyl ether and methyl formate column densities are about 3--10 times lower than that of acetaldehyde.

\textit{H-MM1} is an intriguing prestellar core\index{Prestellar cores}, located in the eastern side of the L1688 cloud of the $\rho$ Ophiuchus complex, and illuminated on one side by a strong UV field from nearby B-type stars. 
The asymmetry of the FUV field illumination causes an enhancement of methanol emission in the shielded side \citep{Harju2020}, similarly to what happens in L1544\index[obj]{L1544} (see above).
Remarkably, also in H-MM1, as in L1544\index[obj]{L1544}, ammonia is  depleted towards its center \citep{pineda2022}.

\textit{B1-b}: Finally, the first evidence of efficient production of iCOMs in cold environments was reported by \citet{Oberg2010} and \citet{Cernicharo2012} toward B1-b, a dense core\index{Dense cores} which hosts a Solar-type protostar. 
These authors reported the detection of three iCOMs (methanol, methyl formate and acetaldehyde) towards the quiescent core and not the outflow\index{Molecular outflows} driven by the protostar. 
However, the presence of the outflow\index{Molecular outflows} could affect the observed species, as also discussed by \citet{Fuente2016}, who observed the same source and found enhanced sulphur chemistry, typically an indication of the presence of shocks\index{Shocks!chemistry} \citep[as sulphur is heavily depleted in starless cores\index{Starless cores}; e.g.][]{Ruffle1999}. 

\subsection{iCOMs in starless cores}\label{subsec:sec3-icoms-starless}

Starless cores\index{Starless cores}, which do not present evidence of infalling motion, have a relatively flat density profile and motions sometimes displaying oscillations or even expansion \citep{Lada2003,Tafalla2004}.
Thus, a significant fraction of them could be transient and not evolve toward the prestellar\index{Prestellar cores} phase, i.e. the star formation \citep{Belloche2011}.

We have already mentioned above the study by \cite{Scibelli2020}, which showed a high frequency of targeted starless cores\index{Starless cores} presenting acetaldehyde and all of them methanol emission.
Here we discuss three cases.

\textit{L1521E} is a starless core\index{Starless cores} with no indications of infall motions and, on the contrary, signs of a dynamical and chemical youth: modest central density ($10^5$ cm$^{-3}$) and CO depletion ($\leq 2$).
\cite{Scibelli2021} detected acetaldehyde, dimethyl ether, methyl formate and acrylonitrile in addition to methanol.
The column densities of these iCOMs are about $10^{12}$ cm$^{-12}$, similar therefore to the prestellar cores\index{Prestellar cores}, within the measurement uncertainties.
The youth of L1527E ($\leq 10^5$ yr) and the presence of iCOMs at an early evolutionary stage testifies that the formation of the latter is relatively fast, another constraint for astrochemical models. 

\textit{L1448}\index[obj]{L1448-mm} is also a starless core\index{Starless cores} where \citet{Jimenez2021} found interestingly relatively high abundances of N-bearing iCOMs but not O-bearing ones, except for methanol.

\textit{TMC-1}\index[obj]{TMC-1}, in the Taurus Molecular Cloud\index{Taurus Molecular Cloud} complex, is by far the most studied starless core\index{Starless cores}, in particular the position known as the "cyanopolyyne peak".
TMC-1\index[obj]{TMC-1} has a volume density similar to that in the external layer of L1544\index[obj]{L1544} (at $\sim 10^4$ au), with n(H$_2$) of a few $\times$10$^4$ cm$^{-3}$. 
What made it famous was the detection of abundant large cyanopolyynes, up to 11 carbons (HC$_{11}$N) \citep{Loomis2021}.
This has been interpreted by many authors as an indication of a chemically young dense core\index{Dense cores}  \citep[e.g.][]{Agundez2013}, where carbon is not yet largely locked into CO and, therefore, available for the observed rich carbon chemistry.

Indeed, TMC-1\index[obj]{TMC-1} is the site of many first detections of C-bearing large molecules, such as HC$_4$NC \citep{Cernicharo2020, Xue2020}, HC$_5$NH$^+$ \citep{Marcelino2020}, H$_2$CCCHC$_3$N \citep{Shingledecker2021}, indene (C$_9$H$_8$) \citep{Burkhardt2021,Cernicharo2021a}, propargyl radical (CH$_2$CCH) \citep{Agundez2021}, propargyl cyanide (HCCCH$_2$CN) \citep{McGuire2020}, allenyl acetylene (H$_2$CCCHCCH) \citep{Cernicharo2021d}, cyclic aromatic molecules such as benzonitrile (c-C$_6$H$_5$CN) \citep{McGuire2018}, cyclopentadiene (c-C$_5$H$_6$) \citep{Cernicharo2021a} and cyanocyclopentadiene (c-C$_5$H$_5$CN) \citep{McCarthy2021}. 
Remarkably, the detection of benzyne (ortho-C$_6$H$_4$) led \citet{Cernicharo_benzyne2021} to hypothesise that, in TMC-1\index[obj]{TMC-1}, PAHs are formed bottom-up from small carbon chains, via gas-phase reactions of a few radicals with abundant gaseous hydrocarbons, rather than top-down from carbonaceous-grain fragments. 

\subsection{Astrochemical models}\label{subsec:sec3-chem-models}

As mentioned in the Introduction (Sec. \ref{sec1:intro}), the discovery in 2012 of iCOMs in prestellar cores\index{Prestellar cores} (\S ~\ref{subsec:sec3-icoms-psc}) was unexpected and not predicted by the astrochemical models.
It still poses serious challenges to the current models for two major reasons.
The first one is the route of formation of iCOMs for the low temperatures and high densities of the prestellar cores\index{Prestellar cores}.
The second challenge is connected with the very presence of iCOMs in the gas-phase, as the BEs of iCOMs are generally very large and close to that of water \citep[e.g.][see also Sec. \ref{sec2:chemistry}]{Ferrero2020,Ferrero2022,Corazzi2021}.

In 2012, iCOM formation models were prevalently based on the \cite{Garrod2006} scheme, described in Sec. \ref{sec2:chemistry} (see also Fig. \ref{fig:sec2-chem-scheme}).
Briefly, iCOMs were expected to form during the warm-up phase of the dust grains when the protostar switches on and starts to heat the surrounding material \citep[e.g.][]{Viti1999,Garrod2008}, allowing mobility of the radicals trapped within the icy mantles.
These radicals could then react to form iCOMs, which would then be released back in the gas-phase when the dust temperature exceeds the sublimation temperature of the specific species.
However, at the low temperatures found in prestellar cores\index{Prestellar cores}, only hydrogen (and deuterium) atoms can sweep across dust grain surfaces, thus allowing them to react with other species. 
Depletion of carbon, oxygen and nitrogen atoms as well as CO, as evidenced in prestellar cores\index{Prestellar cores} (\S ~\ref{subsec:sec3-depl-duet}), leads to the production of stable species such as methane, water, ammonia and methanol in the icy mantles (see Sec. \ref{sec2:chemistry}). 
These species might be broken into radicals by the external UV photons in the external shells of the core, or by the secondary UV photons produced across the whole core by CR interacting with the abundant H$_2$ molecules. 
The radicals which are not photodesorbed and remain in the icy mantles are, however, too slow at these temperatures to effectively meet together and, therefore, react to produce heavier iCOMs. 
Therefore, it appears difficult for iCOMs other than methanol to be synthesized in the solid phase in the cold prestellar cores\index{Prestellar cores} via this scheme. 

Two alternative routes have since been proposed (see Sec. \ref{sec2:chemistry}): the so-called "non-diffusive grain-surface chemistry" \citep{Jin2020} (see below) and  iCOM formation in the gas-phase \citep{Vasyunin2013,Balucani2015}.
In the latter scheme, iCOMs form in the gas-phase upon release of methanol and other smaller iCOMs bricks from the grain surfaces.
We stress again the difficulty of having iCOMs (including methanol) in the gas-phase, where they are observed (see the discussion in \S ~\ref{subsec:sec2-mantle-desorption}).

Detailed astrochemical models have been so far almost exclusively developed to reproduce the iCOMs observed towards L1544\index[obj]{L1544} \citep{Vasyunin2017,Goto2021}.
These models use the core physical structure described by \citet{Keto2010} and \cite{Chacon2019}, and start with the production of methanol on icy mantles rich in CO, at the edge of the catastrophic freeze-out zone of prestellar cores\index{Prestellar cores} \citep{Caselli1999}. 
Surface CO molecules are then efficiently hydrogenated into CH$_3$OH, while the CO-rich ice below the newly formed CH$_3$OH molecule allows its efficient reactive desorption \citep{Vasyunin2017} following the recipe by \citet{Minissale2016b} (but see the discussion in Sec. \ref{sec2:chemistry}).
Finally, once methanol and other radical/molecules are released into the gas-phase, neutral-neutral reactions can proceed to produce larger iCOMs \citep[e.g.][]{Vasyunin2013,Balucani2015,Vasyunin2017}.
These models predict that the peak of methanol and other iCOMs abundance is in a layer at about 8000 au, in agreement with the observations (see \S ~\ref{subsec:sec3-l1544}).
Besides, the predicted and observed abundances of gaseous methanol and iCOMs agree within a factor 10, which is reasonable within the various uncertainties.
However, the predicted abundance of solid methanol is significantly higher, by about a factor 5, than that observed at the edge of the L1544\index[obj]{L1544} condensation by \citet{Goto2021}.

\begin{figure*}[tb]
 \epsscale{1.4}
 \plotone{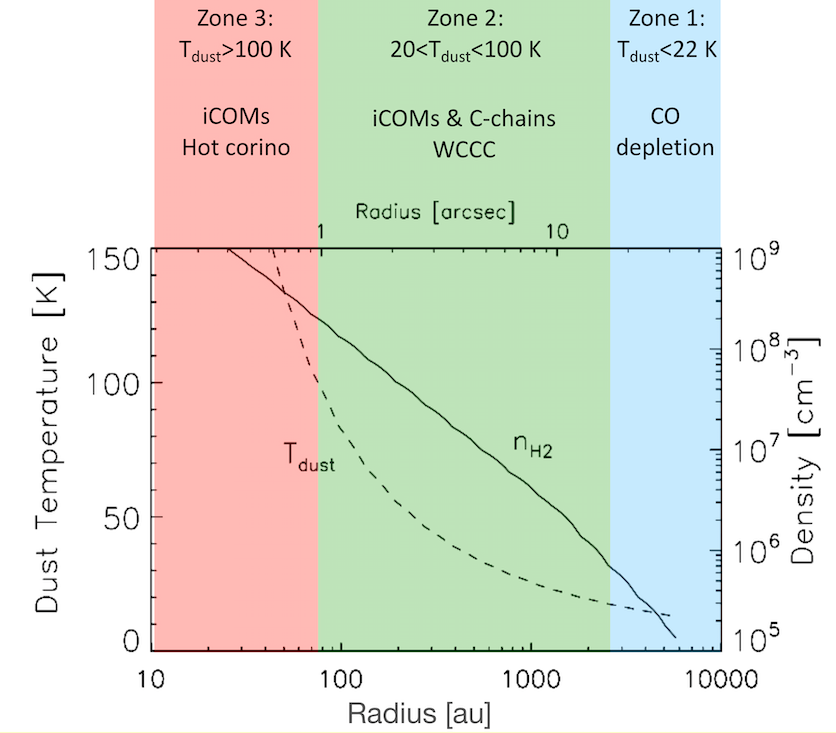}
 \caption{\small Physical and chemical structure of a typical Solar-type Class 0/I protostar. 
 It can be roughly divided into three zones, which are defined by the status of the grain icy mantles formed during the prestellar phase\index{Prestellar cores} (Sec. \ref{sec3:psc}). 
 In zone 1, the dust temperature is lower than the sublimation one of CO-rich ices ($\sim 22$ K); in zone 3 it is higher than the sublimation temperature of H$_2$O-rich ices ($\sim 100$ K); zone 2 is the intermediate, lukewarm envelope region.
 The upper x-axis reports the radius in arcsec, assuming a distance of 120 pc.}
 \label{fig:sec4-physchem-structure}
\end{figure*}

Other models aiming at reproducing the iCOM
abundances observed towards prestellar\index{Prestellar cores} and, more generally, starless cores\index{Starless cores} have also been published.
The model by \citet{Jin2020} is based on the non-diffusive grain-surface chemistry, described in \S ~\ref{subsec:sec2-onlygrainchem}.
This model, which uses a denser condensation than the one used by \citet{Goto2021}, approximately reproduces the abundances of acetaldehyde and methyl formate observed towards L1544\index[obj]{L1544} at about 9000 au, but it fails to reproduce methyl formate by several orders of magnitude.
The model by \citet{Holdship2019b} is based on the Rapid Radical Association, described in \S ~\ref{subsec:sec2-other-formation-theories}.
These authors tried to reproduce the abundances of the iCOMs observed towards TMC-1\index[obj]{TMC-1} and concluded that the Rapid Radical Association is not the dominant mechanism for iCOM formation.

As a final note, the authors of the above models caution that "a chemical model is built on numerous parameters that we do not yet understand well" \citep{Goto2021}, both those regarding the microphysics of the involved processes (e.g., efficiencies of grain-surface reactions, rate constants of gas-phase reactions, binding energies, diffusion efficiencies on the surface: Sec. \ref{sec2:chemistry}) and the macroscopic ones, namely the profiles of the density, dust and gas densities and temperatures, dust grain sizes, UV illumination, and CR irradiation.
For example, in the initial work of \citet{Vasyunin2017}, acetaldehyde was produced via a gas-phase reaction  with an impossibly large reaction rate \citep{Vazart2020}, and reduced in  successive works by \citet{Jimenez2021} and \citet{Scibelli2021}.

Much progress is needed on all these fronts and the next few years promise to be crucial ones. 
For example, on the observational point of view, JWST\index{JWST} is expected to provide maps of methanol ice abundance in prestellar cores\index{Prestellar cores}, such as that towards L1544\index[obj]{L1544} by \cite{Goto2021} but more sensitive, and perhaps maps of other iCOMs.
Likewise, ALMA\index{ALMA}, NOEMA\index{NOEMA} and VLA\index{VLA} will provide more detections of iCOMs on a larger number of prestellar cores\index{Prestellar cores} in different environments allowing a better understanding of the influence of the environment on the early organic chemistry.

%

\section{\textbf{PROTOSTARS: THE RETAIL SHOPS OF ORGANICS}}\label{sec4:protostars}

\begin{figure*}[h]
 \epsscale{2.0}
 \plotone{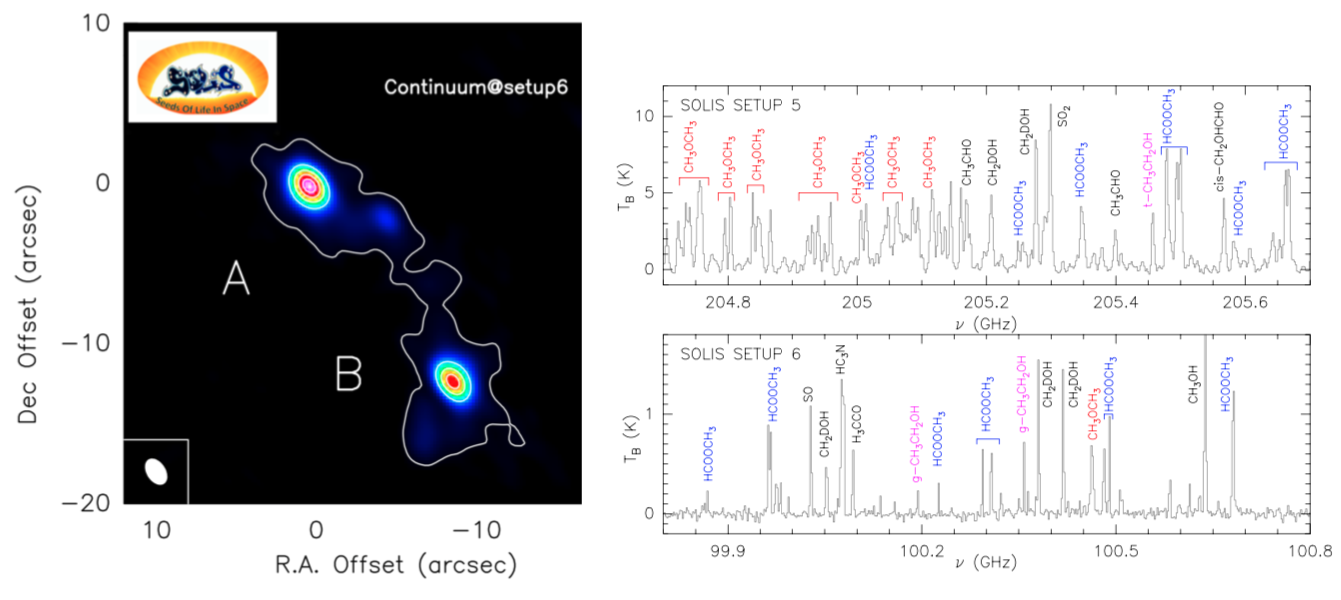}
 \caption{\small Image (left panel) of the hot corino of the Class I protostellar binary system NGC1333-SVS13 \citep{Bianchi2022-svs13} and the spectrum towards SVS13-A (right panel) in the mm obtained by the NOEMA\index{NOEMA} Large Program SOLIS \citep[Seeds Of Life In Space: ][]{Ceccarelli2017}.}
 \label{fig:sec4-hotcorinoSVS13}
\end{figure*}

\subsection{Overview of the physical and chemical structure of protostars}\label{subsec:sec4-overview}
When the free collapse of matter towards the center starts, a protostar is formed.
Briefly, a protostar is made up of four main components: the central object that will eventually become the new star, the infalling envelope from which the central object accretes matter, the circumstellar disk in which eventually planets and comets\index{Comets} will form, and an outflow\index{Molecular outflows} of ejected material.
The latter two components, the disk and the outflow\index{Molecular outflows}, will be discussed in Secs. \ref{sec5:disks} and \ref{sec6:outflows}. 
Here we focus on the infalling envelope and the organic chemistry during the early phases of the protostar, when the observed sources are called Class 0\index{Class 0 sources} and I. 

As for the prestellar cores\index{Prestellar cores} (Sec. \ref{sec3:psc}), the chemical structure and evolution of the matter are tightly linked to its physical structure and evolution (e.g. the density and temperature profiles), as well as the presence of ionising agents (e.g. UV photons and CR).
Given the presence of a central source of heat and centrally peaked infalling envelope, the general physical-chemical structure of a protostar can be divided into three major zones, illustrated in Fig. \ref{fig:sec4-physchem-structure}, as follows:\\
\textit{1-} A cold envelope, characterised by a dust temperature lower than about 22 K, where several molecules, including CO, are largely frozen into the grain mantles, as in the prestellar cores\index{Prestellar cores} described in Sec. \ref{sec3:psc}.\\
\textit{2-} A lukewarm region, having a dust temperature between the CO and H$_2$O sublimation temperatures (approximately 22 and 100 K, respectively), where organics can be present either in the form of carbon chains, iCOMs or both.
In this region, iced methane sublimates: if abundant enough (see \S ~ \ref{subsec:sec4-model}), the injection of methane gives rise to a zone enriched with unsaturated carbon-chains, such as C$_4$H or large cyanopolyynes. 
The sources where this zone is dominating the chemical appearance are called WCCC (Warm Carbon Chain Chemistry) sources \citep{Sakai2013}.\\
\textit{3-} A hot region, characterised by a dust temperature higher than about 100 K, where the water-rich ices sublimate injecting into the gas-phase whatever species is trapped in them.
If frozen methanol and other relatively complex species are abundant enough, they give rise to a zone very enriched with iCOMs. 
This zone is called hot corino which, in its original definition, is a compact source ($\leq 100$ au in radius) possessing high temperatures ($\geq 100$ K), high densities ($\geq 10^7$ cm$^{-3}$) and showing the presence of iCOMs in the region where the water-rich grain mantles sublimate \citep{Ceccarelli2004}.
The typical spectrum of a hot corino is shown in Fig. \ref{fig:sec4-hotcorinoSVS13}.

WCCC sources and hot corinos\index{Hot corinos} are two chemically distinct classes of Solar-type Class 0\index{Class 0 sources}/I\index{Class I sources} protostars\index{Protostars}, meaning that, observationally, their spectra are either dominated by unsaturated carbon-chains or iCOMs.
It is not by chance that the first two distinct chemical classes of protostars\index{Protostars}, WCCC sources and hot corinos\index{Hot corinos}, are based on which C-bearing species dominate their spectra, as carbon is the element that allows the richest and most varied chemistry in space (and on Earth!).
At present, these two classes are the only ones recognised to be present in Class 0\index{Class 0 sources}/I\index{Class I sources} protostars\index{Protostars}, even though there could probably be a differentiation also with respect to N- and S-bearing molecules.
As often happens, hot corinos\index{Hot corinos} and WCCC sources are the two extremes, and hybrid sources, where the WCCC and hot corino chemistry appear on different scales, have been found \citep[e.g. B335 and CB68:][]{Imai2016,Oya2017,Higuchi2018} \citep[see also][for a discussion on the contribution from the surrounding environment]{bouvier2020}. 
As we will discuss in \S ~\ref{subsec:sec4-sources-census}, the number of known protostars\index{Protostars} with a well enough studied chemical composition is very small at present, so that we are strongly biased in our knowledge towards the hot corinos\index{Hot corinos} and WCCC sources.
With the improving observational facilities and the increasing number of observational surveys, the situation promises to evolve in the next few years. 
For example, the ALMA\index{ALMA} Large Program FAUST\footnote{{\it http://faust-alma.riken.jp}} \citep[Fifty AU STudy of the chemistry in the disk/envelope system of Solar-type protostars:][]{Codella2021} is designed to reveal the chemical structure of several Class 0\index{Class 0 sources}/I\index{Class I sources} sources.

\subsection{The census of hot corinos and WCCC sources}\label{subsec:sec4-sources-census}

\paragraph{Hot corinos}
The first official detection of a hot corino, IRAS16293-2422\index[obj]{IRAS16293-2422} (hereinafter IRAS16293), dates back to 2003 \citep{Cazaux2003}, although its existence was suggested earlier by the observed formaldehyde abundance jump in its inner region \citep{Ceccarelli2000b} and the tentative identification of several iCOMs \citep{Ceccarelli2002c}.
Later interferometric observations showed that actually IRAS16293\index[obj]{IRAS16293-2422} harbors two hot corinos\index{Hot corinos}, A and B, respectively \citep{Kuan2004,Bottinelli2004b,Jorgensen2016}.
Soon after the first discovery, hot corinos\index{Hot corinos} were also found towards NGC1333-IRAS4A\index[obj]{NGC1333-IRAS4A} \citep{Bottinelli2004a}, NGC1333-IRAS2A\index[obj]{NGC1333-IRAS2A}
\citep{Jorgensen2005b} and NGC1333-IRAS4B\index[obj]{NGC1333-IRAS4B} \citep{sakai2006,Bottinelli2007}.
Then for several years no new hot corinos\index{Hot corinos} have been discovered, despite studies of formaldehyde and methanol emission had suggested hot corinos\index{Hot corinos} also towards L1448-mm\index[obj]{L1448-mm}, L1448-N\index[obj]{L1448-N} and L1157-mm\index[obj]{L1157-mm} \citep{Maret2005,Jorgensen2005aJ}.
Only in 2014, the number of hot corinos\index{Hot corinos} with the detection of at least one iCOM other than methanol almost doubled, adding to the list B1-a, SVS4-5 and B5-IRS1 \citep{Oberg2014}.
These first seven hot corinos\index{Hot corinos} were all discovered using the single-dish telescope IRAM-30m\index{IRAM}.
The advent of ALMA\index{ALMA} and IRAM\index{IRAM}/NOEMA\index{NOEMA} spurred new studies and, finally, new surveys, such as CALYPSO \citep{Belloche2020}, PEACHES \citep{Yang2021} and ORANGES \citep{bouvier2021,Bouvier2022} are steadily increasing the number of known hot corinos\index{Hot corinos}.
Another breakthrough was obtained with the VLA observations at $\sim25$ GHz of the binary system NGC1333-IRAS4A, that showed that the dust is a major obstacle to the hot corino detection \citep{Desimone2020a}.
At present, about two dozens hot corinos\index{Hot corinos} are confirmed (meaning that they present line emission from more than three iCOMs) and more are candidates (mostly when only methanol is detected).

Before 2015, the fraction of known hot corinos\index{Hot corinos} in Class 0\index{Class 0 sources}/I\index{Class I sources} protostars\index{Protostars} was $\geq$50\%, which led to think that almost all Solar-type protostars\index{Protostars} went trough the hot corino phase and that our Solar System may have not been an exception \citep{Ceccarelli2007,Oberg2014}.
However, the number was heavily biased by the search criteria as, given the difficulty of discovering hot corinos\index{Hot corinos}, the most probable sources were searched for.
The new surveys provide some contrasting results in terms of the fraction of hot corinos\index{Hot corinos} in Class 0\index{Class 0 sources}/I\index{Class I sources} protostars\index{Protostars}:
CALYPSO found $\sim30$ \% in a list of 26 protostars\index{Protostars} with a distance $\leq420$ pc, when only sources with more than 3 detected iCOMs, referred in the article as "confirmed" hot corinos\index{Hot corinos}, are counted \citep{Belloche2020}, PEACHES found ($56\pm14$)\% in a large unbiased sample of 50 protostars\index{Protostars} in Perseus \citep{Yang2021}, and ORANGES found ($26\pm23$)\% in an unbiased sample of 19 protostars\index{Protostars} in Orion OMC-2/3 \citep{Bouvier2022}, the last two surveys counting the sources with detected methanol lines.
When comparing the two unbiased samples, ORANGES and PEACHES, which were carried out with comparable spatial resolutions (in au) and sensitivities, there is a marginal, but potentially very interesting evidence that the rate of hot corinos\index{Hot corinos} is affected by the environment.
Specifically, ORANGES studied the Orion region, where both high- and low- mass stars are presently forming, whereas PEACHES studied the Perseus region, where only low-mass protostars\index{Protostars} are present.
In Orion, the presence of high-mass forming stars makes the regions permeated by strong UV and CR irradiation fields \citep[e.g.][]{Ceccarelli2014a,Favre2017}, which may reduce the onset of a complex organic chemistry leading to iCOMs in Solar-type protostars\index{Protostars} (see \S ~\ref{subsec:sec4-model}).
Perseus, on the contrary, is not exposed to high UV/CR irradiation even though it is known to be the theater of titanic clashes between bubbles and clouds \citep{Dhabal2018,Desimone2022}.
Obviously, more regions need to be studied to draw firm conclusions concerning environmental effects, and to what extent they affect the chemical nature of Solar-type protostars\index{Protostars}.

\vspace{-0.3cm}
\paragraph{WCCC sources}
The discovery of the first WCCC protostar arrived in 2008 \citep{Sakai2008}, with the unexpected detection of abundant C$_4$H$_2$, C$_4$H and CH$_3$CCH towards IRAS04368+2557 in L1527\index[obj]{L1527} and no iCOMs containing heavier atoms than C.
Subsequent observations showed that L1527\index[obj]{L1527} is enriched with a variety of unsaturated carbon-chains, including cyanopolyynes with up to 9 carbons (HC$_9$N).
As for the hot corinos\index{Hot corinos}, the early hunt for new WCCC sources ended up in a frustrating addition of only two other protostars\index{Protostars}: B228 \citep{Sakai2009a} and TMC-1A\index[obj]{TMC-1A} \citep{Sakai2016} and a few other candidates.

Again, new surveys were carried out with single-dish telescopes, with the aim of finding more WCCC sources and understanding the phenomenon, and ended up with some contrasting results.
\cite{Graninger2016} studied a sample of 15 Class 0\index{Class 0 sources}/I\index{Class I sources} protostars\index{Protostars} with measured (Spitzer) ice abundances and found the co-existence and even correlation between C$_4$H and CH$_3$OH line emission, suggesting that the observed protostars\index{Protostars} possess both a WCCC and a hot corino zone.
However, \cite{lindberg2016} studied the same species towards 16 protostars\index{Protostars} in the $\rho$ Ophiuchus molecular cloud complex and found no correlation between C$_4$H and CH$_3$OH.
On the contrary, they found a net distinction between WCCC sources and (known) hot corinos\index{Hot corinos}.
The extensive survey of 36 sources in Perseus by \cite{Higuchi2018} used the CCH and CH$_3$OH line emission to discriminate between the two classes.
Based on the two prototypes of each chemical class, IRAS16293\index[obj]{IRAS16293-2422} and L1527\index[obj]{L1527}, \cite{Higuchi2018} assumed that an abundance ratio [CCH]/[CH$_3$OH]$\geq 2$ would define a WCCC source, while [CCH]/[CH$_3$OH]$\leq 0.5$ a hot corino.
With this definition, they found four new WCCC candidates.
The most recent single-dish survey, carried out by \cite{bouvier2020}, used again CCH and CH$_3$OH line emission, towards nine sources in the Orion OMC-2/3 clouds.
All nine sources present a [CCH]/[CH$_3$OH]$\geq 2$, which would lead to conclude that they are all WCCC sources.
However, a careful analysis demonstrates that the single-dish CCH and CH$_3$OH line emission is likely dominated by the Photo-Dissociation Region (PDR) enveloping the OMC-2/3 hosting clouds, making the criterion flawed, when used in combination with single-dish observations only.

%
\begin{figure}[t]
 \epsscale{1.0}
 \plotone{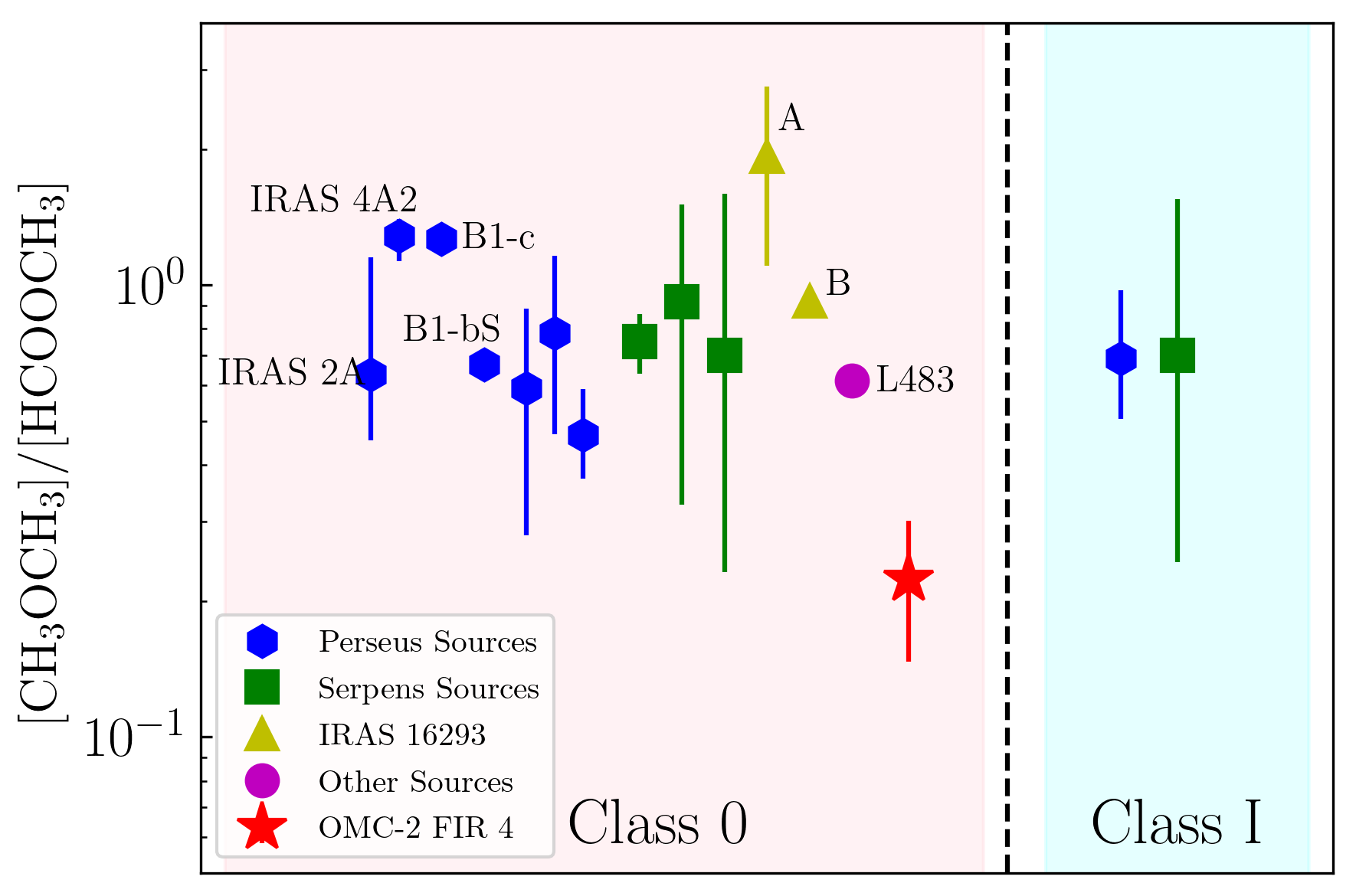}
 \plotone{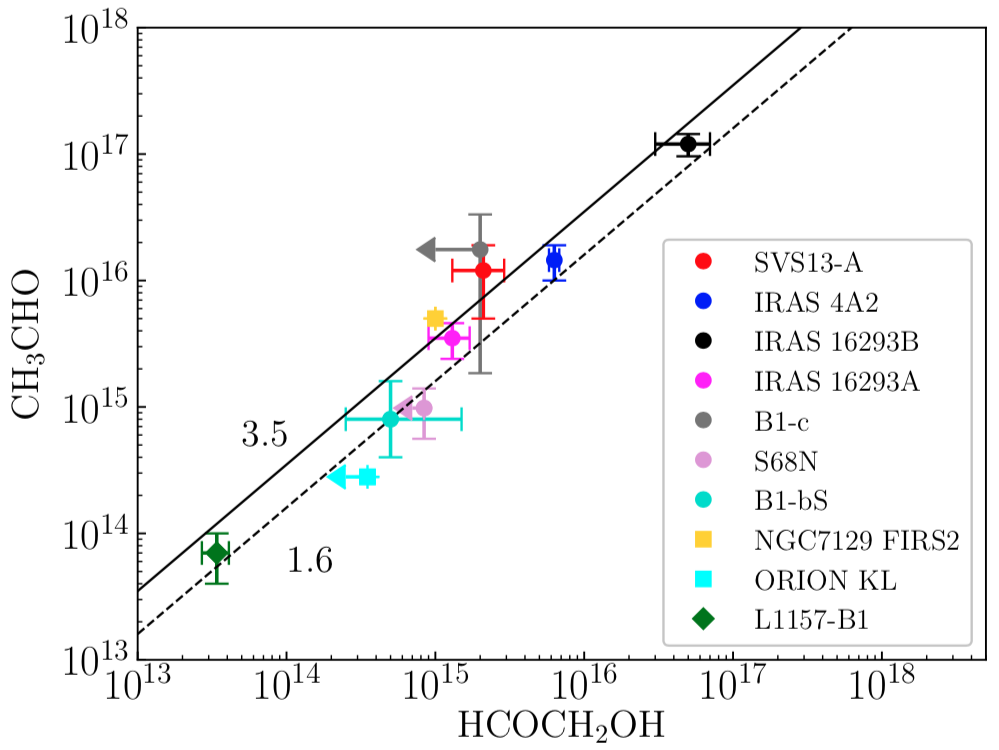}
 \caption{\small iCOMs in Class 0/I hot corinos.
 \textit{Top panel:} Abundance ratio of dimethyl ether over methyl formate \citep[from][]{Chahine2022}.
 \textit{Bottom panel:} Column densities of glycolaldehyde and acetaldehyde \citep[from][]{Vazart2020}. 
 The two lines are predictions obtained assuming that both species are formed in the gas-phase by a chain of reactions starting from ethanol \citep{Skouteris2018}.}
 \label{fig:sec4-DME=MF}
\end{figure}

\subsection{The census and abundances of detected iCOMs}\label{subsec:sec4-icom-census}

Hot corinos\index{Hot corinos} are the Class 0\index{Class 0 sources}/I\index{Class I sources} protostars\index{Protostars} that present the richest chemical composition, so that in this section we will focus on the iCOMs detected there.

\vspace{-0.3cm}
\paragraph{iCOM census and abundances}
At present, more than twenty iCOMs have been detected in hot corinos\index{Hot corinos}, mostly due to the spectacular spectral survey carried out with the ALMA\index{ALMA} interferometer in the 300 GHz band by the PILS survey \citep[Protostellar Interferometric Line Survey:][]{Jorgensen2016}.
However, the vast majority of iCOMs has only been detected towards IRAS16293\index[obj]{IRAS16293-2422} A and B \citep[e.g.][]{Jorgensen2016,Ligterink2017,Jorgensen2018,Calcutt2018,zeng2019,Manigand2020,Manigand2021}, whereas only less than ten iCOMs have been detected in other hot corinos\index{Hot corinos} \citep[e.g.][]{Taquet2015,Imai2016,Lopez2017,Marcelino2018,Bianchi2019a,Bergner2019,Belloche2020,Ligterink2021,Nazari2021,MartinDomenech2021,Chahine2022}.

In general, measuring absolute abundances of iCOMs provides relatively unreliable numbers, because of the uncertain H$_2$ column density (for example, the continuum does not necessarily trace the same gas where the iCOM emits) and the problem of dust absorption of the iCOM lines \citep{Desimone2020a}.
For example, \cite{Zamponi2021} found that the dust opacity towards the IRAS16293\index[obj]{IRAS16293-2422} B peak exceeds 100 at 300 GHz, the frequency range covered by the PILS survey.
Consequently, even considering that the PILS survey provided abundance one beam offset from the center, the absolute abundances towards that source are likely substantially underestimated.

For the above reason, it is common use to plot relative iCOM abundance ratios to understand how much chemistry is affected by the hot corino possible different conditions and to provide constrains to the astrochemical models.
As an example, Fig. \ref{fig:sec4-DME=MF} shows the abundance ratio of dimethyl ether over methyl formate in known Class 0\index{Class 0 sources}/I\index{Class I sources} hot corinos\index{Hot corinos} \citep{Chahine2022}.
This ratio, close to unity in all hot corinos\index{Hot corinos}, with the possible exception of HOPS108 in OMC-2 FIR4 \citep{Chahine2022}, suggests a mother-daughter or sister relation between these two species \citep[e.g.][]{Jaber2014,Balucani2015}, even though it is not a proof \citep{Belloche2020}.
Another example is provided by the tight correlation between the measured abundance of glycolaldehyde and acetaldehyde in Class 0\index{Class 0 sources}/I\index{Class I sources} hot corinos\index{Hot corinos}, again suggesting a possible mother-daughter or sister relation, where the two species may be one the mother of the other or they both derive by a common mother, for example ethanol, as suggested by \cite{Vazart2020}.
We will discuss both cases in \S ~\ref{subsec:sec4-model}.

Finally, recent observations towards the hot corino of NGC1333-SVS13A\index[obj]{SVS13-A} have shown a spatial stratification of the iCOM emission over a $0.15''$ ($\sim 45$ au) scale, caused by the combined effect of the dust optical depth and iCOM binding energy \citep{Bianchi2022-svs13}.
Interestingly, the observed stratification supports the idea that the iCOMs presence in the gas-phase is caused by the thermal sublimation\index{Mantle evaporation} of the mantle ices, rather than shocks or other non-thermal processes \citep[as initially supposed:][]{Ceccarelli2004}.
Note this does not necessarily imply a grain-surface origin as iCOMs could also be formed from smaller species released in the gas-phase (Sec. \ref{sec2:chemistry}). 

\begin{figure}[bt]
    \centering
    \includegraphics[width=8cm]{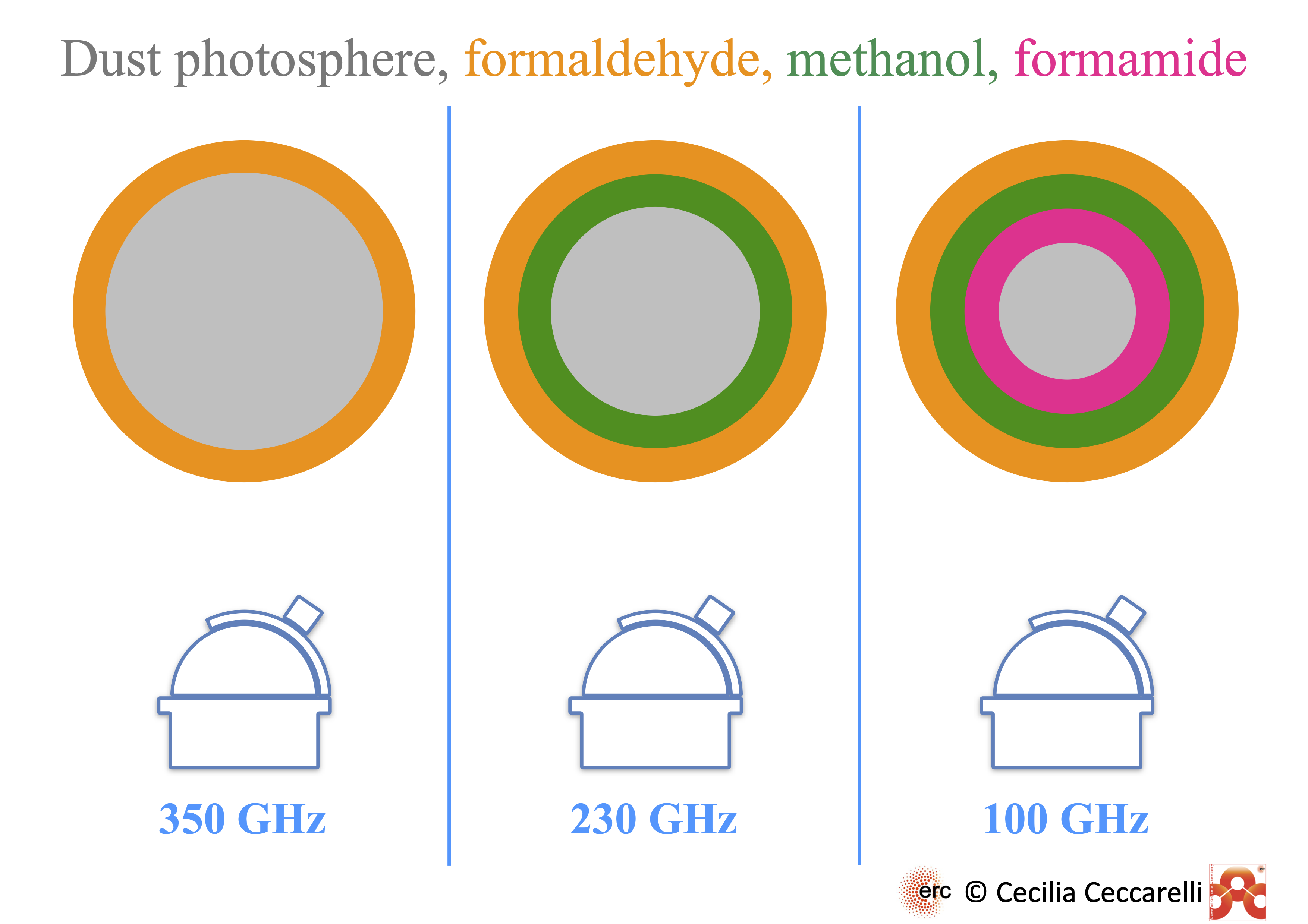}
    \caption{Sketch of the combined effect of dust optical depth and iCOM binding energy (BE) on the iCOM detection towards the NGC1333-SVS13A\index[obj]{SVS13-A} hot corino over a $\sim 45$ au scale \citep{Bianchi2022-svs13}. 
    Observations at 350 GHz (left panel) can only detect formaldehyde, whose BE allows it to be present over the almost entire envelope. 
    Observations at 230 GHz (middle panel), where the dust optical depth is smaller than at 350 GHz, can detect also methanol, whose BE is larger than that of formaldehyde. 
    Finally, observations at 100 GHz (right panel) can penetrate the most through the envelope because the dust optical depth is the lowest and, therefore, detect formamide whose BE is the largest one.}
    \label{fig:sec4-dust-BE-scheme}
\end{figure}

\vspace{-0.3cm}
\paragraph{D fractionation in iCOMs}
A recent and extremely important breakthrough has been the detection of D-bearing iCOMs, other than methanol, where the D fractionation could be measured: formamide, acetaldehyde, dimethyl ether, methyl formate, glycolaldehyde, ethanol and ethyl cyanide \citep{richard2013,Coutens2016,Jorgensen2018,Persson2018,Manigand2019,richard2021,Bianchi2022-ch2dcn}.
The detection and abundance determination of deuterated\index{Deuteration} iCOMs is important because it has the great diagnostic power to discriminate the route of formation of the species, as it will be explained in \S ~\ref{subsec:sec4-model}.
In general, iCOMs show a deuterium fractionation (2--10 \%) larger than that measured in water ($\leq 1$\%) towards the same sources.
As already pointed out by previous works \citep[see e.g.][and the references there]{Ceccarelli2014-PP6}, this systematic difference is probably caused by the different formation time of water with respect to iCOMs.
Indeed, water is the first ice component to be formed whereas iCOMs are formed after the depletion of CO onto the icy grain mantles, which enhances the H$_2$D$^+$/H$_3^+$ ratio in the gas-phase, at the base of the methanol and formaldehyde deuteration\index{Deuteration}.

\vspace{-0.3cm}
\paragraph{iCOMs in the cold envelopes}
Although with a much lower abundance, iCOMs are also present in the more extended and cold envelopes of Class 0\index{Class 0 sources}/I\index{Class I sources} protostars\index{Protostars}, where the dust temperature is lower than the icy mantle sublimation\index{Mantle evaporation} one.
\cite{Oberg2010} found weak extended emission from iCOMs associated with the lukewarm envelope of B1-b.
Later, \cite{Jaber2014} found methyl formate, dimethyl ether, acetaldehyde and formamide towards the cold envelope of IRAS16293\index[obj]{IRAS16293-2422}.
Their abundances are about a factor 100 lower than those measured in the hot corinos\index{Hot corinos} A+B, but still they are present.
As mentioned in Sec. \ref{sec3:psc}, iCOMS are also found in prestellar cores\index{Prestellar cores}, in the outer layers where the conditions are similar to the lukewarm envelopes of the protostars\index{Protostars}, so that it is probable that the same process is responsible for the presence of iCOMs in both environments.

\subsection{Astrochemical models}\label{subsec:sec4-model}

As the general ideas about the chemistry of iCOMs have been already discussed in Sec. \ref{sec2:chemistry}, here we only summarise some specific examples of a few iCOMs.

\vspace{-0.3cm}
\paragraph{Hot corino versus WCCC source chemistry}
As discussed in \S ~\ref{subsec:sec4-sources-census}, Class 0\index{Class 0 sources}/I\index{Class I sources} protostars\index{Protostars} can be roughly classified into two classes, hot corinos\index{Hot corinos} and WCCC sources, with respect to their organic contents. 
Note, however, that this classification is not rigorous, because there are also hybrid sources. 
What determines the chemical nature of a protostar is a subject of debate.
At present, two major reasons have been invoked in the literature: a different time scale of the formation of the grain mantles\index{Mantle formation} \citep{Sakai2013}, a different illumination from UV photons \citep{Spezzano2016}, or both.
The above two hypothesis have in common the idea that the differentiation is caused by the different major reservoir of the frozen carbon: methanol versus methane.
When methanol is more abundant, it gives rise to the hot corino chemistry with the productions of iCOMs, whereas when methane is more abundant the chemistry is dominated by the formation of unsaturated carbon chains originating the WCCC sources.
Therefore, the above two hypothesis differ on what species dominates the frozen carbon, methanol rather than methane.
Methanol is thought to be formed on the grain surfaces by the hydrogenation of CO, whereas methane by hydrogenation of C.
Therefore, the basic point is whether CO is abundant or not on the grain surfaces with respect to C.
The first possibility is that, since methanol is formed on the grain surfaces after CO freezes-out at relatively late times ($\geq 10^5$ yr), the formation time of the icy grain mantles\index{Mantle formation} determines whether they are enriched with methanol rather than methane.
The second possibility is that CO is less abundant than C because of the UV illumination that would photo-dissociate CO.
Both mechanisms can explain the existence of sources having the hybrid nature.

Models have been developed to quantitatively address this issue.
\citet{Aikawa2020} carried out a parameter study to understand what affects most the appearance of a hot corino and WCCC source.
They concluded that the chemical appearance of the protostars\index{Protostars} depends on a complex function of time (i.e. mantle formation\index{Mantle formation} duration), extinction (i.e. exposure to UV illumination), and thermal history (i.e. the dust temperature at the formation of the icy mantles\index{Mantle formation}).
On the same vein, \cite{kalvans2021} studied in detail the influence of the UV illumination and CR irradiation, concluding that the WCCC sources are the results of an enhancement of either of the two.


\vspace{-0.3cm}
\paragraph{Deuterated iCOMs}
The iCOM deuteration\index{Deuteration} follows a different path than the standard one initiated by and completely dependent on the H$_2$D$^+$ ion at the moment of the molecule formation \citep[e.g.][]{Ceccarelli2014-PP6}.
Indeed, since iCOMs are "second-generation" species their deuteration\index{Deuteration} is inherited from their precursors (and not directly from H$_3^+$).
With second generation here we mean that iCOMs are not the products of hydrogenation, as  methanol, whose deuteration\index{Deuteration} depends exclusively on the D/H atomic ratio on the grain surfaces, which, in turn, depends on the H$_3^+$/H$_2$D$^+$ ratio.
For example, acetaldehyde (CH$_3$CHO) is thought to be formed by the combination of HCO and CH$_3$ on the grain surfaces \citep{Garrod2006} (\S ~\ref{subsubsec:sec2-radical-reactions}) or by a series of reactions starting from ethanol (CH$_3$CH$_2$OH) in the gas-phase \citep{Vazart2020} (\S ~\ref{subsubsec:sec2-gas-neutneut}).
Therefore, the production of its two D-isomers, CH$_3$CDO and CH$_2$DCHO, depends on the ratios of DCO/HCO and CH$_2$D/CH$_3$, if formed on the grain surfaces, and of CH$_2$DCH$_2$OH/CH$_3$CH$_2$OH, CH$_3$CHDOH/CH$_3$CH$_2$OH and CH$_3$CH$_2$OD/CH$_3$CH$_2$OH, if formed in the gas-phase.

In addition, the deuteration\index{Deuteration} of the precursors is not just plainly passed on to the iCOM: the reaction leading to the iCOM itself can alter the deuteration\index{Deuteration}, because of the so-called kinetic isotope effect (KIE), a well-known effect in chemistry, which is different from the thermodynamic isotope effect (TIE).
The latter dominates the deuteration\index{Deuteration} of the species directly linked to H$_3^+$, and it is due to the difference in the zero point energy of H$_3^+$ and H$_2$D$^+$ which makes that H$_3^+$ + HD $\rightarrow$ H$_2$D$^+$ + H$_2$ is slightly exothermic, while the reverse process is slightly endothermic.
In iCOMs other than methanol, the final deuteration\index{Deuteration} depends on the passed moiety and how this is obtained, so that the iCOM deuteration\index{Deuteration} depends on the deuteration\index{Deuteration} of its precursors \textit{and} its formation route.
If one has the possibility to measure the deuteration\index{Deuteration} of the iCOM and of its precursors, then this is potentially an extremely powerful way to constrain the iCOM formation route.

\begin{figure*}[t]
    \centering
    \includegraphics[width=11.5 cm]{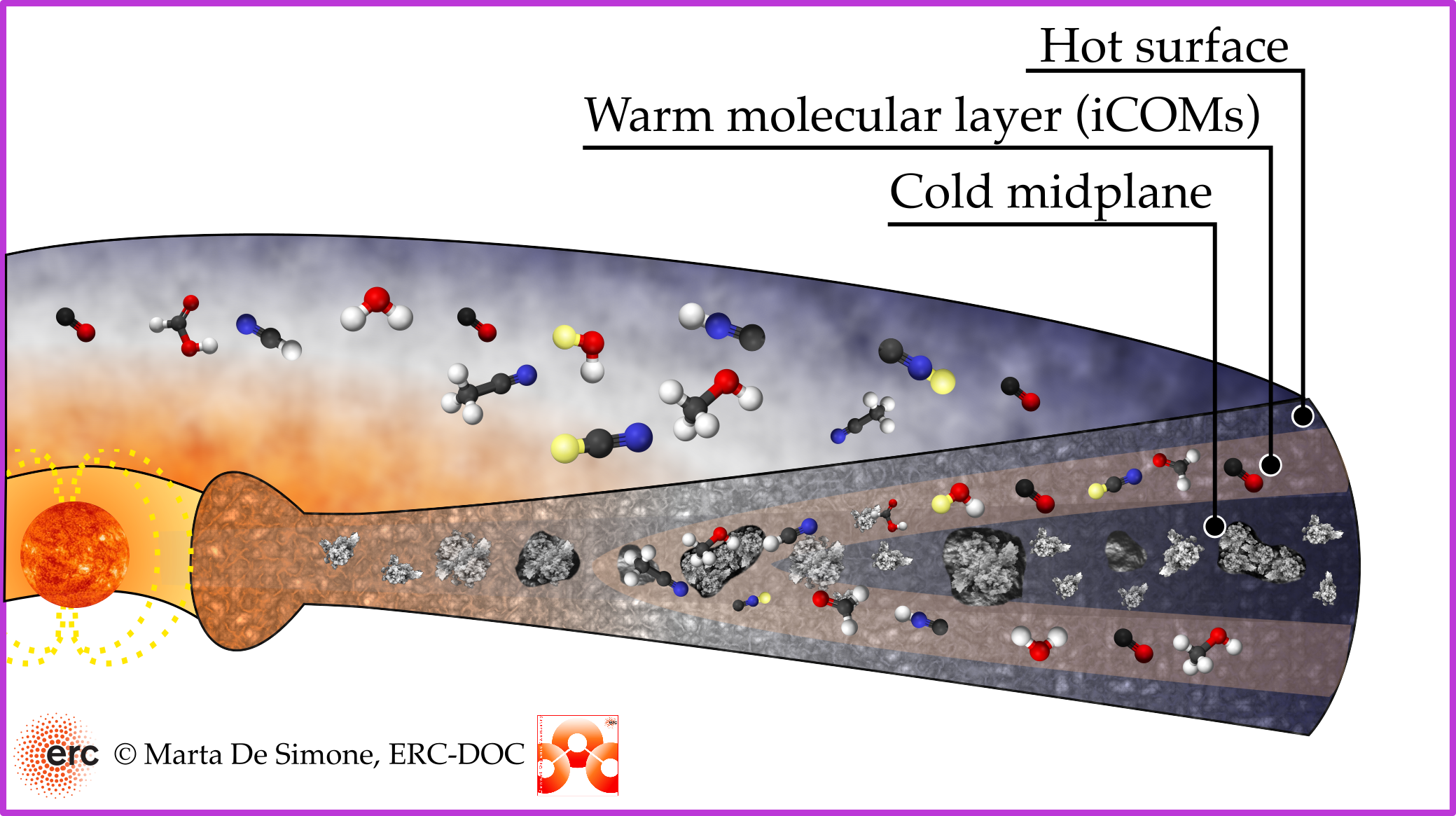}
    \caption{Sketch of the chemical structure of a generic protoplanetary disk.
    In the hot surface layer, UV photons ionise atoms and photo-dissociate molecules. 
    In the warm molecular layer, where the UV photons are absorbed (by dust and gas), gaseous molecules, including iCOMs, can survive.
    In the cold midplane, all species, except those containing only H and D atoms, freeze-out onto the grain mantles.
    }
    \label{fig:sec5-disk-structure}
\end{figure*}

A very nice example is provided by the case of formamide (NH$_2$CHO), a species that can in principle be formed both on the grain surfaces by the combination of NH$_2$ and HCO \citep{Garrod2006} and in the gas-phase by the reaction H$_2$CO + NH$_2$ \citep{Barone2015}.
In IRAS16293\index[obj]{IRAS16293-2422} B, the measured abundances of the formamide deuterated\index{Deuteration} isomers are in agreement with the theoretical predictions of the gas-phase formation route \citep{Skouteris2017}, which suggests that formamide is formed in the gas-phase in this source.
Unfortunately, no theoretical predictions exist so far on the predicted deuteration\index{Deuteration} in case of formamide formation on the grain surfaces so that it is not possible to perform a similar significant comparison with the observations.
In the same vein, \cite{vazart2022} found that the deuterated\index{Deuteration} forms of glycolaldehyde are also in agreement with the theoretical predictions of its formation in the gas, following the scheme starting from ethanol \citep{Skouteris2018}.


\section{\textbf{PROTOPLANETARY DISKS: WHERE ORGANICS ARE BURIED}}\label{sec5:disks}

\subsection{Overview of the physical and chemical structure of protostars}\label{subsec:sec5-overview}

Circumstellar disks\index{Disks}  are expected to be present since the earliest phases of the star forming process, i.e. the protostellar Class 0\index{Class 0 sources} protostar phase described in the previous section, and evolve into what are traditionally called protoplanetary disks\index{Protoplanetary disks}, in the Class I\index{Protostars} and II\index{Class II sources} protostars\index{Protostars} phases.
Protoplanetary disks\index{Protoplanetary disks} are complex structures composed by a mixture of dust and gas and the chemistry occurring in the gas-phase and/or on the grain surfaces reflects the diversity, the inhomogeneity and the evolutionary stage of these objects.

Although their physical and chemical structure changes with time, from gas-rich and with small ($\sim 0.1~ \mu$m) grains to gas-poor and pebbles ($\sim 0.1-1$ m) (see, e.g., Chapters by Bae et al. and Miotello et al.), the chemical structure can be approximately subdivided in three different zones, shown in Fig. \ref{fig:sec5-disk-structure} \citep[e.g.][]{Aikawa1999c,Dullemond2007,Dutrey2014,Walsh2014}:\\
\textit{1-} The hot surface layer, where the chemistry is dominated by the UV photons. 
As in the Photo-Dissociation Regions (PDRs), only ionised and neutral atoms are present in this zone, because molecules are photo-dissociated.\\
\textit{2-} The warm molecular layer, where molecules are in the gas-phase.
In this zone, the UV photons are largely absorbed and the dust temperature is large enough for several molecules to remain gaseous.
This is the layer where iCOMs can be and are detected \citep[e.g.][]{Oberg2014}.\\
\textit{3-} The cold (outer) midplane, where molecules freeze-out onto grain mantles\index{Mantle formation} and only H and D bearing species are in the gas-phase, e.g. H$_2$D$^+$ \citep{Ceccarelli2005b}. 

\begin{figure}[tb]
    \centering
    \includegraphics[width=7cm]{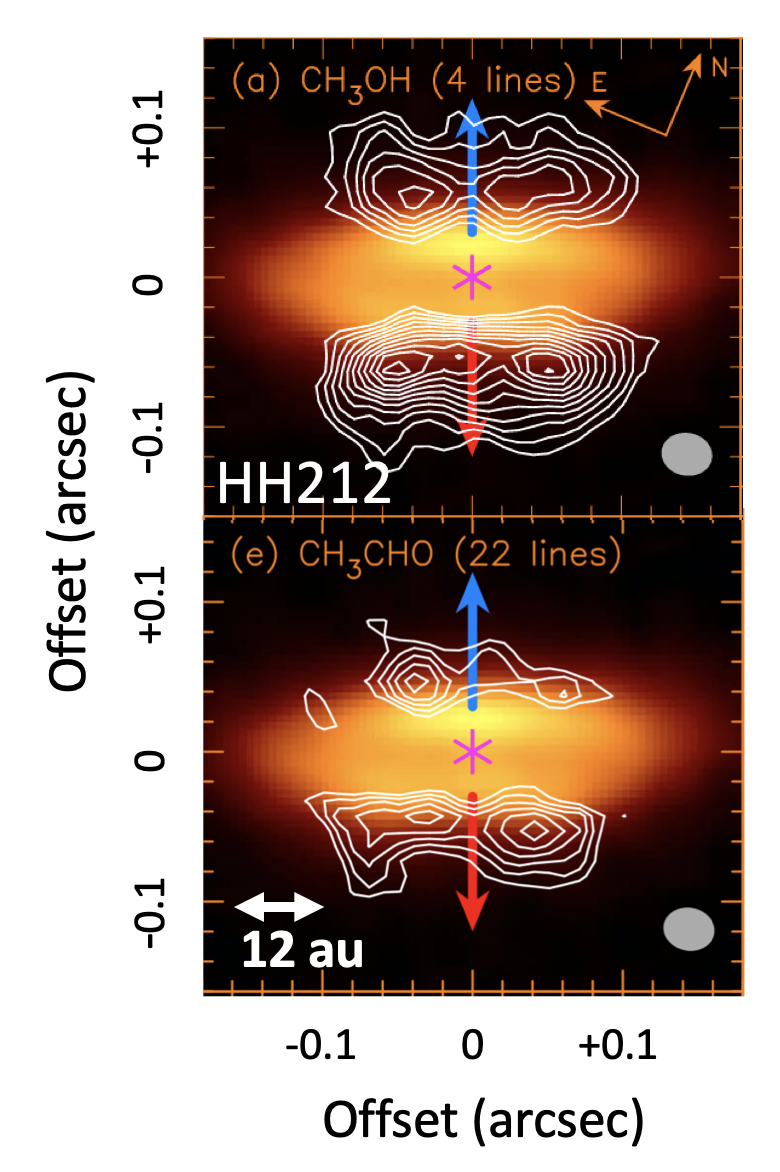}
    \caption{Methanol (upper panel) and acetaldehyde (bottom panel) line emission (white contours) overlapped to the dust continuum emission (color image) towards the protoplanetary disk associated with the Class 0 HH 212-mm object.
    Each species line image has been obtained by stacking several lines (as indicated in the parentheses).
    The blue and red arrows indicate the axes of the blue- and red- shifted jet components, respectively.
    Adapted from \cite{Lee2019}.}
    \label{fig:sec5-HH212}
\end{figure}

\begin{figure}[bt]
    \centering
    \includegraphics[width=8cm]{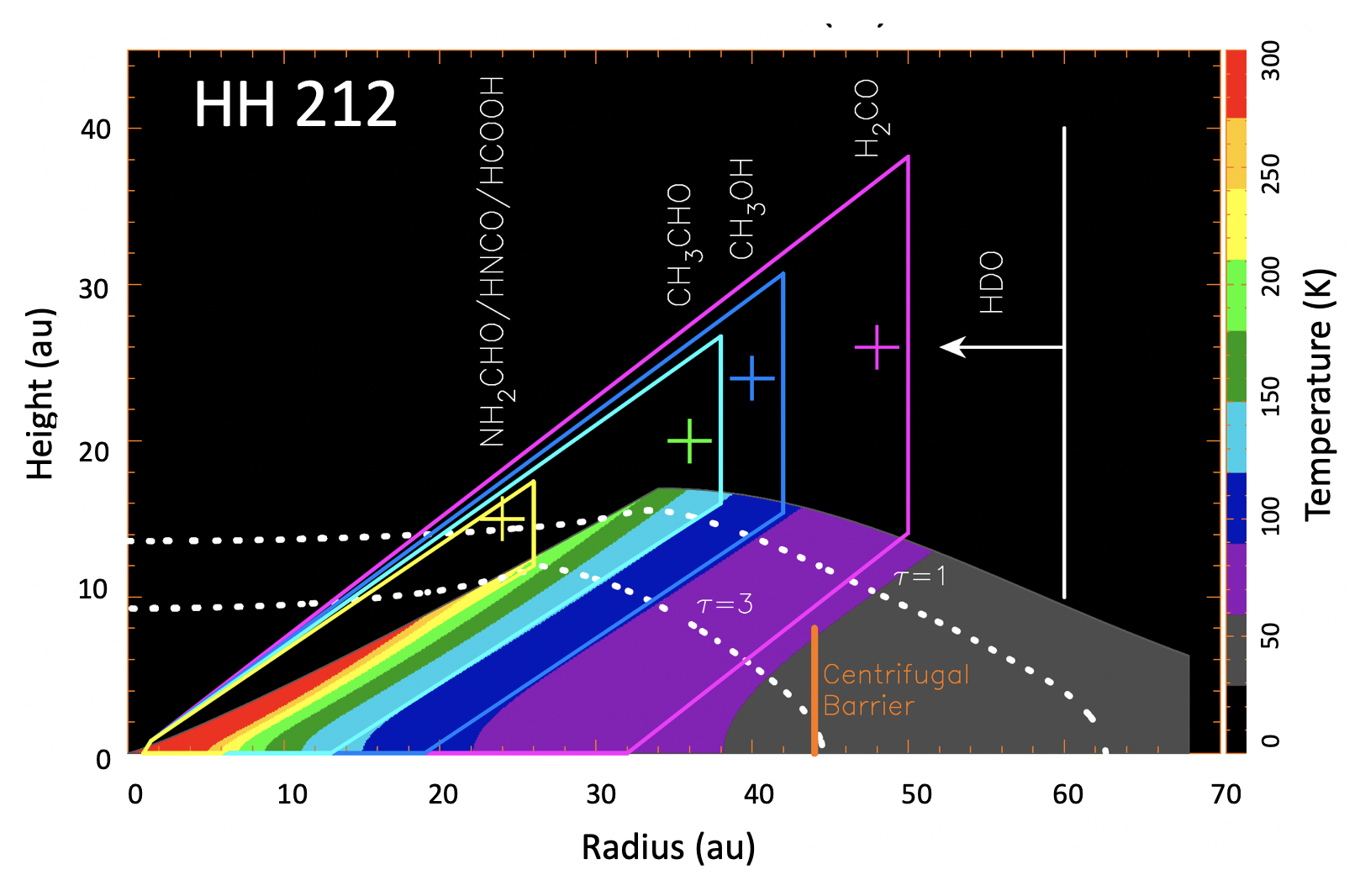}
    \caption{Stratification of several iCOMs (CH$_3$OH, CH$_3$CHO, and NH$_2$CHO among them) in the HH 212-mm protostellar disk, plotted on the temperature structure.
    The crosses mark the outer radius and vertical height as derived from the ALMA observations. Adapted from \citet{Lee2022-h212}.}
    \label{fig:sec5-HH212-2022}
\end{figure}

The freeze-out of different molecular species occurs at the disk radius and height where the dust temperature is lower than the molecule freeze-out temperature, which in turn depends on its BE.
The loci of the gaseous-solid transition of a molecule is called snowline and it is specific to each molecule.
Although in the regions beyond the snowline molecules are prevalently frozen into the grain mantles\index{Mantle formation}, a small fraction of them can be in the gas-phase thanks to the UV and X-ray photon illumination or the CR irradiation \citep[e.g.][]{Walsh2014,Loomis2015}, via the non-thermal processes described in detail in \S ~\ref{subsec:sec2-mantle-desorption}.

In general, the detection of iCOMs in disks\index{Disks!chemistry}  is very challenging because (i) the dust along the line of sight might heavily absorb the line emission, and (ii) the most chemically active regions are actually expected to have low (beam-averaged) column densities because of the small vertical size of the outer molecular layer and the small size of the warm molecular inner zone.
As a consequence, high-angular resolution and high sensitivities are crucial to detect organic molecules \citep[e.g.][]{Walsh2014}.
Remarkably, protoplanetary disks\index{Protoplanetary disks} show gaps in the dust continuum emission maps, possibly testifying of forming planets \citep[e.g.][]{Fedele2016,Sheehan2017,Sheehan2018,Segura-Cox2020} (see also Chapters by Lesur et al., Benisty et al and Pinte et al.).
Therefore, despite this complex scenario and the observational challenge, it is essential to probe the chemical composition of the young disks\index{Disks!chemistry}  to understand whether more evolved disks\index{Disks!chemistry} inherit the chemical complexity observed at the protostellar stage (Sec. \ref{sec4:protostars}), as suggested by \citet{Bianchi2019a} and \citet{Droz2019}.

\subsection{iCOMs in Class 0 protoplanetary disks}\label{subsec:sec5-class0}

Many recent observations show that circumstellar disks\index{Disks} are present in Class 0\index{Class 0 sources} protostars\index{Protostars}, still embedded in infalling-rotating envelopes.
The dust absorption as well as the line emission from the infalling envelopes easily mask the line emission from the disk itself, in particular weak iCOM lines.

A remarkable exception is represented by the archetypal protostellar disk around HH 212-mm\index[obj]{HH 212-mm}, located in Orion B.
The HH 212\index[obj]{HH 212-mm} system has been observed with ALMA\index{ALMA} with a spatial resolution of up to $\sim 10$ au, revealing a dusty edge-on disk with a radius of $\sim 60$ au.
The disk is obscured along the equatorial plane because of the high dust optical depth \citep[Fig. \ref{fig:sec5-HH212}:][]{Lee2017a,Lee2019}.
Several iCOMs have been detected towards the HH 212\index[obj]{HH 212-mm} disk \citep[e.g.][]{Codella2018,Codella2019,Lee2017b,Lee2019} with the following abundances (with respect to H$_2$): [CH$_3$OH] $\sim 10^{-7}$, [HCOOH] $\sim 10^{-9}$, [CH$_3$CHO] $\sim 10^{-9}$, [HCOOCH$_3$] $\sim 10^{-9}$ and [NH$_2$CHO] $\sim 10^{-10}$.
More specifically, iCOMs have been observed only in the upper outer disk layers, at $\pm$ 40 au from the equatorial plane \citep[Fig. \ref{fig:sec5-HH212}:][]{Lee2019}. 
In addition, \citet{Lee2022-h212} found a stratified radial distribution, with the outer emission radius increasing from 24 au for NH$_2$CHO, to 36 au for 
CH$_3$CHO, 40 au for CH$_3$OH, and to 48 au for H$_2$CO 
(Fig. \ref{fig:sec5-HH212-2022}).
The increasing outer radii are consistent with the decreasing molecules BEs, suggesting that these species are thermally desorbed from the dust icy mantles\index{Mantle evaporation} at the different radii because of the dust heating from the central object and, therefore, not because of shocks\index{Shocks} or other non-thermal processes (\S ~\ref{subsec:sec2-mantle-desorption}).
The HH 212\index[obj]{HH 212-mm} iCOM stratification is similar to that also observed towards NGC1333-SVS13A\index[obj]{SVS13-A} \citep{Bianchi2022-svs13} and described in \S ~\ref{subsec:sec4-icom-census} (see also Fig. \ref{fig:sec4-dust-BE-scheme}). 
Finally, the lack of iCOM emission towards the equatorial plane is probably due to the dust absorption at sub-mm wavelengths (Fig. \ref{fig:sec5-HH212}).
Observations at lower frequencies, e.g. in the cm, would be necessary to reveal the gas chemical composition in the equatorial plane.

\subsection{iCOMs in Class I/II protoplanetary disks}\label{subsec:sec5-classIandII}

In the more evolved disks\index{Disks} surrounding Class I\index{Class I sources} and Class II\index{Class II sources} protostars\index{Protostars}, most of the molecules are frozen onto the dust icy mantles\index{Mantle formation} in the disk midplane and only a relatively low column density remains gaseous.
iCOM detections in Class I\index{Class I sources}/II disks\index{Disks!chemistry} have been rare, but the few detections that do exist provide vital information about the iCOM reservoir available to forming planets, the evolution of iCOM chemistry during star and planet formation, and the specific origins of iCOMs in disks\index{Disks!chemistry}.

\begin{figure*}[tb]
    \centering
    \includegraphics[width=13cm]{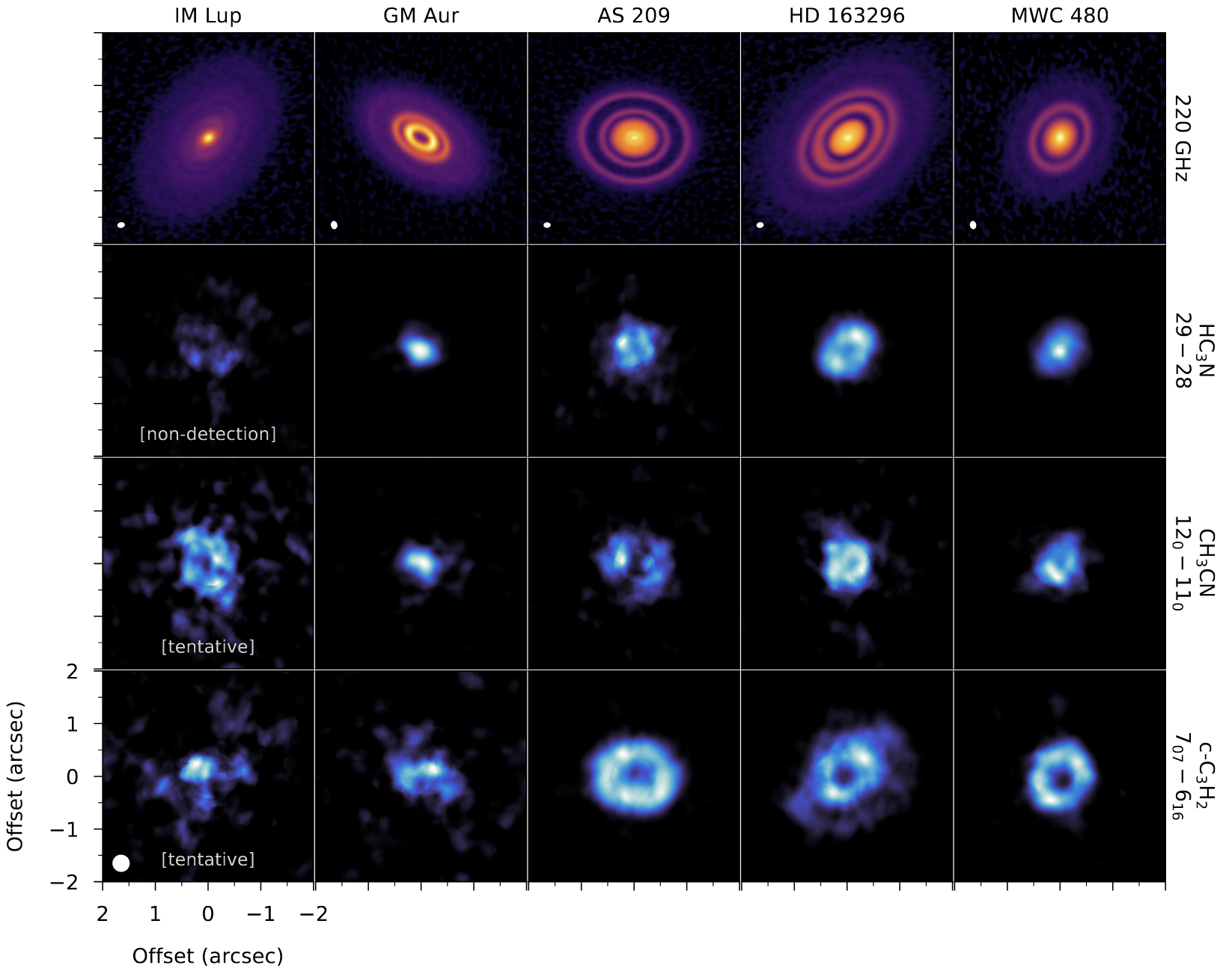}
    \caption{Continuum (first row) and molecular (bottom three rows) emission distribution of the five Class II protoplanetary disks target of the ALMA\index{ALMA} Large Project MAPS \citep{Oberg2021-maps}.
    The CH$_3$CN emission (third row) is distributed in rings or peaked towards the center and it is different from the other imaged molecules HC$_3$N (second row) or c-C$_3$H$_2$ (bottom row). 
    Adapted from \citet{Ilee2021}.} 
    \label{fig:sec5-MAPS}
\end{figure*}

Up to date, few detections of iCOMs towards protoplanetary disks\index{Protoplanetary disks} have been reported.
The detection of methanol \citep{Walsh2016} and formaldehyde \citep{Oberg2017}, which are key species for the formation of other iCOMs, along with that of methyl cyanide \citep{Loomis2018} and formic acid \citep{Favre2018} towards the disk surrounding the very close (60 pc) young Solar-type  T-Tau star TW Hya\index[obj]{TW Hya} strongly suggests an active organic gas-phase chemistry taking place also at this stage (as discussed later) and/or non-thermal desorption from the grain mantles\index{Mantle evaporation} (see \S ~\ref{subsec:sec2-mantle-desorption}) of previously formed organic molecules\index{Organic molecule formation} \citep[see e.g.][]{Walsh2016}.
The recent survey of protoplanetary disks\index{Protoplanetary disks} MAPS \citep[Molecules with ALMA at Planet-forming Scales;][]{Oberg2021-maps} shows that methyl cyanide, for example, is present in four out of the five targets \citep{Ilee2021}.
The emission is concentrated in rings at distances that vary with the source or peaked towards the center (see Fig. \ref{fig:sec5-MAPS}) and it originates in regions with temperatures between 25 and 50 K, relatively close to the midplane, z/r $\sim 0.1-0.2$.
Since the CH$_3$CN binding energy (BE) is similar to that of water, it is formed either via gas-phase reactions between smaller molecules with relatively low BEs or a non-thermal mechanism would be operating (\S ~\ref{subsec:sec2-mantle-desorption}).

Methanol, a key iCOM, has been recently detected in the Class I\index{Class I sources} disk IRAS 04302+2247\index[obj]{IRAS 04302+2247} \citep{Podio2020a,Garufi2021} and towards the disk around the A-type star HD 100546\index[obj]{HD 100546} \citep{Booth2021}.
Methyl cyanide and formaldehyde have been imaged towards a dozen of disks\index{Disks}, such as e.g. TW Hya\index[obj]{TW Hya}, DG Tau B\index[obj]{DG Tau B}, DM Tau\index[obj]{DM Tau}, HL Tau\index[obj]{HL Tau}, MWC 480\index[obj]{MWC 480} \citep{Oberg2015a,Bergner2018,Loomis2018,Podio2019,Podio2020b,Pegues2020,Garufi2021}.
Typical abundances are: $10^{-12}-10^{-10}$ (H$_2$CO), $10^{-12}-10^{-11}$ (CH$_3$OH, HCOOH), and $10^{-13}-10^{-12}$ (CH$_3$CN).
The formaldehyde spatial distribution can be either centrally-peaked or centrally-depressed showing an outer ring at a radius around 50 au \citep{Podio2019,Podio2020b,Pegues2020,Garufi2021}. 
The emission in the inner few au can indeed be masked by dust optically depth effects. 
However, the detection of considerable formaldehyde emission beyond the CO snowline shows that formaldehyde can be produced in the gas-phase from species with lower BE than CO.
Finally, the abundance of methyl cyanide (and also that of HC$_3$N) is relatively high compared to its simpler relative, HCN \citep{Oberg2015a,Bergner2018}.

In general, in protoplanetary Class I\index{Class I sources}/II disks\index{Protoplanetary disks} , N-bearing iCOMs are much easier to detect than other iCOMs observed in the protostellar stage (Sec. \ref{sec4:protostars}): methyl cyanide lines are considerably brighter towards the nearby Class II\index{Class II sources} disk around TW Hya\index[obj]{TW Hya} compared to either methanol or formic acid \citep{Loomis2018,Walsh2016}.
This suggests, first, that, at least in this disk, the iCOM inventory is not simply/completely inherited from the prestellar\index{Prestellar cores} or protostellar stages and, second, that the environmental conditions in Class I\index{Class I sources}/II disks\index{Disks} favor the production of N-bearing iCOMs.
The link between the N- versus O- bearing iCOMs is discussed in \S ~\ref{subsec:sec5-model}.

Finally, it is worth noting that the very recent detection of dimethyl ether and, tentatively, methyl formate towards the Herbig IRS 48 transition disk by \citet{Brunken2022} promises an era where it will be possible to observe more iCOMs in protoplanetary disks\index{Protoplanetary disks} and, consequently, their chemistry.

\begin{figure*}[bt]
    \centering
    \includegraphics[width=12cm]{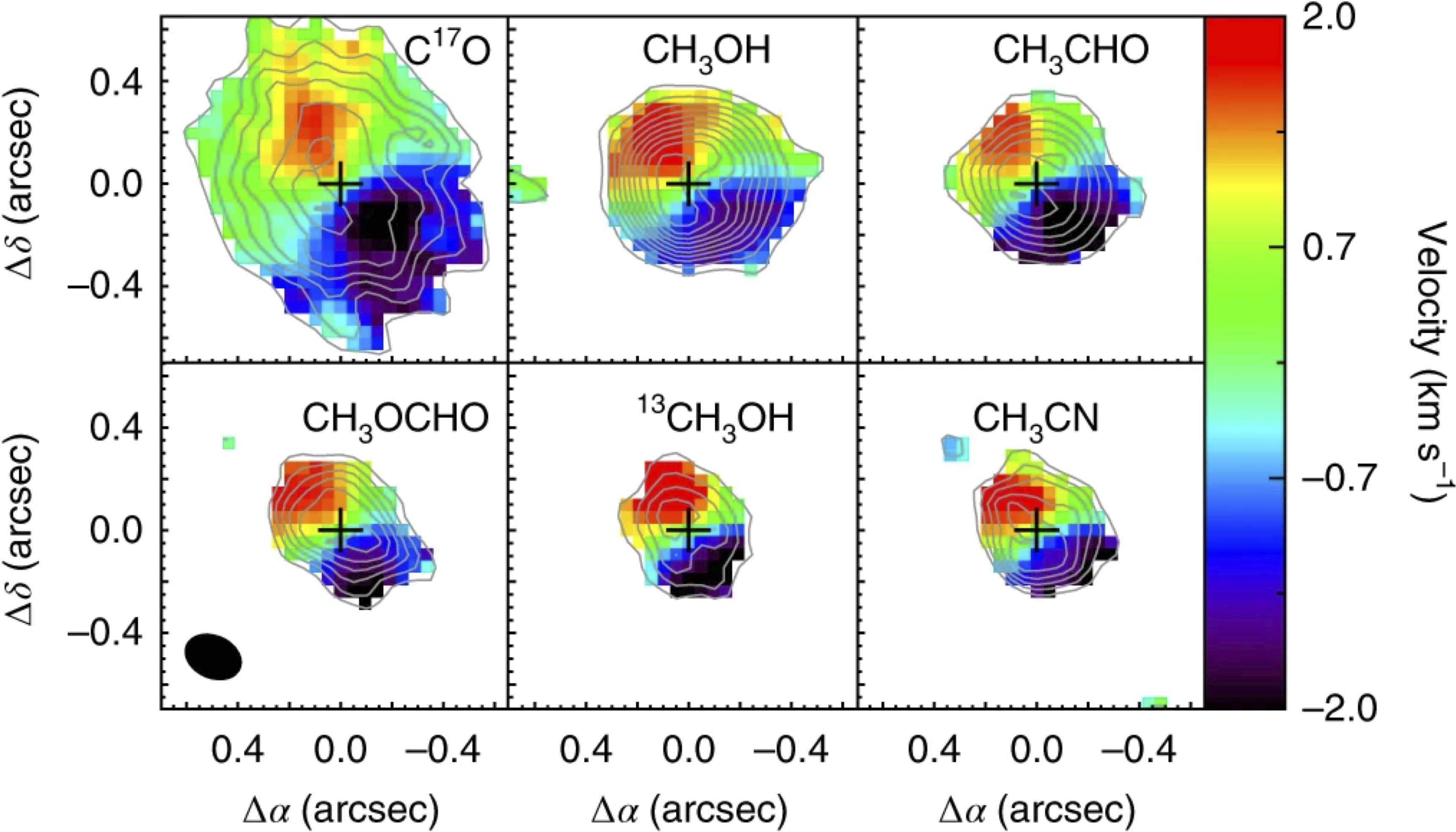}
    \caption{The rotating disk around the Fu Ori object V833 Ori, as mapped by iCOM line emission.
    The panels show the intensity weighted velocity images (colours) and integrated intensity maps (contours) of the species reported in the upper-right corner.
    Adapted from \cite{LeeJ2019-v833}.}
    \label{fig:sec5-FuOri}
\end{figure*}

\subsection{iCOMs in protoplanetary disks of Fu Ori stars}\label{subsec:sec5-fuori}

Protoplanetary disks\index{Protoplanetary disks} surrounding FU Orionis\index[obj]{Fu Ori} stars (hereafter Fu Ori) are precious laboratories where to study the disk chemical composition.
Indeed, in Fu Ori stars\index[obj]{Fu Ori}, the luminosity of the central Solar-type young star fluctuates so that the illumination of the surrounding disk can vary by up to 2-3 orders of magnitude.
The increased luminosity leads to the dust warming up and it causes the sublimation of the dust icy mantles in relatively large portions of the disk.
In turn, this makes it possible to observe the molecular content just released from the grains (see also Sec. \ref{sec7:comets}) and detect iCOMs in the disk. 

The case of the disk surrounding the V883 Ori star\index[obj]{V883 Ori} is an illustrative one.
V883 Ori is a Class I\index{Class I sources} protostar, where the surrounding infalling envelope is not yet completely dissipated, with a prominent disk.
A burst of the star luminosity was first announced in 1888 \citep[but not measured: ][]{Connelley2018} and, later, \citet{Strom1993} have measured a bolometric luminosity of $\sim400$ L$_\odot$. 
The luminosity has steadily decreased in the past years up to the present value of 218 L$_\odot$  \citep{Fisher2019}.
Five iCOMs (methanol, formic acid, acetaldehyde, methyl cyanide and methyl formate) have been detected using ALMA\index{ALMA} observations \citep{vantHoff2018,LeeJ2019-v833} (Fig. ~\ref{fig:sec5-FuOri}).
The relatively large emitting region ($\sim 0.4\arcsec$--0.8$\arcsec$ in diameter, equivalent to about 150--300 au) coupled with the ALMA\index{ALMA} high spatial resolution permit to resolve the iCOM line emission and to map the kinematics of the emitting gas.
There are no doubts that the detected iCOMs originate in a rotating disk \citep[Fig. \ref{fig:sec5-FuOri}][]{LeeJ2019-v833}.
Interestingly enough, the derived gas temperature, 100--120 K, and the measured molecules column densities, 10$^{16}$--10$^{17}$ cm$^{-2}$ \citep{LeeJ2019-v833}, are similar to the ones observed in typical hot corinos\index{Hot corinos} (Sec. \ref{sec4:protostars}).

\subsection{Astrochemical models}\label{subsec:sec5-model}

The formation of iCOMs in disks\index{Disks!chemistry} has been modeled in a few studies, probably because of the scarcity of iCOM detections to date. 
One of the most comprehensive study was carried out by \citet{Walsh2014}, who predicted that disk chemistry can produce large quantities of iCOMs (mostly via grain-surfaces reactions: see the discussion below and in Sec. \ref{sec2:chemistry}). 
In this model, whether these iCOMs are observable in the gas-phase largely depends on their non-thermal sublimation efficiencies, which are poorly constrained for iCOMs (see \S ~\ref{subsec:sec2-mantle-desorption}).
In general, the desorption efficiencies are suspected to be quite low, based on the experiments and computations on the paradigmatic iCOM, CH$_3$OH, \citep[e.g.][]{Martin-Domenech2016,Minissale2016b,Chuang2018,pantaleone2020} and the relatively large BE of the iCOMs.
Whatever the mechanism of the iCOM formation, gas-phase (\S ~\ref{subsec:sec2-gas-phase}) or grain-surface (\S ~\ref{subsec:sec2-onlygrainchem})), and the non-thermal desorption efficiency (\S ~\ref{subsec:sec2-mantle-desorption}), the vast majority of produced iCOMs end up frozen into the grain mantles\index{Mantle formation} due to their short freeze-out timescales. 

Two recent models focused on the most commonly observed iCOM in disks, methyl cyanide \citep{Loomis2018,LeGal2019-nitrile}. 
They both consider gas-phase (mainly by the radiative association reaction CH$_3^+$ + HNC $\rightarrow$ CH$_3$CNH$^+$ + $h\nu$ followed by the recombination of CH$_3$CNH$^+$) and grain-surface (mainly hydrogenation of C$_2$N) formation of CH$_3$CN.
\cite{Loomis2018} found that gas-phase and grain-surface reactions regulate the CH$_3$CN formation in different regions of the disk: the former at high ($z/r\sim 0.5$) heights from the midplane while the latter, which dominates the overall CH$_3$CN column density, closer to the midplane ($z/r\sim 0.3$).
In addition, \citet{LeGal2019-nitrile} and \citet{Fedele2020} found that an enhanced C/O ratio facilitates the CH$_3$CN production, because of the increased availability of carbon atoms to form molecules other than CO.
This may explain why this particular iCOM is so abundant in disks\index{Disks!chemistry}.
Needless to say, all the uncertainties on the chemical processes incorporated in the astrochemical models and discussed in \S \ref{sec2:chemistry} apply to the chemical disk models too, so that a word of caution is mandatory on the interpretation of these theoretical results.
Specifically, as also discussed by \cite{Loomis2018}, it is important to emphasize that the efficiency of the two above routes of CH$_3$CN formation are highly uncertain, as neither of them has been studied in laboratory or theoretically.
In addition, the models assume a reactive and photo- desorption equal to 1\% and $10^{-3}$, respectively, and that the desorption leaves CH$_3$CN intact, which are very likely optimistic values (see discussion in \S ~\ref{subsubsec:sec2-nonthermdes}).

\begin{table*}[bt]
\begin{flushleft}
    \caption[]{List of iCOMs detected in protostellar outflow shocks and ordered according to the presence of only O atoms (upper half of the table), N and S atoms (bottom half).  
    The iCOM maps were all obtained towards the shocked site L1157-B1 and the outflows driven by NGC-1333 IRAS4 A1 and A2.
    Note that we did not include pure hydrocarbon species such as CH$_3$CCH, as done for the other sections of this review.
    Numbers in brackets refer to the references in the Notes.} 
    \label{tab:sec6-table-iCOMs}
\begin{tabular}{lll} \hline
        &          \\  
Element & Detected       &    Imaged  \\ 
    &             \\  \hline
O   & CH$_3$OH [1], $^{13}$CH$_3$OH [2,3], CH$_2$DOH [4] & CH$_3$OH [10], CH$_2$DOH [11]\\   
    & CH$_3$CHO [4], CH$_3$OCH$_3$ [7]                   & CH$_3$CHO [12], CH$_3$OCH$_3$ [14]\\ 
    & CH$_3$CH$_2$OH [5], HCOOCH$_3$ [5], HCO(CH$_2$)OH [7], HCOOH [5]  & \\ \hline
N   & NH$_2$CHO [3], CH$_3$CN [9], HC$_5$N [8]            & NH$_2$CHO [13], CH$_3$CN [9] \\
S   & CH$_3$SH [6]                                        & \\
\hline
\end{tabular}

Notes: We give as reference the first detection reported in the literature both for detection and interferometric imaging. [1] \citet{Bachiller1997}; [2] \citet{Codella2012}); [3] \citet{Yamaguchi2012}; [4] \citet{Sugimura2011}; [5] \citet{Arce2008} [6] \citet{Lefloch2018}; [7] \citet{Lefloch2017}; [8] \citet{Mendoza2018}; [9] \citet{Codella2009}; [10] \citet{Benedettiini2007}; [11] \citet{Fontani2014}; [12] \citet{Codella2015}; [13] \citet{Codella2017}; 
[14] \citet{Desimone2020b}.
\end{flushleft}
\end{table*}

Finally, both models and observations show clear evidence for substantial inheritance of iCOMs into disks\index{Disks!chemistry} from the preceding evolutionary stages.
Most of the time, however, the inherited reservoir is expected to be frozen into the icy mantles of the dust grains oi the midplane and, therefore, not observable in the gas-phase. 
For example, this scenario has been confirmed by the methanol detection towards two Herbig Ae disks\index{Disks!chemistry}, which are too warm to have in situ CH$_3$OH formation from frozen CO hydrogenation \citep{Booth2021,vanderMarel2021}. 
In summary, both inheritance and in situ chemistry are responsible for the presence of iCOMs in disks. 
On the one hand, iCOMs formed during the prestellar and protostellar phases are incorporated into Class II\index{Class II sources} disks\index{Protoplanetary disks} and constitute a rich chemical reservoir of O- and, at a less extent, N- rich organics.
Most of this reservoir is hidden from sight, in the freezing cold disk midplane. 
On the other hand, in the warmer disk layers, the gas is likely carbon enriched (because oxygen is frozen in the dust icy mantles), which leads to an in situ gas-phase and grain-surface O-poor iCOM chemistry, testified by the presence of abundant complex nitriles. 
Therefore, inherited and freshly formed iCOMs will eventually seed the forming planetesimals, asteroids, comets\index{Comets} and planets.


\section{\textbf{MOLECULAR OUTFLOW SHOCKS: THE SPACE CHEMICAL LABORATORIES}}\label{sec6:outflows}

\subsection{The census of iCOMs in molecular shocks}\label{subsec:sec6-iCOMs}

The first observational evidence of iCOMs in protostellar outflows\index{Molecular outflows} dates back to the mid-1990s with the detection and the spectacular map of methanol in the prototypical outflow \index{Molecular outflows} of L1157\index[obj]{L1157-B1} (and two other outflows\index{Molecular outflows}) by \citet{Bachiller1995}.
More studies of methanol emission followed up towards a few Solar-type Class 0\index{Class 0 sources} sources such as  BHR71, NGC133-IRAS2 and confirmed the association of bright CH$_3$OH emission with shocked\index{Shocks!chemistry} gas, related to non-steady mass-loss ejections \citep{Bachiller1997,Bachiller1998,Bachiller2001,Jorgensen2004}. 
Multi-line radiative transfer modelling of the CH$_3$OH lines allowed to constrain the gas density and temperature to be higher than the ambient gas, supporting the interpretation of methanol being associated with shocked\index{Shocks} gas.
The next significant step was made by \citet{Arce2008}, who reported the first detection, towards the B1 shock\index{Shocks} of the 
L1157\index[obj]{L1157-B1} outflow\index{Molecular outflows}, of other iCOMs than methanol: methyl formate, formic acid, methyl cyanide and ethanol.
Soon after, \citet{Codella2009} reported the first high resolution map of an iCOM other than methanol, that of the methyl cyanide, again towards L1157-B1\index[obj]{L1157-B1}, showing a distribution similar, but not exactly the same, to that of the methanol one and definitively different from that of SiO, another classical shock\index{Shocks!chemistry} tracer.
These results definitely made the shock\index{Shocks} region L1157-B1\index[obj]{L1157-B1} the prototype for studies on the shock-driven\index{Shocks!chemistry} chemical complexity. 

The new generation of broad-band heterodyne receivers has revolutionized the field by significantly reducing the observation time cost of unbiased spectral line surveys in the sub-mm domain making it possible to detect new molecules towards L1157-B1\index[obj]{L1157-B1}: methanol and its D-isotopomer CH$_2$DOH, acetaldehyde, formamide \citep{Sugimura2011,Yamaguchi2012}, dimethyl ether, ketene (H$_2$CCO) and glycolaldehyde \citep{Lefloch2017}, within the IRAM\index{IRAM} Large Program ASAI \citep[see \S ~\ref{subsec:sec3-l1544}:][]{Lefloch2018}. 
Remarkably, the same ASAI survey towards the source driving the L1157\index[obj]{L1157-B1} outflow\index{Molecular outflows}, L1157-mm\index[obj]{L1157-mm}, showed that the latter possesses a much poorer spectrum, dominated by C-bearing species (see \S ~\ref{subsec:sec4-sources-census}) in contrast with the iCOM rich spectrum at the shocked\index{Shocks!chemistry} site B1 \citep{Bachiller2001}.

The list of iCOMs detected in shocks\index{Shocks!chemistry} is given in Table ~\ref{tab:sec6-table-iCOMs}.
The ASAI spectral survey provides source-averaged ($\sim 20\arcsec$) column densities towards L1157-B1\index[obj]{L1157-B1} \citep{Lefloch2017}: $\sim 10^{15}$ cm$^{-2}$ (CH$_3$OH),
2--5 $\times$ 10$^{13}$ cm$^{-2}$ (CH$_3$COH, HCOOCH$_3$, HCOCH$_2$OH,
CH$_3$OCH$_3$, C$_2$H$_5$OH), and $\sim 5\times 10^{12}$ cm$^{-2}$ (NH$_2$CHO).
Thanks to a multiline and profile CO analysis, \citet{Lefloch2012} estimated in the same region an H$_2$ column density equal to $\sim 2\times 10^{21}$ cm$^{-2}$. 
In turn, this implies iCOMs abundances in the range from a few 10$^{-9}$ to $\sim 10^{-7}$ (CH$_3$OH).
The spatial distributions obtained thanks to IRAM\index{IRAM} NOEMA\index{NOEMA} \citep{Codella2017,Codella2020} reveals higher values at the peak regions (not beam-diluted): 
$\sim 10^{16}$ cm$^{-2}$ (CH$_3$OH), 7 $\times$ 10$^{13}$ cm$^{-2}$ (CH$_3$COH), $\sim 10^{13}$ cm$^{-2}$ (NH$_2$CHO).
In the same vein, imaging the NGC1333-IRAS4A\index[obj]{NGC1333-IRAS4A} shocked\index{Shocks} regions, \cite{Desimone2020b} derived column densities up to 5 $\times$ 10$^{15}$ cm$^{-2}$ for methanol, $\sim 10^{14}$ cm$^{-2}$ for acetaldehyde and dimethyl ether, and 5 $\times$ 10$^{12}$ cm$^{-2}$ for formamide.

When considering the case of L1157-B1\index[obj]{L1157-B1}, where the largest number of iCOMs is detected, there is a predominance of O-bearing iCOMs, seven detected, against only three N- and and one S- bearing iCOMs (Tab. \ref{tab:sec6-table-iCOMs}).
In general, the abundances of iCOMs other than methanol are similar, $\sim 10^{-8}$, within a factor of 3, and represent about 2--5\% of the methanol abundance. 
Puzzlingly, relatively large iCOMs such as acetone, ethylene glycol, acrylonitrile and ethyl cyanide have not been detected so far.
Note that, on average, iCOM abundances in L1157-B1\index[obj]{L1157-B1} are higher than those measured in  hot corinos\index{Hot corinos} by a factor 2-10, depending on the species, with the possible exception of dimethyl ether \citep{Lefloch2017}. 
\citet{Arce2008} were the first to notice that the iCOM abundances relative to methanol are similar to those found in massive hot cores and molecular clouds in the Galactic Center region.


Finally, \citet{Holdship2019} analysed the iCOM line profiles of a sample of eight molecular outflow\index{Molecular outflows} shocks\index{Shocks}.
They found an approximately constant abundance ratio of methanol over acetaldehyde in the lower-velocity gas and suggested that this is in favor of a grain surface scenario for acetaldehyde formation.
However, we note that, if acetaldehyde is formed from ethanol \citep{Vazart2020}, which is probably a grain-surface product \citep{Perrero2022-ethanol}, a constant abundance ratio [CH$_3$OH]/[CH$_3$CHO] would likely result as well.
\citet{Holdship2019} also found evidence for the destruction of methanol and acetaldehyde in the post-shock\index{Shocks} gas of fast outflows\index{Molecular outflows} \citep[see also][]{Suutarinen2014}. 
We caution, though, that these results, based on the line profile analysis, need to be confirmed by high-spatial-resolution maps of shocked\index{Shocks} regions.

\begin{figure*}[bt]
    \centering
    \includegraphics[width=16cm]{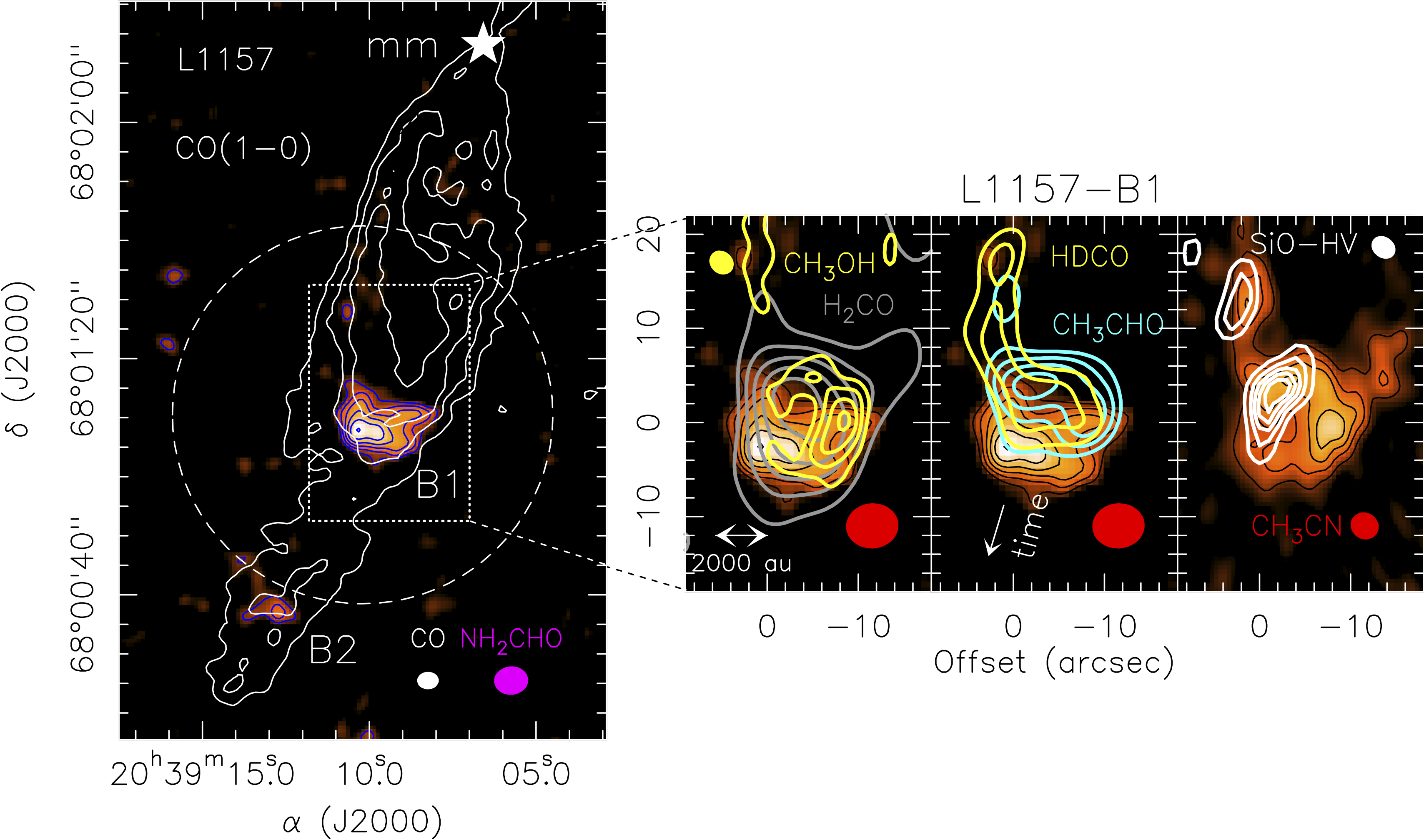}
    \caption{Images of the L1157 southern outflow lobe, showing a chemical segregation in L1157-B1.
    {\it Left panel:} The map of the CO (1-0) line emission \citep[white contours][]{Gueth1996} shows the morphology of the outflowing gas. 
    The (precessing) jet ejected by the central object L1157-mm excavated several clumpy cavities, the brightest one being B1. 
    The emission map of the formamide line 4$_{\rm 1,4}$–-3$_{\rm 1,3}$ is shown in colour \citep{Codella2017}. 
    The dashed circle indicates the primary beam of the NH$_2$CHO image. 
    The magenta and white ellipses depict the synthesised beams of the CO (white) and formamide (magenta) observations.
    {\it Right panels:} The formamide spatial distribution is compared with those of methanol \citep{Codella2020} and formaldehyde \citep{Benedettini2013}(left), HDCO \citep{Fontani2014} and acetaldehyde \citep{Codella2015,Codella2017} (middle), high-velocity SiO \citep{Gueth1998} and methyl cyanide \citep{Codella2009} (right). 
    The colour images refer to formamide for the left and middle panels and to methyl cyanide for the rightmost one. 
    Red circles show the colour images synthetic beams. 
    The beams of the HDCO, H$_2$CO, and CH$_3$CHO observations are equal to that of formamide.
    Time increases and chemistry evolves going from North to South. 
    The youngest shocked\index{Shocks} region (within L1157-B1) is that imaged by the SiO emission at high velocity.}
      \label{fig:sec6-l1157-B1}
\end{figure*}

\subsection{The post-shock gas chemical structure}\label{subsec:sec6-structure}

It has recently become possible with radio to mm interferometers to resolve the molecular gas emission down to spatial scales of a few 100 au or less in the nearest star forming regions.
This instrumental progress has allowed to peer into the time-dependent chemical structure of the shocked\index{Shocks} gas and to set stringent constraints on the possible iCOM formation pathway(s) at work in shocked\index{Shocks} regions. 

As already mentioned, the best characterized source is the L1157-B1\index[obj]{L1157-B1} shock\index{Shocks} region, in which several iCOMs and closely related species were mapped at sufficiently high angular resolution to constrain their possible formation pathways (Fig. \ref{fig:sec6-l1157-B1} and Table \ref{tab:sec6-table-iCOMs}). 
Dynamical modelling of the outflow\index{Molecular outflow} and its precessing jet has provided a detailed picture of the mass-ejection history and the associated shock\index{Shocks} regions, which trace the jet impact against the outflow\index{Molecular outflow} cavity \citep{Gueth1996,Podio2016}. 
Based on these studies, the kinematical age of the L1157-B1\index[obj]{L1157-B1} region is $\sim 1000-2000$ yr.
Figure \ref{fig:sec6-l1157-B1} shows the chemical differentiation between formamide and acetaldehyde, observed towards L1157-B1\index[obj]{L1157-B1} the NOEMA\index{NOEMA} SOLIS Large Project.
A key to interpret the L1157-B1\index[obj]{L1157-B1} observed chemical structure is SiO, expected to form in high-velocity ($\geq$ 10 km s$^{-1}$) shocks\index{Shocks}, which can sputter dust mantles\index{Mantle evaporation} as well as part of the refractory core \citep[e.g.][]{Guillet2011}.
The high-velocity SiO emission identifies the region where the youngest shock\index{Shocks} events in B1 occurred \citep{Gueth1998}. 
Therefore, time increases and chemistry evolves going from North to South in Fig. \ref{fig:sec6-l1157-B1}. 
Three main classes of organic molecules can be identified, as follows.

\vspace{0.1cm}
\noindent
{\em 1. Tracers of the entire bow shock\index{Shocks} structure:}
The line emission of simple organic species such as H$_2$CO, HNCO or HC$_3$N encompasses the whole outflow\index{Molecular outflow} lobe, including the bow shock\index{Shocks} structure of B1 \citep[e.g.][]{Benedettini2013}. 
These species are also relatively abundant in molecular clouds so that the enhancement of their emission in correspondence to the bow shock\index{Shocks} is due to the combination of two effects: (i) the increase of the density and temperature of the shocked\index{Shocks} gas and (ii) the increase of their abundances caused by gas-phase reactions with temperature barriers which become accessible in the warm shocked\index{Shocks} gas \citep{Burkhardt2016,Mendoza2018}.

\vspace{0.1cm}
\noindent
\textit{2. Tracers of freshly shocked\index{Shocks} material:}
Species such as methanol \citep[e.g.][]{Codella2020} and the deuterated\index{Deuteration} isotopologue of formaldehyde \citep[e.g. HDCO][]{Fontani2014} trace only the youngest shocked\index{Shocks} regions of B1.
They are (prevalently) formed on the icy mantles\index{Mantle formation} of dust grains in the initial, quiescent pre-shock\index{Shocks} gas of the natal molecular cloud (see \S ~\ref{subsec:sec2-mantle-form}), and subsequently released in the initial phase of velocity decoupling between the neutral and ionic fluids. 
They trace the shocked\index{Shocks!chemistry} material freshly injected into the gas-phase and are primarily located at the head of the bow shock\index{Shocks}. 
Interestingly, acetaldehyde and methyl cyanide show a spatial distribution similar to that of methanol and HDCO \citep{Codella2009,Codella2015,Codella2020}, so that they are likely tracers of freshly shocked\index{Shocks!chemistry} gas too.

\vspace{0.1cm}
\noindent
{\em 3. Tracers of chemically evolved shocked\index{Shocks} material:}
Formamide represents a peculiar case: it is detected downstream of the freshly shocked\index{Shocks!chemistry} material region in L1157-B1\index[obj]{L1157-B1} and the high-temperature peak region revealed by the high-velocity SiO \citep{Codella2017}. 
Note that the anti-correlation between formamide and acetaldehyde is
not due to excitation effects; both species are detected through low-lying upper level energy transitions (13-26 K).
When considering the sequence of ejection events (see the time arrow in Fig. \ref{fig:sec6-l1157-B1}), this implies that formamide is a tracer of chemically more evolved shocked\index{Shocks!chemistry} material.
We will show in the next subsection that the segregation between formamide and acetaldehyde provides a crucial information that helps to constrain the route of formation of formamide.

\vspace{0.1cm}
The above structure refers to the only shocked\index{Shocks} region so far investigated in detail, L1157-B1\index[obj]{L1157-B1}. 
Even in this case, only a few iCOMs have been studied with mm interferometers until now and a lot remains to be done. 
Several iCOMs detected with single-dish surveys are still not imaged with interferometers.
As we will discuss in the next subsection, having more imaged iCOMs and understanding whether they are associated with fresh or evolved shocked\index{Shocks!chemistry} material would provide stringent tests to current iCOM formation models.  

Furthermore, it is time to have new "chemical laboratories" in addition to L1157-B1\index[obj]{L1157-B1}.
This is mandatory to carry on the search for iCOMs in shocked\index{Shocks} regions as well as to confirm on a statistical basis what has been found so far. 
Recently, \citet{Desimone2020b} imaged the two
outflows\index{Molecular outflows} driven by the 
NGC1333-IRAS4A\index[obj]{NGC1333-IRAS4A} binary system in the context of the NOEMA\index{NOEMA} Large Program SOLIS \citep{Ceccarelli2017} reporting, beside methanol, the detection of acetaldehyde, dimethyl ether, and formamide.

\begin{figure*}[tb]
    \centering
\includegraphics[width=17cm]{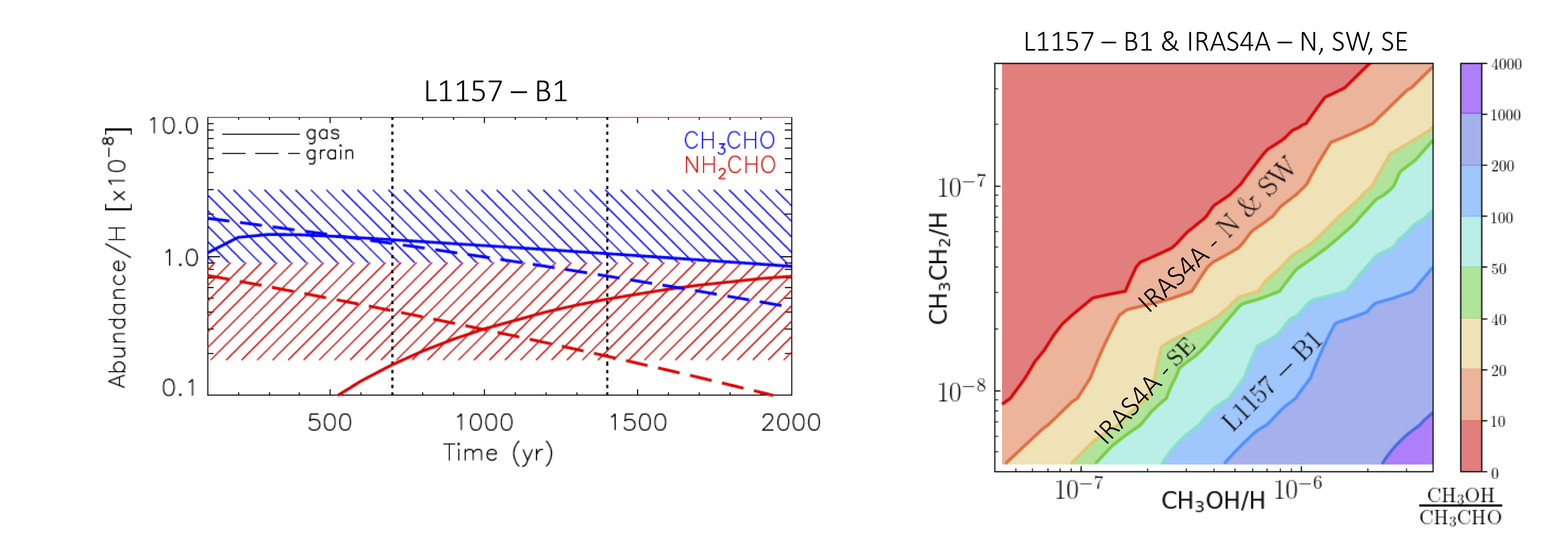}
    \caption{Astrochemical model predictions compared with the SOLIS interferometric observations of CH$_3$OH, CH$_3$CHO, and NH$_2$CHO of the shocked\index{Shocks!chemistry} regions in the L1157-B1\index[obj]{L1157-B1} and the two NGC1333-IRAS4A\index[obj]{NGC1333-IRAS4A} outflow\index{Molecular outflows}.  
    {\it Left panel:} Model predictions of acetaldehyde and formamide abundance as a function of the time after the shock\index{Shocks!chemistry} passage in L1157-B1\index[obj]{L1157-B1} \citep[from][]{Codella2017}. 
    Solid lines refer to a model in which both species are synthesised in the gas-phase, while dashed lines assume the species are injected into the gas-phase directly from the grain mantles\index{Mantle evaporation}. 
    The dashed blue and red regions show the measured abundances from the maps shown in Fig. \ref{fig:sec6-l1157-B1}.
    {\it Right panel:} [CH$_3$OH]/[CH$_3$CHO] abundance ratio at 1000 yr after the passage of a shock\index{Shocks!chemistry} as a function of the injected abundances of methanol and ethyl radical CH$_3$CH$_2$ (assumed to be the parent molecule of CH$_3$CHO), respectively \citep{Desimone2020b}. 
    The [CH$_3$OH]/[CH$_3$CHO] ratios measured towards three positions of the NGC1333-IRAS4A\index[obj]{NGC1333-IRAS4A} outflows\index{Molecular outflows} \citep[N and SW in orange, and NE in green][]{Desimone2020b}, and towards L1157-B1 \citep[blue;][]{Codella2020} are reported.}
    \label{fig:sec6-models}
\end{figure*}

\subsection{Chemical Modelling}\label{subsec:sec6-modeling}

While it is established that methanol forms from frozen CO  hydrogenation on the dust grain icy surfaces \citep{Watanabe2002,Hidaka2008} (see also \S ~\ref{subsec:sec2-mantle-form}), the cases of formamide and acetaldehyde are less clear, as for both species routes of formation in the gas-phase and on the grain surfaces have been evoked in the literature.
Specifically, formamide can be the product of the gas-phase reaction NH$_2$ + H$_2$CO \citep{Barone2015,Skouteris2017} or the grain-surface reaction HCO + NH$_2$ \citep{Garrod2008} (see the discussion in Sec. \ref{sec2:chemistry}).
Likewise, acetaldehyde can be the product of the gas-phase O + CH$_3$CH$_2$ reaction \citep{Charnley2004} and the chain of reactions starting from ethanol \citep{Skouteris2018} or the grain-surface reaction HCO + CH$_3$ \citep{Garrod2008}.

The chemical stratification of these two species observed in L1157-B1\index[obj]{L1157-B1}, which corresponds to a chemical evolution (see \S ~\ref{subsec:sec6-structure}), can be compared with the predicted time-dependent abundance of these species and, hence, can constrain their formation routes.
The SOLIS observations by \cite{Codella2017} show that the [CH$_3$CHO]/[NH$_2$CHO] abundance ratio decreases with time (equivalent to a more distant position from the central star: see Fig. \ref{fig:sec6-l1157-B1}). 
To reproduce this trend either acetaldehyde is destroyed faster than formamide or formamide takes more time to be formed.
This already rules out that both species are both sputtered from the grain mantles\index{Mantle evaporation} because they would be injected at the same time and destroyed by similar reactions with molecular ions: hence, their abundance ratio should be constant across the B1 region \citep{Codella2017}.

Figure \ref{fig:sec6-models} shows the model predictions obtained by \cite{Codella2017} assuming that formamide and acetaldehyde are formed in the gas-phase or grain surfaces.
The observed [CH$_3$CHO]/[NH$_2$CHO] decrease with time can only be reproduced if NH$_2$CHO forms in the gas-phase at later times than CH$_3$CHO. 
This is because NH$_2$ is formed in the gas-phase from ammonia, which is sputtered from the grain mantles\index{Mantle evaporation}, and the conversion takes time.
With a different and more complete modeling, \citet{Burkhardt2019} reached the same conclusion, i.e. that formamide attains high abundances in the post shock\index{Shocks!chemistry} phase at the B1 kinematical age ($\sim 10^3$ yr) only if formed in the gas-phase.

On the contrary, so far observations towards L1157-B1\index[obj]{L1157-B1} have been unable to put constrain to the formation routes, grain-surface and gas-phase chemistry, of acetaldehyde.
\cite{Codella2020} observed a substantial overlap between the spatial distribution of acetaldehyde and methanol towards L1157-B1\index[obj]{L1157-B1} (Fig. \ref{fig:sec6-l1157-B1}).
Using methanol as reference of grain-surface product sputtered in the shock, this indicates that either acetaldehyde is formed on the grain surfaces or that it is quickly ($\leq$ 10$^3$ yr) formed in the gas-phase using a parent molecule released from dust mantles\index{Mantle evaporation}.
Unfortunately, the relatively large errors in the estimates of the methanol and acetaldehyde column densities do not allow to discriminate between these two hypothesis \citep{Codella2020}.
That said, if the major formation route of acetaldehyde is the gas-phase reaction  O + CH$_3$CH$_2$ \citep{Charnley2004}, then CH$_3$CH$_2$ would be formed from C$_2$H$_6$, released from dust mantles\index{Mantle evaporation}.
Figure \ref{fig:sec6-models} shows the model predicted [CH$_3$OH]/[CH$_3$CHO] abundance ratio as a function of the methanol and the CH$_3$CH$_2$ abundance.
In order to reproduce the observed abundance ratio and the large ($\geq 10^{-6}$) methanol abundance, very high CH$_3$CH$_2$ abundances are requested, i.e. about 0.1\% of carbon would be locked into iced hydrocarbons \citep{Codella2020}. 
A similar conclusion is reached by \cite{Desimone2020b} in their study of the IRAS4A outflows\index{Molecular outflows}, also reported in Fig. \ref{fig:sec6-models}.
As for formamide, using a sophisticated modeling, \citet{Burkhardt2019} concluded that the CH$_3$CHO formation in the gas-phase is predominant in the post-shock\index{Shocks} region of L1157-B1\index[obj]{L1157-B1}.

\subsection{Future challenges}

Having seen the ability of interferometric maps of shocked\index{Shocks} regions to efficiently constrain the chemistry at work, it is essential to obtain images of more iCOMs, for example those already detected using single-dish telescopes, such as CH$_3$OCH$_3$, CH$_3$OCHO, and HCOCH$_2$OH.

However, in order to fully take advantage of the observations, a better understanding of the iCOM formation and destruction routes in shocks\index{Shocks} is mandatory. 
This requires a detailed description of both the structure and chemistry of the shocked\index{Shocks!chemistry} dust and gas, as well as the microphysics entering into the chemical process (Sec. \ref{sec2:chemistry}).
Different approximations are usually made regarding the dust grain surface chemistry and reactivity, from considering only dust sputtering but neglecting the shock\index{Shocks} impact on the dust structure, to modelling the dust behaviour (heating, reactivity, condensation) across the shock\index{Shocks}. 
The knowledge of the initial conditions in the pre-shock\index{Shocks} phase is also essential, given that it determines both the abundances of iCOMs present on the mantles\index{Mantle formation} and their precursors, which come into play once released into the gas-phase. 

Species produced by H atom addition reactions, such as CH$_3$OH, are predicted to be abundant in icy mantles\index{Mantle formation} in the pre-shock\index{Shocks} gas. 
Conversely, the formation of iCOMs on the grain surfaces via radical-radical combination \citep{Garrod2008} would have a very low efficiency at low dust temperature (\S \ref{subsec:sec2-onlygrainchem}), so that no iCOM is expected to be formed in this way. 
Likewise, the non-diffusive iCOM formation route proposed by \citet{Jin2020} would also be inefficient, as it needs a prevalently CO covered surface to take place (\S \ref{subsec:sec2-onlygrainchem}), while the shocked\index{Shocks} regions are usually far away from the centers and, therefore, the mantles are rather CO-poor.
To support this point, the methanol abundance is a small fraction ($\sim 1$\%) of the CO one in those shocked\index{Shocks!chemistry} regions.
On the other hand, at high temperatures, it appears that some species released from grain mantles are chemically processed in the hot post-shock\index{Shocks!chemistry} gas. 
This is well illustrated by formamide, whose abundance is predicted to efficiently increase in the post-shock\index{Shocks!chemistry} region, in agreement with the L1157-B1\index[obj]{L1157-B1} observations. 

The observation of a similar/different behaviour in other iCOMs (e.g. methyl formate) would be a good test for their formation routes.
The sensitivity to the initial gas and dust conditions, the role of the CR ionization rate and an external radiation field, the duration and elemental composition of the gas-phase, the influence of the shock\index{Shocks} parameters (density, velocity, B-field) on iCOM chemistry, and the possible O/N differentiation, also remain to be investigated. 

Finally, low-velocity shocks\index{Shocks} deserve attention as they can release molecules from the grain mantles without destroying most molecular bonds. 
They are also the best sites to test the occurrence of a short-lived dust warm-up phase before sputtering takes over, which would result from collisional heating and could induce a rapid formation of iCOMs on ice mantles. 
Sputtering efficiency in the low-velocity regime has been poorly studied and observational/experimental constraints to models are needed.

\section{\textbf{COMETS: GET THE ORGANICS OUT OF THE REFRIGERATOR}}\label{sec7:comets}
\subsection{Overview}\label{subsec:sec7-overview}
Comets\index{Comets} are among the most primitive and unprocessed bodies in the Solar System. 
They likely formed at large distances from the Sun, where volatile ices could condense and remain stable, and retain a cosmochemical record of the composition of, and the processes in, the Solar Nebula\index{Solar nebula} (aka, the Solar System protoplanetary disk\index{Protoplanetary disks}). 
They have been stored in cold, distant orbits, either in the Kuiper Belt region at 35--100 au (which is the source of short-period comets of the Jupiter family), or in the Oort cloud  at $\sim 20,000$ au (the source of long period comets). 
The primitive nature of cometary material is confirmed by, e.g., a near-solar bulk elemental composition including the light elements C and O \citep{Bardyn2017,Rubin2019}, no clear  evidence of aqueous alteration after accretion unlike asteroids \citep{Capaccioni2015,Quirico2016}, and the presence of highly volatile ices such as N$_2$ and CO, e.g. see \cite{Rubin2015} for measurements in comet 67P/Churyumov-Gerasimenko (hereafter 67P\index[obj]{67P}).

The present knowledge of the composition of cometary ices is essentially based on remote sensing and in situ investigations of cometary atmospheres. 
A little more than two dozen molecules (not including isotopologues, molecular ions, atoms and radicals) have been identified from spectroscopic investigations in the radio (20--600 GHz), near-IR (2.9--5 $\mu$m), and, to a lesser extent, UV domains.
The major volatiles are water (about 80\% by number) followed by CO$_2$, CO, CH$_3$OH, H$_2$CO, CH$_4$, H$_2$S and CH$_4$ \citep[Table~\ref{tab:sec7-comet-abundances},][]{Bocke2017,Rubin2020}. 

\subsection{iCOMs in comets}\label{subsec:sec7-icoms}
Most of the iCOMs, including methyl formate, acetaldehyde, ethylene glycol and formamide, have been first identified in the great comets\index{Comets} C/1996 B2 (Hyakutake\index[obj]{Hyakutake}) and C/1995 O1 (Hale-Bopp\index[obj]{Hale Bopp}) from emission lines in the mm range \citep{Lis1997,Bocke2000,Crovisier2004}. 
More recently, glycolaldehyde and ethanol were discovered in the atmosphere (i.e. coma) of C/2014 Q2 (Lovejoy), together with other previously identified iCOMs \citep{Biver2014,Biver2015}. 
Sensitive upper limits for iCOMs that are found in protostars (e.g., acetic acid, and dimethyl ether: Sec. \ref{sec4:protostars}) have also been derived from spectra of bright comets\index{Comets} \citep{Crovisier2004,Biver2021}.

The ESA/Rosetta mission \citep{Taylor2017} accompanied comet 67P\index[obj]{67P}, a short-period comet of the Jupiter family, for over two years along its orbit around the Sun, and provided a more complete picture of the composition of cometary ices by more than doubling the number of identified molecules \citep{Altwegg2019,Rubin2019,Rubin2020,Hanni2022-iCOMcomets}. 
The payload of the Rosetta mission included several mass spectrometers designed for gas-phase studies that all revealed a chemical complexity of the organics: COSAC (Cometary Sampling and Composition experiment) \citep{Goesmann2015} and Ptolemy \citep{Wright2015} on the Philae lander, which collected information directly from the comet surface, and DFMS (Double Focusing Mass Spectrometer) and RTOF (Reflectron-type Time-Of-Flight Mass Spectrometer) of the ROSINA instrument suite, which observed the coma throughout the two-year encounter. 
Thanks to a high mass resolution ($m$/$\Delta m$ = 3000 at 1\% peak height) and large mass range (12--150 amu), DFMS was perfectly suited for a detailed analysis of the comet's organic composition. 
Since mass spectrometers cannot unambiguously distinguish isomers as they have the same mass, the identification of large cometary iCOMs was made considering the expected fragmentation pattern inside the instrument and the detection of the daughter species in the mass spectra. 
However, unequivocal isomer identification was often difficult \citep{Rubin2019,Schuhmann2019a}. 
For example, the abundances of methyl formate and glycolaldehyde (which have the chemical formula C$_2$H$_4$O$_2$ and, hence, the same mass 60 Da) could not be measured from DFMS mass spectra \citep{Schuhmann2019a}. 
Their isomer, acetic acid, best explains the fragmentation pattern observed in DFMS mass spectra acquired in May 2015 \citep{Schuhmann2019a} but, in spectra acquired in September 2016, most of the C$_2$H$_4$O$_2$ 60 u/e peak was assigned to glycolaldehyde \citep{Altwegg2020}. 
Another example is the detected peak at 62 u/e, which might be a combination of ethylene glycol (detected in several comets\index{Comets} by spectroscopy) and methoxymethanol (CH$_3$OCH$_2$OH), both observed in protostellar sources, including  the Solar-type Class 0\index{Class 0 sources} IRAS 16293-2422 source \citep{Manigand2020}.

\begin{table}
\caption{Abundances relative to H$_2$O (in \%) of C, N, O- bearing molecules identified in comets (excluding P compounds), both remotely and in situ.} \label{tab:sec7-comet-abundances}
\begin{tabular}{lcl}
\hline\noalign{\smallskip}
Molecule$^a$ & Comets$^b$ & 67P$^c$ \\
\hline
O$_2$ & -- & 3.1 $\pm$1.1 \\
CO & 0.4--35 & 3.1 $\pm$0.9 \\
CO$_2$ & 2.5--30 & 4.7 $\pm$1.4 \\
\hline
CH$_4$ & 0.12--1.5 & 0.34$\pm$0.07 \\
C$_2$H$_2$ & 0.04--0.5 & -- \\
C$_2$H$_4$  & 0.2 & --\\
C$_2$H$_6$ & 0.14--2.0 & 0.29$\pm$0.06\\
C$_3$H$_8$ & -- & 0.018$\pm$0.004\\
C$_6$H$_6$ & -- & 0.00069$\pm$0.00014 \\
C$_7$H$_8$ & -- &  0.0062$\pm$0.0012 \\
\hline\noalign{\smallskip}
CH$_3$OH & 0.7--6.1 & 0.21$\pm$0.06 \\
H$_2$CO & 0.13--1.4 &  0.32$\pm$0.10 \\
HCOOH & 0.03--0.18 & 0.013$\pm$0.008 \\
CH$_3$CHO & 0.05--0.08 & \vline~0.047$\pm$0.017 \\
c-C$_2$H$_4$O & $<$0.006& \vline~\\
CH$_2$CHOH & --&  \vline~\\\noalign{\smallskip}
CH$_2$CO & $<$0.008 & -- \\ 
HCOOCH$_3$ & 0.06--0.08 & \vline\\
CH$_2$OHCHO & 0.02--0.04 & \vline \\
CH$_3$COOH$^d$ & $<$0.03 & \vline~0.0034$\pm$0.0020 \\\noalign{\smallskip}
C$_2$H$_5$OH & 0.1--0.2 & \vline~0.039 $\pm$0.023 \\
CH$_3$OCH$_3$ & $<$0.025 & \vline \\\noalign{\smallskip}
(CH$_2$OH)$_2$ & 0.07--0.35 & \vline~0.011$\pm$0.007 \\
CH$_3$OCH$_2$OH & -- & \vline~ \\\noalign{\smallskip} 
CH$_3$(CH$_2$)$_2$CHO & -- & \vline~0.010$\pm$0.003 \\
CH$_3$CH$_2$OCH=CH$_2$ & -- & \vline~\\\noalign{\smallskip} 
CH$_3$COOCH$_3$ & -- & \vline~0.0021$\pm$0.0007 \\
CH$_3$CH$_2$COOH & -- & \vline\\
CH$_3$CCH$_2$OH & -- & \vline \\\noalign{\smallskip} 
CH$_3$CH$_2$CHO & -- & \vline~0.0047$\pm$0.0024 \\
(CH$_3$)$_2$CO & $<$0.01 & \vline\\\noalign{\smallskip} 
\hline\noalign{\smallskip}
NH$_3$ & 0.3--0.7 & 0.67$\pm$0.20 \\
N$_2$ & -- & 0.089$\pm$0.024 \\
HCN & 0.08--0.25 & \vline~0.14$\pm$0.04 \\
HNC & 0.002--0.035 &  \vline \\\noalign{\smallskip}
HNCO & 0.009--0.08 & \vline~0.027$\pm$0.016 \\
HCNO & $<$0.0016 & \vline \\\noalign{\smallskip}
CH$_3$CN & 0.008--0.054 & \vline~0.0059$\pm$0.0023 \\
CH$_3$NC & -- & \vline~\\\noalign{\smallskip}
HC$_3$N & 0.002--0.068 & 0.00040$\pm$0.00023 \\
HC$_5$N & $<$ 0.003 & --\\
NH$_2$CHO & 0.016--0.022 & 0.0040$\pm$0.0023 \\
CH$_2$NH & $<$0.03 & -- \\
CH$_3$NH$_2$ & $<$0.06 & -- \\
NH$_2$CN & $<$0.004 & -- \\
C$_2$H$_3$CN & $<$0.003& -- \\
C$_2$H$_5$CN & $<$0.004 & -- \\
\hline
\end{tabular}
\footnotesize{
NOTES: $^a$ Isomers contributing to the mass peak in DFMS mass spectra are listed.  
$^b$ Range of values and upper limits from remote sensing spectroscopic observations in a large ($>$ 50) sample of comets \citep{Bocke2017,Biver2021,Biver2022,Crovisier2004}. 
$^c$ Measurements with the DFMS/Rosetta instrument from data acquired in May 2015 at $\sim 1.5$ au from the Sun \citep{Rubin2019}. 
$^d$ Best explains the DFMS data acquired in May 2015.
}
\end{table}


\begin{figure*}
    \includegraphics[angle=0,width=7.5cm]{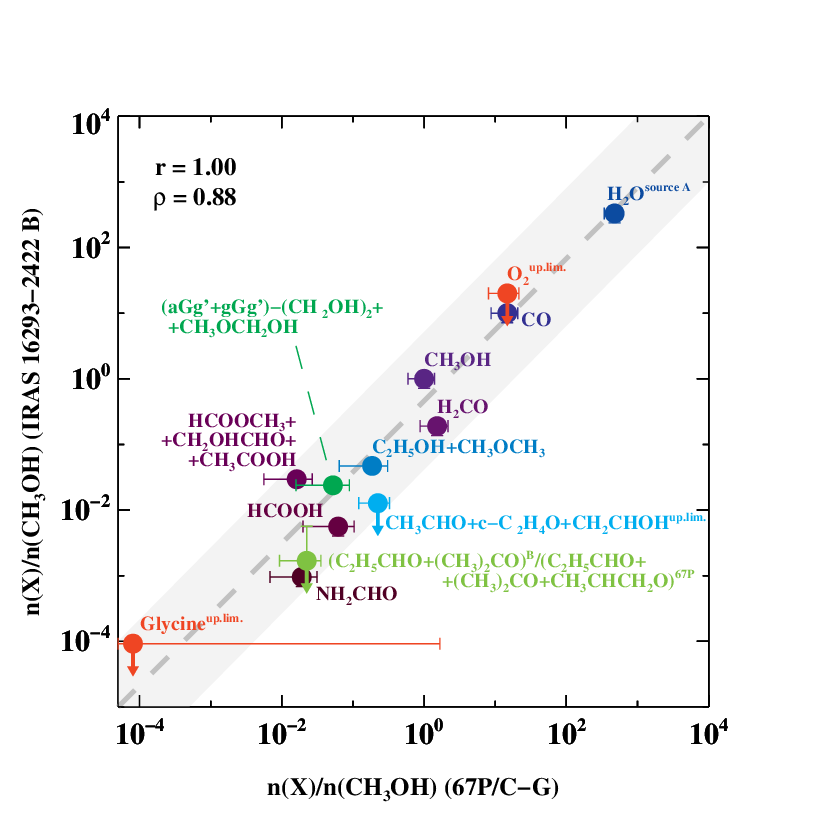}
    \includegraphics[angle=0,width=9cm]{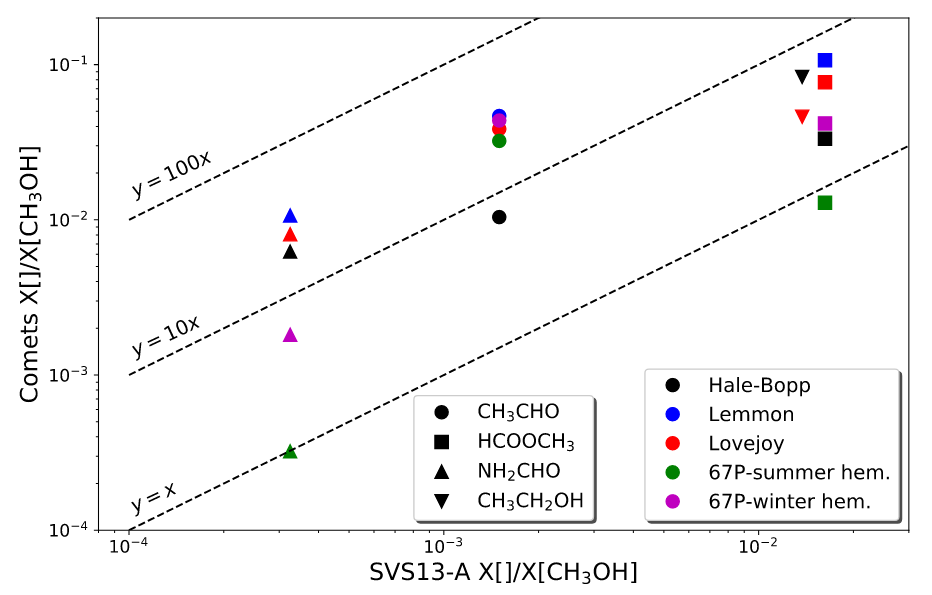}
    \caption{Comparison between the iCOM abundances, normalized to methanol, observed in comets and Class 0 (left) and Class I (right) protostars.
    \textit{Left panel}: iCOM abundances in the Class 0 protostar IRAS 16293-2422 B versus those measured in comet 67P (arrows indicate upper limits) \citep[from][]{Droz2019}.
    \textit{Right panel}: iCOM abundances in the Class I protostar SVS13A versus those in various comets \citep[from][]{Bianchi2019a}.
    }
    \label{fig:sec7-iCOMs-comets}
\end{figure*}

A list of organics identified by ROSINA/DFMS, which includes a great variety of CH-, CHN-, CHS-, CHO-, CHO$_2$-, and CHNO- bearing saturated and unsaturated species, is given by \citet{Altwegg2019} and \citet{Hanni2022-iCOMcomets}. 
Among the newly identified cometary iCOMs are compounds at mass of 58 Da, that could be acetone or propanal, or a combination of both. 
These two species have been detected in protostellar sources, with the former dominating the latter by a factor of $\sim10$ in IRAS 16293-2422 B  \citep{Lykke2017}. 
Acetone is indicated as firmly identified in comet 67P\index[obj]{67P} by \cite{Altwegg2017}. 
A peak at 72 u/e is identified as C$_4$H$_8$O, and is provisionally attributed to butanal (CH$_3$(CH$_2$)$_2$CHO) \citep{Schuhmann2019a}. 
Another peak at 74 u/e is best assigned to methyl acetate (CH$_3$COOCH$_3$) \citep{Schuhmann2019a}. 
The detected hydrocarbons include the full series of alkanes up to C$_9$H$_{20}$, and  aromatic hydrocarbons such as benzene, C$_6$H$_6$, toluene, C$_7$H$_8$, xylene, C$_8$H$_{10}$  and naphtalene, C$_{10}$H$_{8}$ \citep{Altwegg2019,Schuhmann2019b,Hanni2022-iCOMcomets}. 
Complex alcohols are detected (e. g. propanol and pentanol) as well as complex acids as benzoic acid (C$_7$H$_6$O$_2$).
Besides HNCO and NH$_2$CHO, first detected in cometary mm spectra, two other CHON compounds are identified in 67P\index[obj]{67P} mass spectra: acetamide (CH$_3$CONH$_2$), and glycine (C$_2$H$_5$NO$_2$) \citep{Altwegg2016}, with several lines of evidence that glycine is released from warm dust icy grains. 
The concentration of glycine in water ice\index{Water ice} is estimated to 170 ppb in mass \citep{Hadraoui2019}. 
Methylamine (CH$_3$NH$_2$) and ethylamine (C$_2$H$_5$NH$_2$) are seen in the mass spectra only when glycine is detected, with relative abundances with respect to glycine of 1.0$\pm$0.5 and 0.3$\pm$0.2 \citep{Altwegg2016}, respectively, suggesting a  formation pathway for glycine, e.g., via the photochemistry of methylamine and CO$_2$ ice \citep{Bossa2010}. 
Glycine and these two amines were previously detected in the dust particles collected by the {\it Stardust} sample return mission \citep{Glavin2008}. 
Finally, new organo-sulfur compounds have been identified : methyl mercaptan (CH$_3$SH), a combination of ethyl mercaptan (C$_2$H$_5$SH) and dimethyl sulfide (CH$_3$SCH$_3$), and methyl hydrogen sulfide (CH$_4$S$_2$) \citep{Calmonte2016,Altwegg2019}. 


The ROSINA/DFMS spectra collected on 3 August 2015, when comet 67P\index[obj]{67P} was just about to reach its perihelion, showed signals of C-bearing species up to $m/z$ = 140 u/e \citep{Hanni2022-iCOMcomets}. 
The DFMS likely measured the outgassing from both the nucleus and small dust particles. 
The mass spectra are dominated by the signatures of hydrocarbons, with the next dominant group being CHO-bearing species.
A plethora of chained-based, cyclic, and aromatic hydrocarbons are present, at a ratio of 6:3:1. 
The ensemble has an average composition of C$_1$O$_{0.134}$N$_{0.046}$S$_{0.017}$, not considering the abundant and major C-bearing cometary gases \citep[e.g. CO, CO$_2$][]{Hanni2022-iCOMcomets}.     


Tables \ref{tab:sec7-comet-abundances} and \ref{tab:sec7-comet-abundances-S} present a compilation of the abundances measured using spectroscopy in a large sample of comets\index{Comets} \citep{Bocke2017}. 
It also lists abundances in comet 67P's\index[obj]{67P} coma derived from DFMS May--2015 data at 1.5 au from the Sun \citep{Rubin2019}; quantification is still missing for a number of detected species. 
Abundances vary by a factor of about 3 to 100 (for CO), both in the Oort-cloud and Kuiper-Belt comet populations  \citep{DelloRusso2016,Bocke2017}. 
There is increasing evidence that this chemical diversity is primarily primitive (related to comet formation environment), and secondarily the result of evolutionary processing history after formation \citep{DelloRusso2016,A'Hearn2012,Rubin2020}. 
Whereas the majority of the detected molecules are presumably released from the nucleus ices, there are lines of evidence that some of them have significant contributions from extended sources in the coma. 
The best documented species are H$_2$CO and HNC, whose behaviors with heliocentric distance and spatial distributions suggest that they are produced by the thermal degradation of organic grains \citep{Lis2008,Cordiner2014}. 
Several ammoniated salts (NH$_4$Cl, NH$_4$CN, NH$_4$OCN,  NH$_4$HCOO, and NH$_4$CH$_3$COO) have been identified in a grain of comet 67P\index[obj]{67P} \citep{Altwegg2020}, and are present as well on the whole nucleus surface \citep{Poch2020}.  
Their sublimation in the coma should be the source of some of the detected cometary species (e.g., HCOOH, HNCO), the contribution of which is still unclear. 

Altogether, based on the coma volatile composition, CHO-bearing molecules constitute $\sim 4$\% of the volatiles (in number) in cometary ices, hydrocarbons $\sim 2$\%, N-bearing $\sim 1$\%, and S-bearing $\sim 1.5$\%. 
Saturated aliphatic and aromatic hydrocarbons are in about equal (namely 44:56) proportion \citep{Schuhmann2019a}. 
A huge fraction of the organic material is in the refractory phase (almost 50\% by weight), in the form of very large macromolecular compounds, analogous to the insoluble organic matter (IOM) found in carbonaceous chondritic meteorites \citep{Fray2016}. 
The composition and structural properties of the organics captured by ROSINA/DFMS near 67P\index[obj]{67P} perihelion resemble to meteoritic soluble organic matter (SOM) \citep{Hanni2022-iCOMcomets}. 
Finally, a large fraction of nitrogen is possibly locked in the ammonium salts \citep{Altwegg2020,Poch2020}.

\begin{table}[tb]
\caption{S-bearing abundances relative to H$_2$O in comets.}\label{tab:sec7-comet-abundances-S}
\begin{tabular}{lcl}
\hline\noalign{\smallskip}
Molecule$^a$ & Comets$^b$ & 67P$^c$ \\
\hline
H$_2$S & 0.09--1.5 & 0.11$\pm$0.046 \\
OCS & 0.05--0.40 & 0.041$^{+0.082}_{-0.020}$ \\
CS & 0.02--0.20 & -- \\
CS$_2$ & -- & 0.0057$^{+0.0114}_{-0.0028}$ \\
SO & 0.04--0.30 & 0.071$^{+0.142}_{-0.037}$  \\
SO$_2$ & 0.03-0.23 & 0.127$^{+0.254}_{-0.064}$ \\
S$_2$ & 0.001-0.25 & 0.002$^{+0.004}_{-0.001}$ \\
H$_2$CS & 0.009-0.09 & 0.0027$^{+0.0058}_{-0.0024}$  \\
CH$_3$SH & $<$0.023  & 0.038$^{+0.079}_{-0.028}$ \\
C$_2$H$_5$SH & -- & \vline~0.00058$^{+0.00123}_{-0.00049}$  \\
(CH$_3$)$_2$S & -- & \vline \\\noalign{\smallskip}\
NS & 0.006--0.012 & --\\
(CH$_3$)$_2$S & -- & \vline \\\noalign{\smallskip}\
NS & 0.006--0.012 & --\\
\hline
\end{tabular}

\footnotesize{$^{a,b,c}$ see caption to Table~\ref{tab:sec7-comet-abundances}.}
\end{table}

\subsection{Cometary ices and protostellar environments}

\begin{figure*}[tb]
\vspace{-0.5cm}
    \centering
    \includegraphics[angle=0,width=10cm]{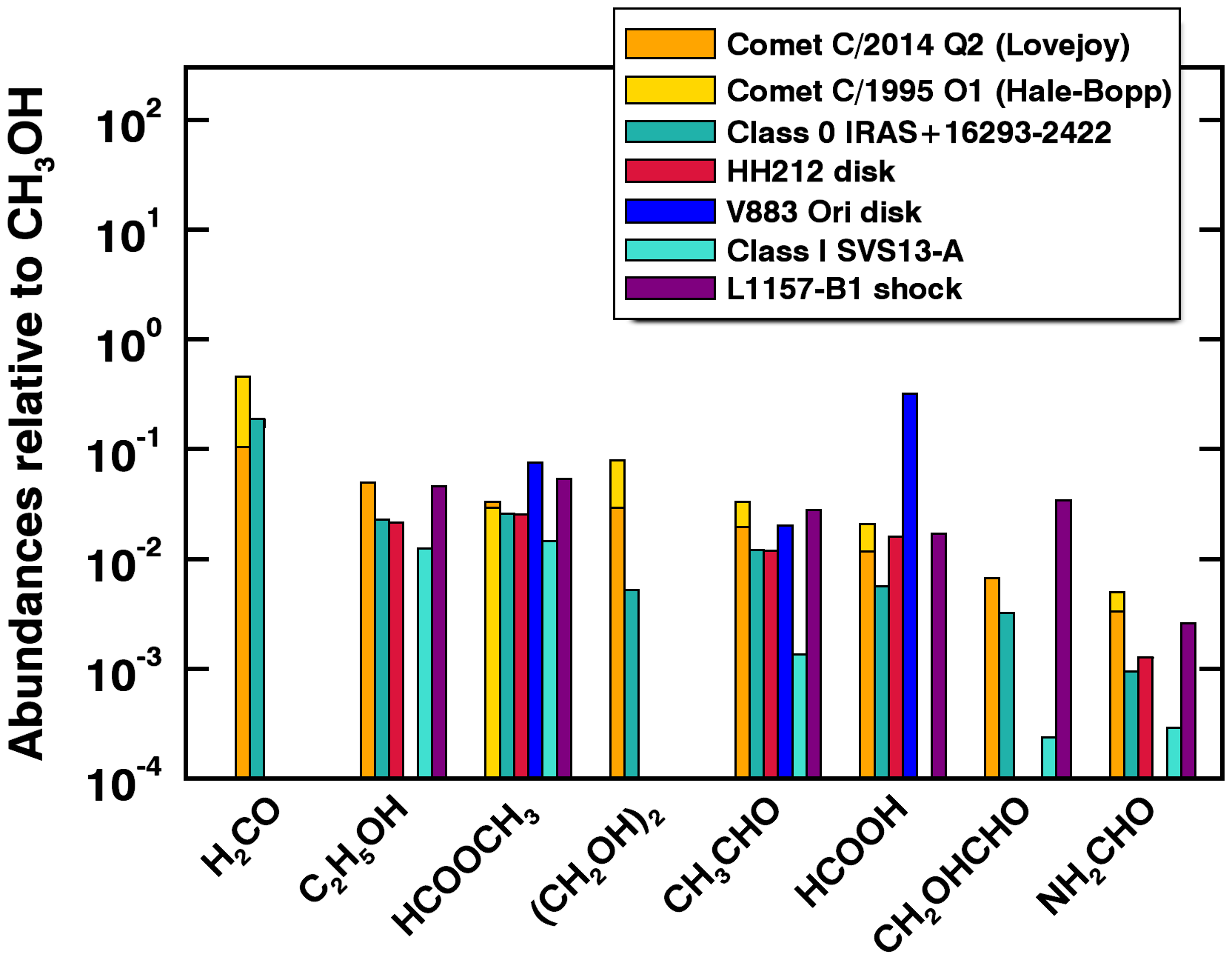}
    \caption{Comparison of iCOMs abundances in comets with respect to Solar-type protostellar objects. Values are taken from  \citet{Bocke2000} (comet Hale-Bopp), \citet{Biver2014,Biver2015} (comet Lovejoy), \citet{Lefloch2017} (L1157),  \citet{Droz2019} (IRAS 16293-2422 B), \citet{Bianchi2019a} (SVS13-A\index[obj]{SVS13-A}), \citet{Lee2019} (HH 212), and \citet{LeeJ2019-v833} (V883 Ori).}
    \label{fig:sec7-comet-ISMsources}
\end{figure*}  

The similarity of cometary ices with interstellar ices and protostellar environments was first highlighted in the early 2000s. 
\citet{Ehrenfreund2000} made comparisons with simple interstellar ices observed with the {\it Infrared Space Observatory} (ISO) and concluded that cometary ices are a mixture of original interstellar material and material that has been processed to some extent in the Solar Nebula\index{Solar nebula}. 
\citet{Bocke2000} found a strong correlation between Hale-Bopp's\index[obj]{Hale Bopp} abundances of CHO-- and N-bearing molecules and those measured in hot cores and bipolar flows, and concluded that similar chemical processes must be at work forming these compounds under comparable conditions, perhaps gas-grain chemistry. 
A similar conclusion was obtained using early  DFMS data of comet 67P\index[obj]{67P} (at 3 au from the Sun) from \cite{LeRoy2015} and comparing with the same ISM abundances as used by \citet{Bocke2000}.
A detailed comparison of comet 67P\index[obj]{67P} (abundances from Tables~\ref{tab:sec7-comet-abundances}--\ref{tab:sec7-comet-abundances-S}) with the Solar-type young Class 0\index{Class 0 sources} protostar IRAS 16293-2422 B was recently performed by \citet{Droz2019}, and shows a close match for CHO-- (Fig.~\ref{fig:sec7-iCOMs-comets}) and N-bearing molecules, as well as for the S-bearing species albeit with a significant scatter. 
Likewise, \cite{Bianchi2019a} found a good agreement between the formamide and methyl formate abundance in the Class I\index{Class I sources} protostar SVS13-A\index[obj]{SVS13-A} and the 67P\index[obj]{67P} ices, within a factor 10.
On the contrary, the comparison with other comets\index{Comets} shows larger difference, in particular for the acetaldehyde and formamide (Fig.~\ref{fig:sec7-iCOMs-comets}).

In Fig.~\ref{fig:sec7-comet-ISMsources}, we summarise the comparison of iCOMs abundances measured in the Oort-cloud comets\index{Comets} C/1995 O1 (Hale-Bopp\index[obj]{Hale Bopp}) and C/2014 Q2 (Lovejoy) with several low mass protostellar objects. 
In addition to IRAS 16293-2422 B, an instructive comparison with the Class 0\index{Class 0 sources} stage can be done with HH 212\index[obj]{HH 212-mm}.  
\citet{Lee2019} and \citet{Lee2022-h212} imaged several iCOMs (e.g. CH$_3$OH, CH$_3$CHO, HCOOCH$_3$, NH$_2$CHO and CH$_3$CH$_2$OH) associated with the rotating accreting disk within $\sim 40$ au from the protostar. 
Figure \ref{fig:sec7-comet-ISMsources} shows that the abundance ratios 
of iCOMs, with respect to methanol, of HH 212\index[obj]{HH 212-mm} and the comets\index{Comets} are, within a factor of a few, consistent.

FU Orionis\index[obj]{Fu Ori} (FU Ori) stars are young stellar objects (\S ~\ref{subsec:sec5-fuori}) that undergo rapid increases of accretion rate, and high luminosity enhancements shifting the snow lines\index{Snow line} to much larger radii. 
They provide a unique and direct probe to study the composition of material freshly sublimated from ices located in the disk midplane. 
Several iCOMs (CH$_3$OH, HCOOH, CH$_3$CHO, HCOOCH$_3$, CH$_3$COCH$_3$, c-H$_2$COCH$_2$, CH$_3$CN, CH$_3$SH) have been detected in the FUor V883 Ori \citep{LeeJ2019-v833}, whose water snow line\index{Snow line} in the disk mid-plane has been estimated to be located at $\sim 40$ au \citep{Cieza2016}. 
The estimated abundances of acetaldehyde and methyl formate agree with those measured in comets\index{Comets} (Fig.~\ref{fig:sec7-comet-ISMsources}). 
The abundance of acetone, relative to methanol, [CH$_3$COCH$_3$]/[CH$_3$OH[ = 0.014 \citep{LeeJ2019-v833}, is comparable with the value measured in comet 67P\index[obj]{67P} (0.022, Table~\ref{tab:sec7-comet-abundances}). 
However, formic acid is more abundant in V883 Ori than in comets\index{Comets}, but the identification in V883 Ori is considered as tentative \citep{LeeJ2019-v833}. 

The comparison of the cometary ices with the chemical composition observed in the shocked\index{Shocks!chemistry} regions induced by protostellar jets\index{Jets} is, in principle, very instructive.
Indeed, in these regions, the chemical content of the gas is enriched with the grain mantle components injected into the gas-phase because of  the sputtering and shattering of the dust (both volatile mantles and refractory cores: see Sec. \ref{sec6:outflows}). 
Thus, Fig. \ref{fig:sec7-comet-ISMsources} reports also the iCOM content of the prototypical L1157-B1\index[obj]{L1157-B1} shocked\index{Shocks} region (\S ~\ref{subsec:sec6-iCOMs}) showing a good agreement with what is obtained in comets\index{Comets}.

In summary, these findings suggest a limited chemical evolution of the iCOMs content, when considering relative abundances, during the evolution from the Class 0\index{Class 0 sources}/I\index{Class I sources} protostellar phases to comets\index{Comets}. 
It is tempting to conclude that the chemical complexity in the regions where planets form has been significantly inherited  from the earliest star forming stages. 
However, this speculation is based on a limited number of objects and needs to be supported by the measurements of a statistically significant number of protostars\index{Protostars}, imaged on Solar System scales (as in the HH 212\index[obj]{HH 212-mm} case: \S ~\ref{subsec:sec5-class0}), as well as of comets\index{Comets}. 

As final remark, it is important to stress that the mere presence of iCOMs in cometary ices is not an evidence of their formation on the grain surfaces (Sec. \ref{sec2:chemistry}) as, even if formed in the gas-phase, iCOMs would anyway freeze-out onto the grain surfaces in the cold regions of the protoplanetary disk\index{Protoplanetary disks} of the Solar Nebula\index{Solar nebula} (see Sec. \ref{sec5:disks} and, more specifically, \S ~\ref{subsec:sec5-model}) and then they would be found frozen in the cometary ices.

\subsection{D fractionation in iCOMs}
It has been shown that molecular deuteration\index{Deuteration} is a good proxy to link the various evolutionary stages of a Solar-type star and planet formation and the small bodies of the Solar System, including comets\index{Comets} \citep[e.g.][and see also the Chapter by Nomura et al.]{Ceccarelli2014-PP6}.

In comets\index{Comets}, only deuterated\index{Deuteration} methanol has been so far detected.
Specifically, D-methanol (CH$_3$OD and CH$_2$DOH combined) and D$_2$-methanol (CH$_2$DOD and CHD$_2$OH combined) have been detected in ROSINA mass spectra of comet 67P\index[obj]{67P}, with abundances of 5.5$\pm$0.5\% and 0.069$\pm$0.014\% relative to normal methanol, respectively \citep{Droz2021}. 
The cometary D-methanol abundance is consistent with the values measured with interferometers in Class 0\index{Class 0 sources} and I\index{Class I sources} hot corinos\index{Hot corinos} \citep{Bianchi2017b,Bianchi2020,Jorgensen2018,Jacobsen2019,Manigand2020,vanGelder2020}, supporting the inheritance scenario also for methanol deuteration\index{Deuteration}. 
Unfortunately, a meaningful comparison between D$_2$-bearing cometary methanol and existing ISM values \citep{Bianchi2017a} is currently not feasible \citep{Droz2021}, because of the various uncertainties in the derivation of the column densities, the major one being linked to spectroscopic data.
The D/H ratio has also been measured in the refractory organic material of 67P\index[obj]{67P} \citep{Paquette2021}. 
It is one order of magnitude lower than the D/H ratio in cometary methanol, but much higher than the D/H protosolar value and also higher than the bulk value in primitive chondritic IOM, suggesting at least partial inheritance from the pre-solar stage \citep{Paquette2021}. 
The macromolecular organics in comets\index{Comets} and carbonaceous meteorites possibly formed from the UV irradiation of ice mixtures, but the exact formation mechanism is unknown.

\section{\textbf{CONCLUSIONS: CONNECTING THE DOTS TO DISCOVER OUR ORIGINS}}\label{sec8:conclusions}


Organic chemistry has a very special role in Chemistry, because it lies at the basis of terrestrial life.
It starts with small molecules and reaches the complexity that makes a living being.
This is chemistry, the bonding of atoms in structures, in a seemingly infinite number of combinations.
Yet, they are not random combinations, because all the chemical structures, from the smallest to the most complex ones, just follow the laws of Physics.
Everything is connected to and is a consequence of the laws of Physics, hence also life.
Take the three most abundant and critical elements for terrestrial life: hydrogen, oxygen and carbon.
The first two make water molecules, while the last is the basis of organic chemistry.
It cannot be a pure chance that H, O and C are also the three most abundant elements in the Solar neighborhood, when excluding He which does not participate in chemistry.
Life, thus, used what was available in the largest quantities. 

Water is known to be very abundant in Solar-type star forming regions \citep[e.g.][]{whittet1988,Ceccarelli1999,Cernicharo2005,vandishoeck2021}, exactly because O and H are abundant and H$_2$O is easily formed on the grain mantles \citep[e.g.][]{tielens1982,oba2009,Mokrane2009,dulieu2010,molpeceres2019}.
Organic chemistry is now known to be active and rich in Solar-type star forming regions, as described throughout this Chapter.
In fact, all interstellar molecules with at least six atoms are organic.
Again, this is due to the laws of Physics, specifically the electronic structure of C, which gives the largest number of possible bonds among the most abundant elements synthesised in the stellar nucleosynthesis.
Silicon would also be potentially a large-molecule maker but, contrary to C, it is mostly trapped into the refractory dust grains and, hence, not available for making molecules, in addition to making much weaker chemical bonds than carbon.

In this Chapter, we have shown that organic chemistry starts to be complex very early in the formation of a Solar-type planetary system, already at the starless core\index{Starless cores} phase, when gravitational collapse has not even started (Sec. \ref{sec3:psc}).
It then progresses at its highest level of known richness during the protostellar phase, in the hot corinos\index{Hot corinos}, which have sizes comparable to the planet forming regions (Sec. \ref{sec4:protostars}).
Although more difficult to observe for practical reasons, it is very likely that the same level of organic complexity, if not higher, is also present in protoplanetary disks\index{Protoplanetary disks}, where planets, asteroids and comets\index{Comets} form (Sec. \ref{sec5:disks}).
We can study the processes responsible for this organic richness better in molecular outflow\index{Molecular outflow} shocks\index{Shocks}, because they provide the precious additional constraint of temporal chemical evolution (Sec. \ref{sec6:outflows}).
As a matter of fact, we find a similar organic chemical composition, within a rough factor of 10, of comets\index{Comets} and various objects during the Solar-type star formation process (Sec. \ref{sec7:comets}).

With the advent of larger or more sensitive facilities the "interstellar organic world" continues to expand and, at present, the limit to new discoveries comes from the lack of spectroscopic data \cite[e.g.][]{Cernicharo2021a}.
However, whatever the progress will be and whatever complex organic molecule will be discovered, there will always be a limit to which molecule we can identify from astronomical observations.
This limit exists because of the intrinsic problem that the larger the molecule the more numerous and, consequently, the weaker the lines.
This will lead to a point where any spectrum will consist of a "grass" of faint lines, where the line coincidence criterion for identifying a species will fail.
Therefore, in order to know the ultimate organic complexity reachable during the Solar-type star forming process, we need a reliable theory for the iCOM formation.
As very briefly discussed in Sec. \ref{sec2:chemistry} and throughout the following sections, we are far from being in that position.
Many more interdisciplinary studies are necessary, where laboratory experiments and quantum chemical computations need to converge to understand the micro-physics processes.
Only then can they reliably be incorporated into astrochemical models and the astronomical observations compared with model predictions (see for example the interdisciplinary projects DOC -the Dawn of Organic Chemistry\footnote{\url{https://doc.osug.fr/}}- and ACO -AstroChemical Origins\footnote{\url{www.aco-itn.org}}).
This is the future for  organic astrochemistry and there are no shortcuts.
Eventually, only when astrochemical models can be trusted in predicting the existence of molecules that we will never be able to detect, will we be in a position to know the ultimate organic complexity reached by our Solar System progenitor, and whether it may have had a role in the emergence of life on Earth.

\bigskip

\noindent\textbf{Acknowledgments} 
This project has received funding within the European Union’s Horizon 2020 research and innovation programme from the European Research Council (ERC) for the project “The Dawn of Organic Chemistry” (DOC), grant agreement No 741002, and from the Marie Sklodowska-Curie for the project ”Astro-Chemical Origins” (ACO), grant agreement No 811312.
We wish to thank Prof. P. Ugliengo, Dr. A. Rimola, Dr. J. Enrique-Romero, Dr. S. Pantaleone, L. Tinacci and Prof. Gretobape for the numerous and stimulating discussions on  interstellar grain-surface chemistry.
We also acknowledge the extremely useful discussions with Dr. A. Lop\'ez-Sepulcre, Dr. L. Podio, Prof. S. Viti, and the invaluable contribution of Dr. E. Bianchi, Dr. M. Bouvier and Dr. M. De Simone.

\bigskip

\bigskip
\bibliographystyle{pp7}
\bibliography{PP7-Chapter11.bib}

\end{document}